# Median Based Unit Weibull (MBUW) A New Unit Distribution: Properties with New Estimation Methods Generalized Methods of Moments and Percentile Estimator

Iman M. Attia *

Imanattiathesis1972@gmail.com ,imanattia1972@gmail.com

**Department of Mathematical Statistics, Faculty of Graduate Studies for Statistical Research, Cairo University, Egypt*

**Abstract: This paper introduces a new two-parameter unit Weibull distribution defined on the interval (0,1). It discusses the methodology for deriving its probability density function (PDF), explores various properties, and presents related functions. Numerous figures illustrate the distribution and demonstrate its effectiveness in fitting a wide range of skewed real data. The parameter estimation process using maximum likelihood estimation (MLE) faced challenges, particularly concerning large variance. To address these issues, the generalized method of moments (GMMs) and percentile estimators are introduced as improved alternatives. The paper elaborates on GMMs, percentile estimators, and new variants of both methods for estimating the parameters of the new distribution, supplemented by illustrative analyses of real data.**

# Introduction:

Waloddi Weibull (1951) was the first to introduce the Weibull distribution, a well-known model for life data and reliability. This distribution can describe cases with increasing and decreasing failure rates. The exponential distribution is a special case of the Weibull distribution when the shape parameter is one, while the Rayleigh distribution is another special case when the shape parameter is two. The Weibull distribution can also be used to assess life expectancy in scenarios involving fatigue-derived failure and to evaluate the reliability of electron tubes and load-handling machines. Its application spans various fields including medicine, physics, engineering, biology, and quality control.

Due to the distribution's inability to adequately represent bathtub or unimodal shapes, many researchers have sought to generalize and transform it in recent decades. Notable contributions include (Singla et al., 2012) on beta-generalized Weibull, (Khan et al., 2017) on transmuted Weibull, (Xie et al., 2002) on modified Weibull, (Lee et al., 2007) on beta Weibull, (Cordeiro et al., 2010) on Kumaraswamy Weibull, (Silva et al., 2010) on beta modified Weibull, (Mudholkar & Srivastava, 1993) on exponentiated Weibull, (Zhang & Xie, 2011) on truncated Weibull, (Khan & King, 2013) on transmuted modified Weibull, and (Marshall & Olkin, 1997) on extended Weibull.

Numerous distributions have been defined on unit intervals by various authors, including the Johnson SB distribution (Johnson, 1949), Beta distribution (Eugene et al., 2002), Unit Johnson (SU) distribution (Gündüz & Korkmaz, 2020), Unit Burr-XII (Korkmaz & Chesneau, 2021), Topp-Leone distribution (Topp & Leone, 1955), Unit-Gompertz (Mazucheli et al., 2019), Unit Logistic distribution (Tadikamalla & Johnson, 1982), Kumaraswamy distribution (Kumaraswamy, 1980), Unit Burr-III (Modi & Gill, 2020), Unit modified Burr-III (Haq et al., 2023), Unit Gamma distribution (Consul & Jain, 1971; Grassia, 1977; Mazucheli et al., 2018b; Tadikamalla, 1981), Unit-Lindley (Mazucheli et al., 2019), Unit-Weibull (Mazucheli et al., 2020), and Unit Muth distribution (Maya et al., 2024). Recently, Mazucheli et al. (2018a) proposed a unit Weibull distribution to model data on the unit interval, using real data sets such as maximum flood levels and petroleum reservoir data. They introduced a quantile regression model for this unit Weibull distribution, demonstrating its superiority over competing distributions like the beta, Kumaraswamy, Unit Logistic, Simplex, Unit Gamma, Exponentiated Topp-Leone, and Extended Arcsine distributions, as it provides a closed formula for the quantile function.

In this paper, the author explores a new unit Weibull distribution based on the probability density function (PDF) of the median order statistics from a sample size of three. The discussion focuses on the newly defined two-parameter distribution, the Median Based Unit Weibull (MBUW), and its basic properties.

The paper is structured into five sections. Methodology contains section (1-3). Section 1 outlines the methodology for obtaining the new distribution. Section 2 elaborates on its PDF, cumulative distribution function (CDF), survival function, hazard function, reversed hazard function, and additional properties. Section 3 details new methods for estimating the parameters, such as the generalized method of moments and percentile estimators. Results contain section 4. Section 4 analyzes various real data sets to illustrate how to fit data to different competing distributions and how to apply the aforementioned estimation methods for the MBUW distribution. Discussion contains section 5. Section 5 discusses the comparison between methods. Conclusions contain section 6. Finally, Section 6 concludes with remarks on the findings and future research directions.

# 1. Methodology:

## 1.1. Section

### Derivation of the MBUW Distribution:

Using the pdf of median order statistics of a sample size equal 3 in equation (1), the CDF and the PDF of the parent distribution Weibull respectively in equation (2), then substitute these functions in equation (1), to give the PDF in equation (3). Both parameters alpha and beta are positive.

$$f_{i:n}(z) = \frac{n!}{(i-1)!\,(n-i)!}\{F(z)\}^{i-1}\{1-F(z)\}^{n-i}f(z), \qquad z>0 \ldots\ldots\ldots\ldots\ldots\ldots\ldots\ldots\ldots..(1)$$

$$f_{2:3}(z) = \frac{3!}{(2-1)!\,(3-1)!}\{F(z)\}^{2-1}\{1-F(z)\}^{3-2}f(z), \qquad z>0$$

$$F(z) = 1-e^{-\left(\frac{z}{\alpha}\right)^{\beta}}\ ,\ \ f(z) = \frac{\beta}{\alpha}\left(\frac{z}{\alpha}\right)^{\beta-1} e^{-\left(\frac{z}{\alpha}\right)^{\beta}}, z>0, \qquad \alpha\ \&\ \beta>0 \ldots\ldots\ldots\ldots\ldots\ldots\ldots\ldots (2)$$

$$f_{2:3}(z) = 3!\left\{1-e^{\frac{-z^{\beta}}{\alpha^{\beta}}}\right\}^{2-1}\left\{e^{\frac{-z^{\beta}}{\alpha^{\beta}}}\right\}^{3-2}\frac{\beta}{\alpha}\left(\frac{z}{\alpha}\right)^{\beta-1} e^{-\left(\frac{z}{\alpha}\right)^{\beta}}, z>0 \ldots\ldots\ldots\ldots\ldots\ldots\ldots\ldots\ldots..(3.A)$$

$$f_{2:3}(z) = \frac{6\ \beta z^{\beta-1}}{\alpha^{\beta}}\left[1-e^{\frac{-z^{\beta}}{\alpha^{\beta}}}\right]\left[e^{\frac{-2z^{\beta}}{\alpha^{\beta}}}\right], \qquad z>0\,, \qquad \alpha\ \&\ \beta>0 \ldots\ldots\ldots\ldots\ldots\ldots\ldots\ldots (3.B)$$

$$f_{2:3}(z) = \frac{6\ \beta z^{\beta-1}}{\alpha^{\beta}}\left[1-e^{\frac{-z^{\beta}}{\alpha^{\beta}}}\right]\left[e^{\frac{-2z^{\beta}}{\alpha^{\beta}}}\right], \qquad z>0\,, \qquad \alpha\ \&\ \beta>0 \ldots\ldots\ldots\ldots\ldots.\ldots\ldots.(3.C)$$

Apply the following transformation $y = e^{-z^{\beta}}$ on equation (3.C):

$$let\ \ y = e^{-z^{\beta}}\ \ \ldots..,\quad so\ \ -ln(y) = z^{\beta}\,,\quad this\ implies\ [-ln(y)]^{\frac{1}{\beta}} = z$$

Using the Jacobian in equation (4) and substitute the transformation and the Jacobian in equation (3.C) with rearrangement, the PDF of the new Median Based Unit Weibull (MBUW) distribution is obtained in equation (5):

$$\frac{dz}{dy} = \frac{1}{\beta}\left[-ln(y)\right]^{\frac{1-\beta}{\beta}}\left(\frac{-1}{y}\right) \ldots\ldots\ldots\ldots\ldots\ldots\ldots\ldots\ldots \ldots (4)$$

$$f(y) = \frac{6}{\alpha^{\beta}}\left[1 - y^{\frac{1}{\alpha^{\beta}}}\right] y^{\left(\frac{2}{\alpha^{\beta}}-1\right)} \ , \quad 0 < y < 1 \, , \qquad \alpha > 0, \qquad \beta > 0 \ldots\ldots\ldots\ldots\ldots\ldots\ldots (5)$$

## 1.2 Section

## ***Some of the properties of the new distribution (MBUW):***

The PDF is shown in equation (5). The CDF, survival function, hazard function, reversed hazard function and quantile function are shown in equations (6-8) respectively.

$$F(y) = 3y^{\frac{2}{\alpha^{\beta}}} - 2y^{\frac{3}{\alpha^{\beta}}} \ , \quad 0 < y < 1 \, , \qquad \alpha > 0, \qquad \beta > 0 \ldots\ldots\ldots\ldots\ldots\ldots\ldots\ldots\ldots\ldots\ldots\ldots (6)$$

$$S(y) = 1 - F(Y) = 1 - \left(3y^{\frac{2}{\alpha^{\beta}}} - 2y^{\frac{3}{\alpha^{\beta}}}\right) \ , \quad 0 < y < 1 \ , \alpha > 0 \, , \beta > 0 \ldots\ldots\ldots\ldots\ldots\ldots\ldots (7)$$

$$h(y) = \frac{f(y)}{S(y)} = \frac{\frac{6}{\alpha^{\beta}}\left(1 - y^{\frac{1}{\alpha^{\beta}}}\right) y^{\left(\frac{2}{\alpha^{\beta}}-1\right)}}{1 - \left(3y^{\frac{2}{\alpha^{\beta}}} - 2y^{\frac{3}{\alpha^{\beta}}}\right)} \ , \quad 0 < y < 1 \ , \alpha > 0, \beta > 0 \ldots\ldots\ldots\ldots\ldots\ldots\ldots (8)$$

$$rh(y) = \frac{f(y)}{F(y)} = \frac{\frac{6}{\alpha^{\beta}}\left(1 - y^{\frac{1}{\alpha^{\beta}}}\right) y^{\left(\frac{2}{\alpha^{\beta}}-1\right)}}{3y^{\frac{2}{\alpha^{\beta}}} - 2y^{\frac{3}{\alpha^{\beta}}}} \ , \quad 0 < y < 1 \ , \alpha > 0, \beta > 0 \ldots\ldots\ldots\ldots\ldots\ldots\ldots (9)$$

The inverse of the CDF is used to obtain y, the real root of this 3rd polynomial function is shown in equation (10):

$$y = F^{-1}(y) = \left\{-.5\left(cos\left[\frac{cos^{-1}(1-2u)}{3}\right] - \sqrt{3}\, sin\left[\frac{cos^{-1}(1-2u)}{3}\right]\right) + .5\right\}^{\alpha^{\beta}} \ldots\ldots (10)$$

In Figures 1-7, for a random variable x distributed as MBUW detailed depictions of the PDF, CDF, hazard rate function and survival functions for specific alpha values (0.5, 1, 1.5, 2, 2.5, 3, 3.5, 4) and beta values (0.6, 3.5) are shown. These comprehensive representations are designed to provide clarity and depth, making it easier to grasp the underlying concepts.

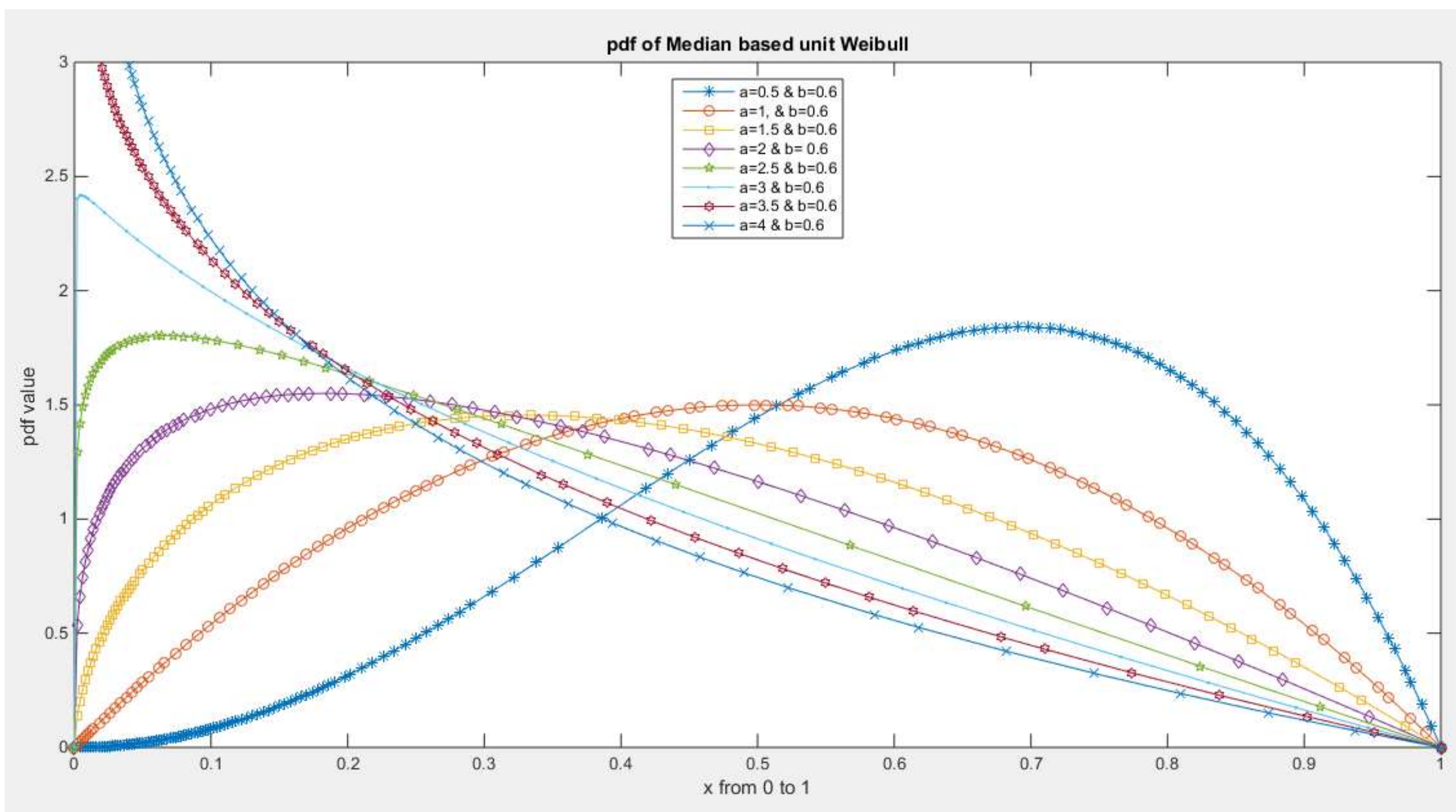

Fig. 1: pdf of (MBUW) distribution, alpha ( 0.5, 1, 1.5, 2, 2.5, 3, 3.5, 4 ) and beta ( 0.6 )

To generate random variable y distributed as MBUW:

1- Generate uniform random variable (0,1): $u \sim uniform(0,1)$.
2- Choose alpha and beta levels
3- Substitute the above values of u (0,1) and the chosen alpha and beta in the quantile function, to obtain y distributed as $y \sim MBUW(\alpha, \beta)$

Equations (11-14) show the first four raw moments of the distribution and the variance.

$$E(y^r) = \frac{6}{(2 + r\alpha^{\beta})(3 + r\alpha^{\beta})} = \int_0^1 y^r \, \frac{6}{\alpha^{\beta}} \left[1 - y^{\frac{1}{\alpha^{\beta}}}\right] y^{\left(\frac{2}{\alpha^{\beta}} - 1\right)} \, dy \; \dots\dots\dots\dots\dots\dots\dots\dots\dots . (11)$$

$$E(y) = \frac{6}{(2 + \alpha^{\beta})(3 + \alpha^{\beta})} \quad , \quad E(y^2) = \frac{6}{(2 + 2\alpha^{\beta})(3 + 2\alpha^{\beta})} \dots\dots\dots\dots\dots\dots\dots . . (12)$$

$$E(y^3) = \frac{6}{(2 + 3\alpha^{\beta})(3 + 3\alpha^{\beta})} \quad , \quad E(y^4) = \frac{6}{(2 + 4\alpha^{\beta})(3 + 4\alpha^{\beta})} \dots\dots\dots\dots\dots\dots\dots . (13)$$

$$\text{var}(y) = E(y^2) - [E(y)]^2 = \frac{78\alpha^{2\beta} + 60\alpha^{3\beta} + 6\alpha^{4\beta}}{(6 + 10\alpha^{\beta} + 4\alpha^{2\beta})(6 + 5\alpha^{\beta} + \alpha^{2\beta})^2} \dots\dots\dots\dots\dots\dots . . (14)$$

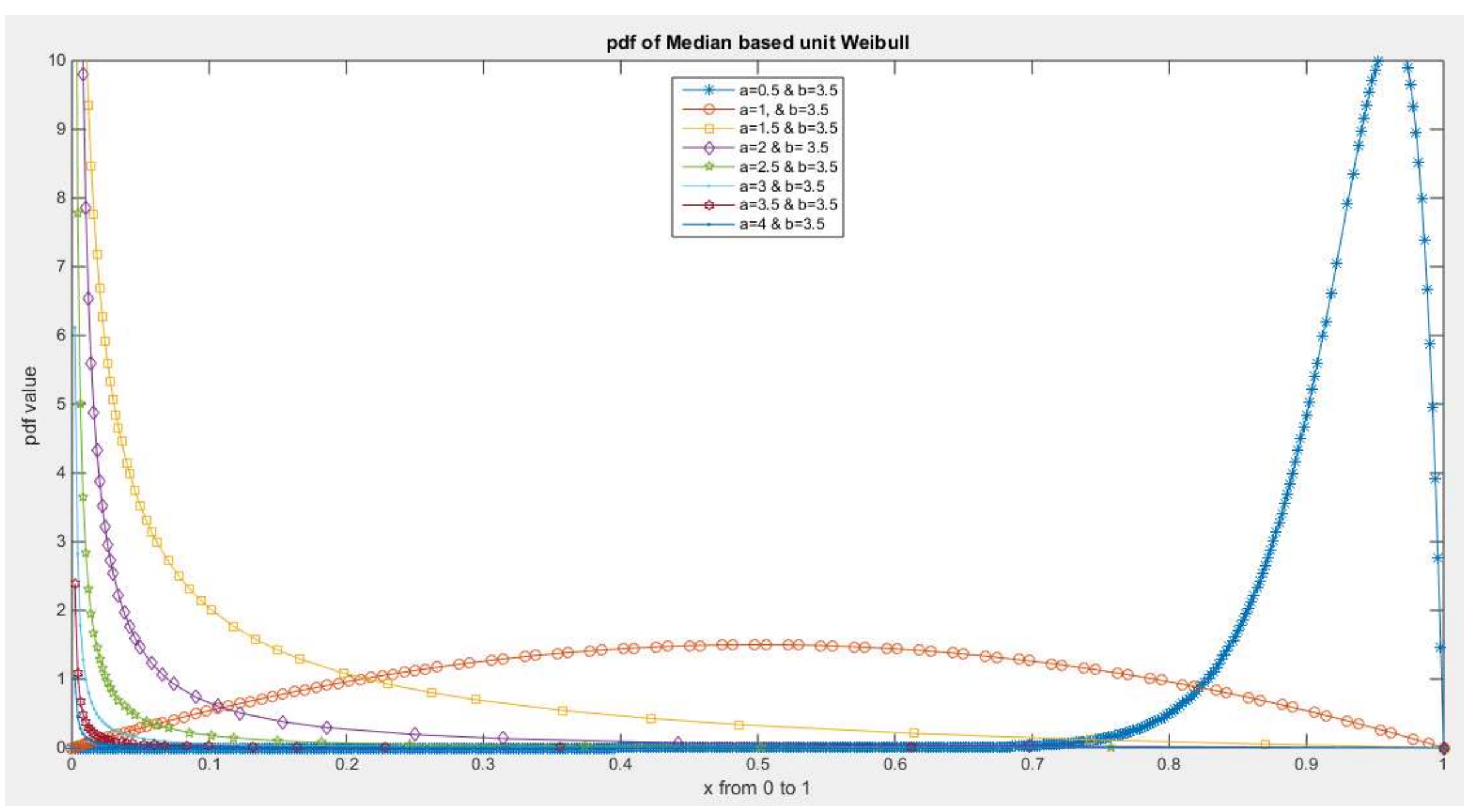

Fig. 2: pdf of (MBUW) distribution, alpha (0.5, 1, 1.5, 2, 2.5, 3, 3.5, 4) and beta (3.5).

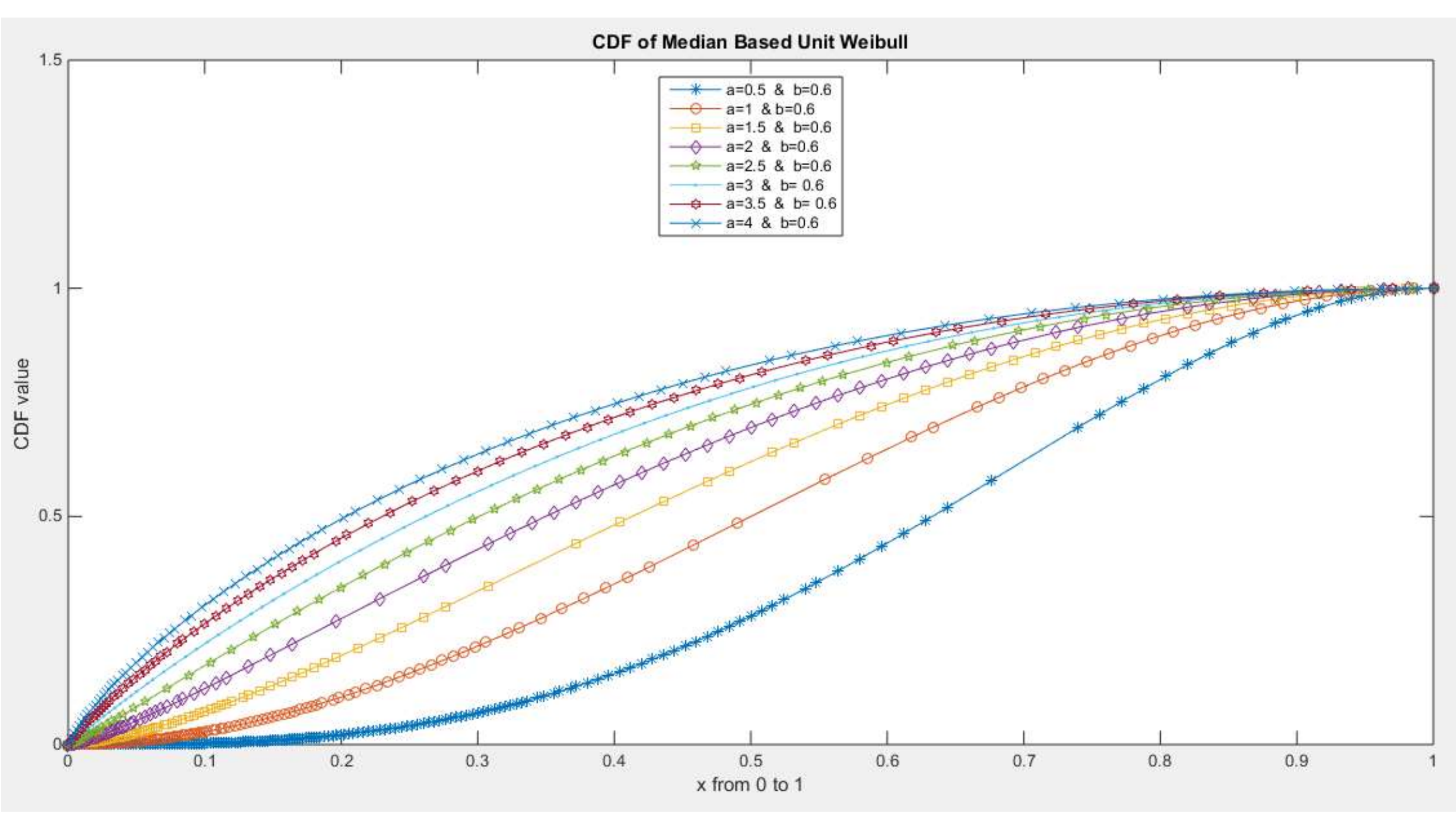

Fig 3: CDF of MBUW distribution, alpha ( 0.5, 1, 1.5, 2, 2.5, 3, 3.5, 4 ) and beta ( 0.6)

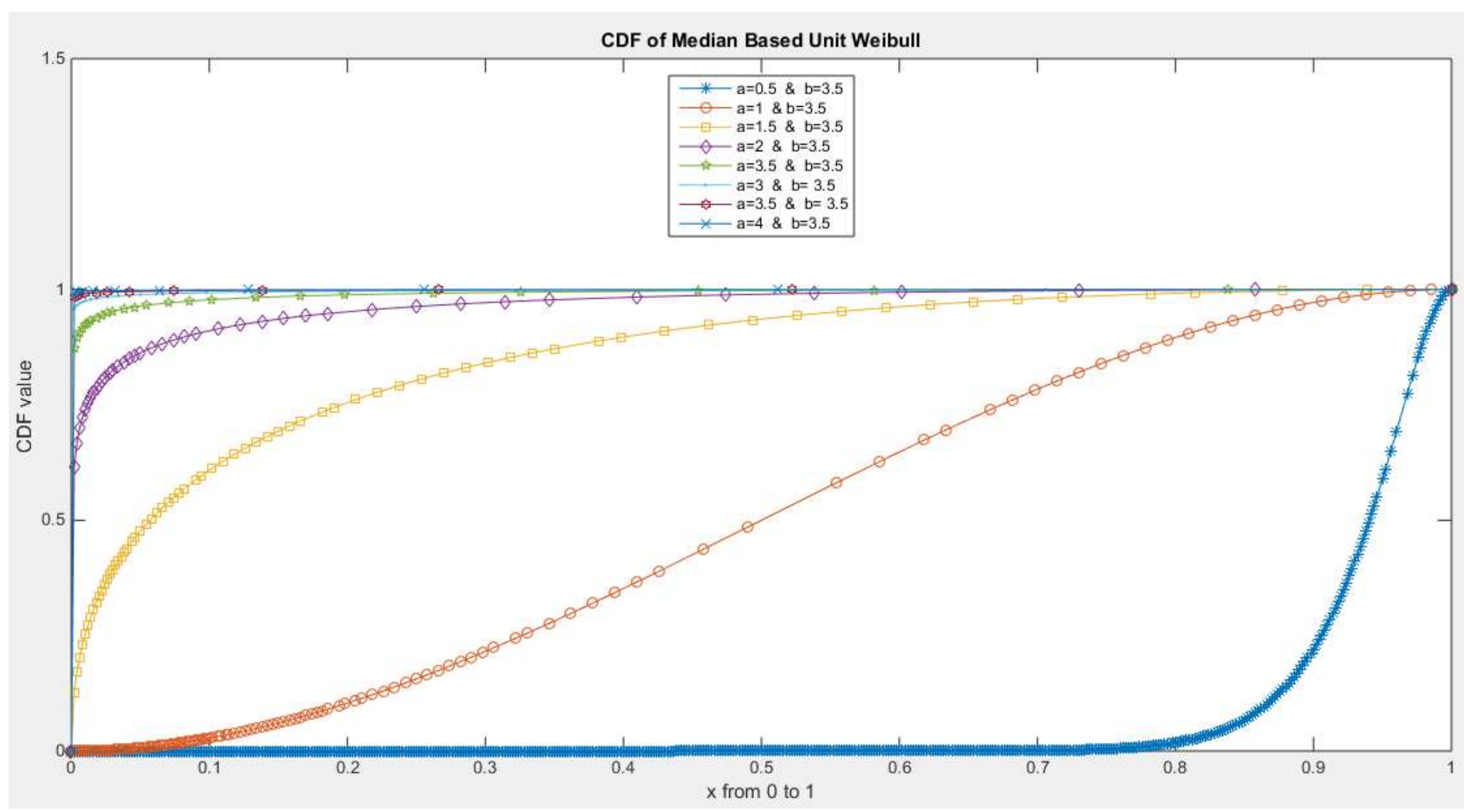

Fig. 4: CDF of (MBUW) distribution, alpha ( 0.5, 1, 1.5, 2, 2.5, 3, 3.5, 4 ) and beta ( 3.5).

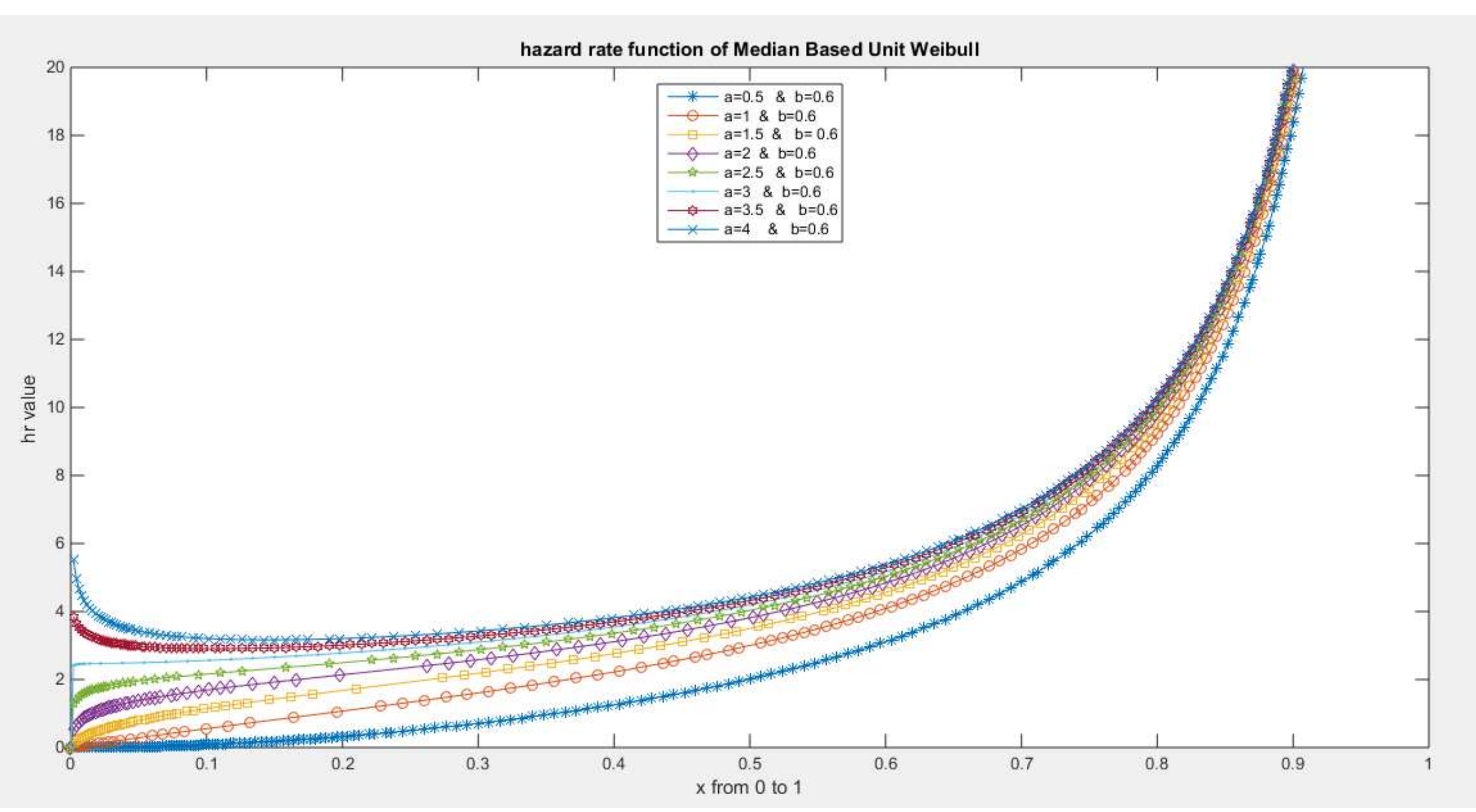

Fig. 5: hazard rate of ( MBUW) distribution, alpha ( 0.5, 1, 1.5, 2, 2.5, 3, 3.5, 4 ) and beta ( 0.6)

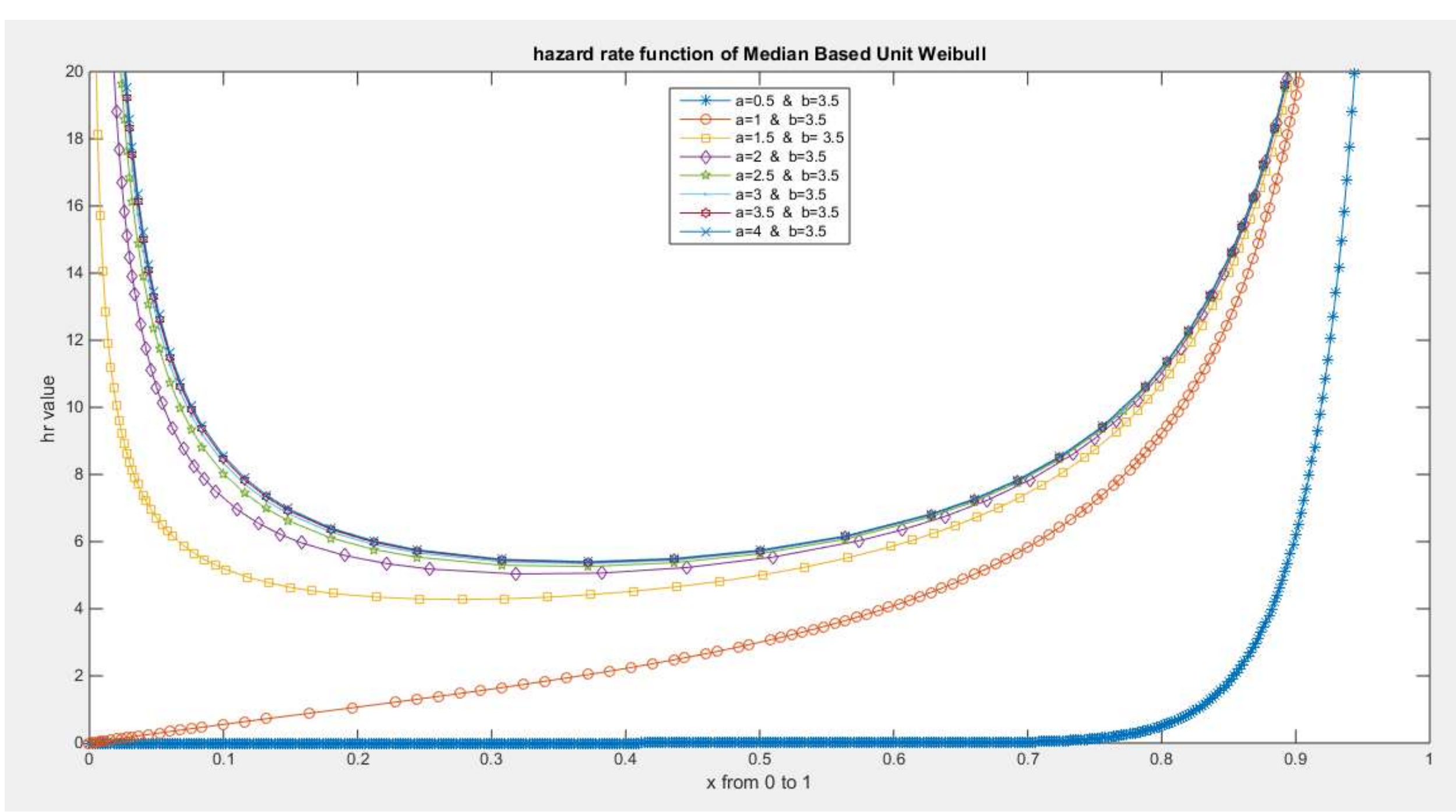

Fig. 6: hazard rate of ( MBUW) distribution, alpha ( 0.5, 1, 1.5, 2, 2.5, 3, 3.5, 4 ) and beta ( 3.5)

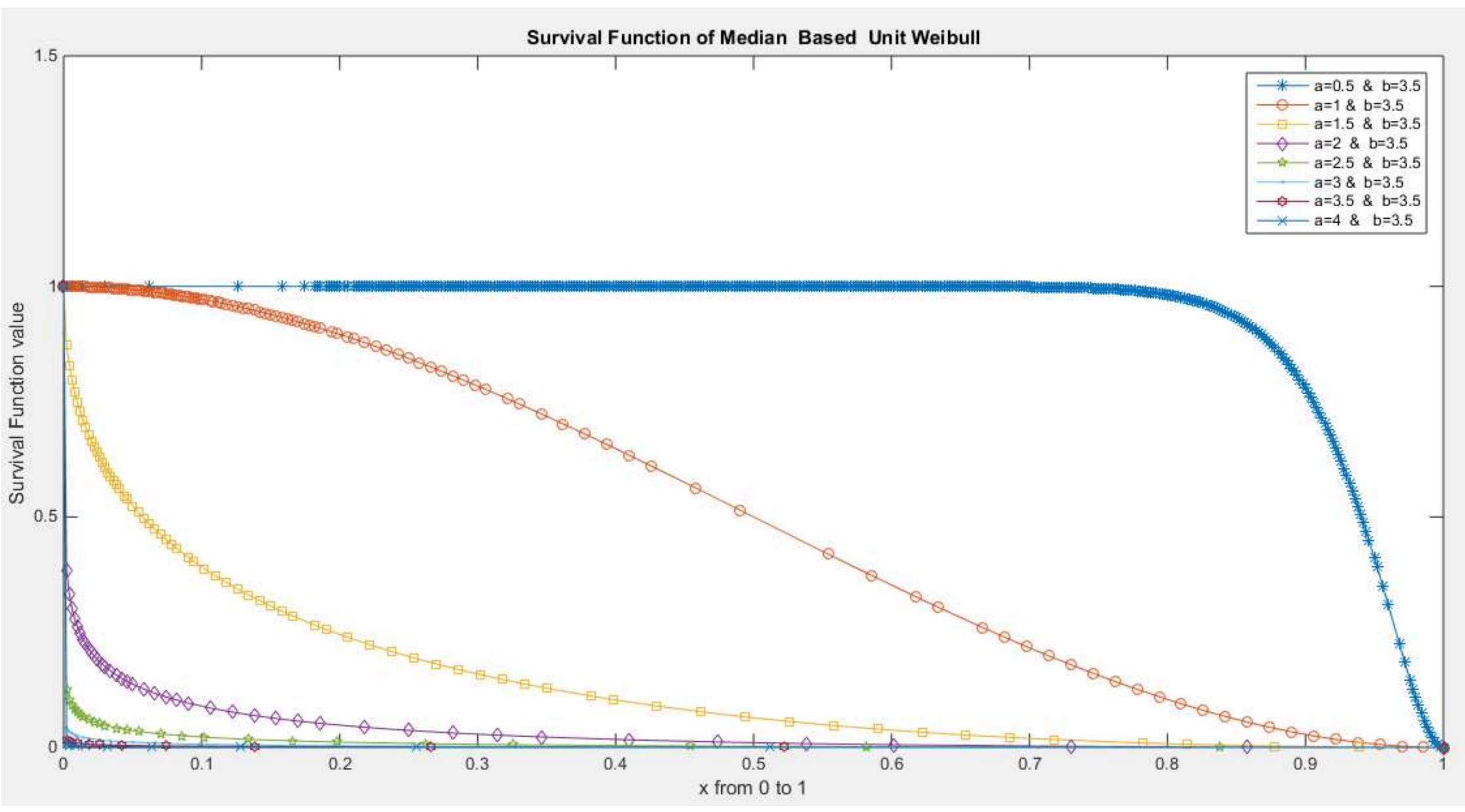

Fig. 7: Survival function of (MBUW) distribution, alpha ( 0.5, 1, 1.5, 2, 2.5, 3, 3.5, 4 ) and beta ( 3.5)

Equation 15 shows the Coefficient of skewness:

$$E\,\frac{(y-\mu)^3}{\sigma^3} = \frac{E(y^3) - 3\mu E(y^2) + 3\mu^2 E(y) - \mu^3}{\sigma^3}$$

$$= \frac{E(y^3) - 3\mu[E(y^2) - \mu E(y)] - \mu^3}{\sigma^3} = \frac{E(y^3) - 3\mu[E(y^2) - \mu\mu] - \mu^3}{\sigma^3}$$

$$= \frac{E(y^3) - 3\mu\sigma^2 - \mu^3}{\sigma^3} = \frac{E(y^3) - \mu(3\sigma^2 + \mu^2)}{\sigma^3} \dots\dots\dots\dots\dots\dots\dots\dots\dots\dots\dots\dots\dots\dots\dots\dots (15)$$

Equation 16 shows the Coefficient of kurtosis:

$$E\frac{(y-\mu)^4}{\sigma^4} = \frac{E(y^4) - 4\mu E(y^3) + 6\mu^2 E(y^2) - 3\,\mu^4}{\sigma^4}$$

$$= \frac{E(y^4) - 4\mu E(y^3) + 6\mu^2[\sigma^2 + \mu^2] - 3\,\mu^4}{\sigma^4} = \frac{E(y^4) - 4\mu E(y^3) + 6\mu^2\sigma^2 + 6\mu^4 - 3\,\mu^4}{\sigma^4}$$

$$= \frac{E(y^4) - 4\mu E(y^3) + 6\mu^2\sigma^2 + 3\mu^4}{\sigma^4}$$

$$= \frac{E(y^4) - 4\mu E(y^3) + 3\mu^2[2\sigma^2 + \mu^2]}{\sigma^4} \dots\dots\dots\dots\dots\dots\dots\dots\dots\dots\dots\dots\dots\dots\dots\dots\dots\dots\dots\dots\dots (16)$$

Equation (17) shows Coefficient of Variation (CV) :

$$\mathrm{CV} = \frac{S}{\mu} \dots\dots\dots\dots\dots\dots\dots\dots\dots\dots\dots\dots\dots\dots\dots\dots\dots\dots\dots\dots\dots\dots\dots\dots\dots\dots\dots\dots\dots\dots\dots\dots\dots\dots\dots\dots (17)$$

where S is standard deviation and mu is the mean. Figures 8-9 depict the variance and the different coefficients mentioned above with different values of alpha and beta=0.6 and 3.5

Equation (18) shows the r[th] incomplete Moments:

$$E(y^r|\, y < t) = \int_0^t y^r\ \frac{6}{\alpha^\beta}\left[1 - y^{\frac{1}{\alpha^\beta}}\right] y^{\left(\frac{2}{\alpha^\beta} - 1\right)}\ dy$$

$$E(y) = \frac{6y^{\frac{2}{\alpha^\beta}+r}}{(2 + r\alpha^\beta)} - \frac{6y^{\frac{3}{\alpha^\beta}+r}}{(3 + r\alpha^\beta)} \dots\dots\dots\dots\dots\dots\dots\dots\dots\dots\dots\dots\dots\dots\dots\dots\dots\dots\dots\dots\dots (18)$$

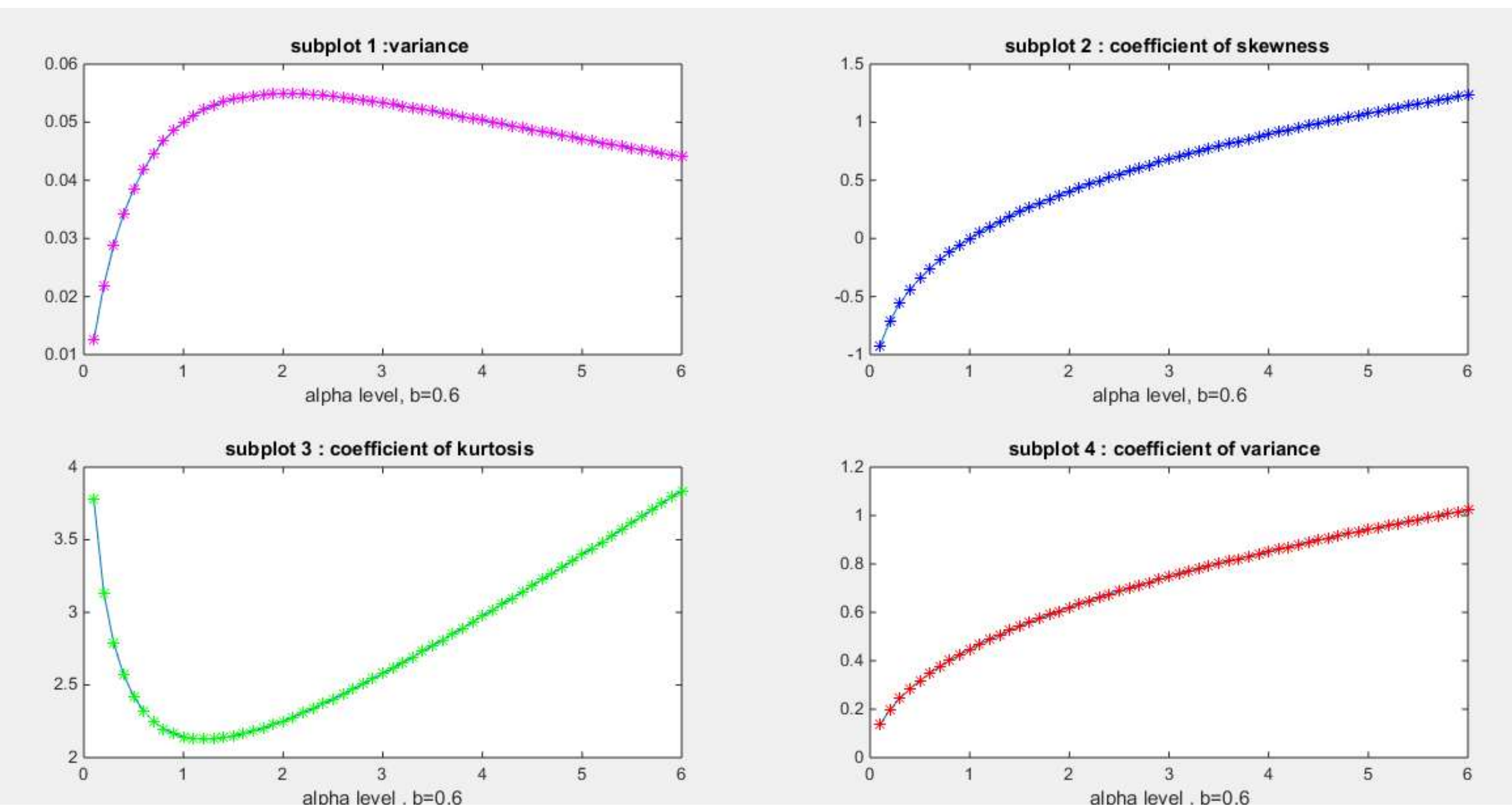

Fig. 8: the variance and different coefficients with alpha (from 0.1 to 6) & beta=0.6

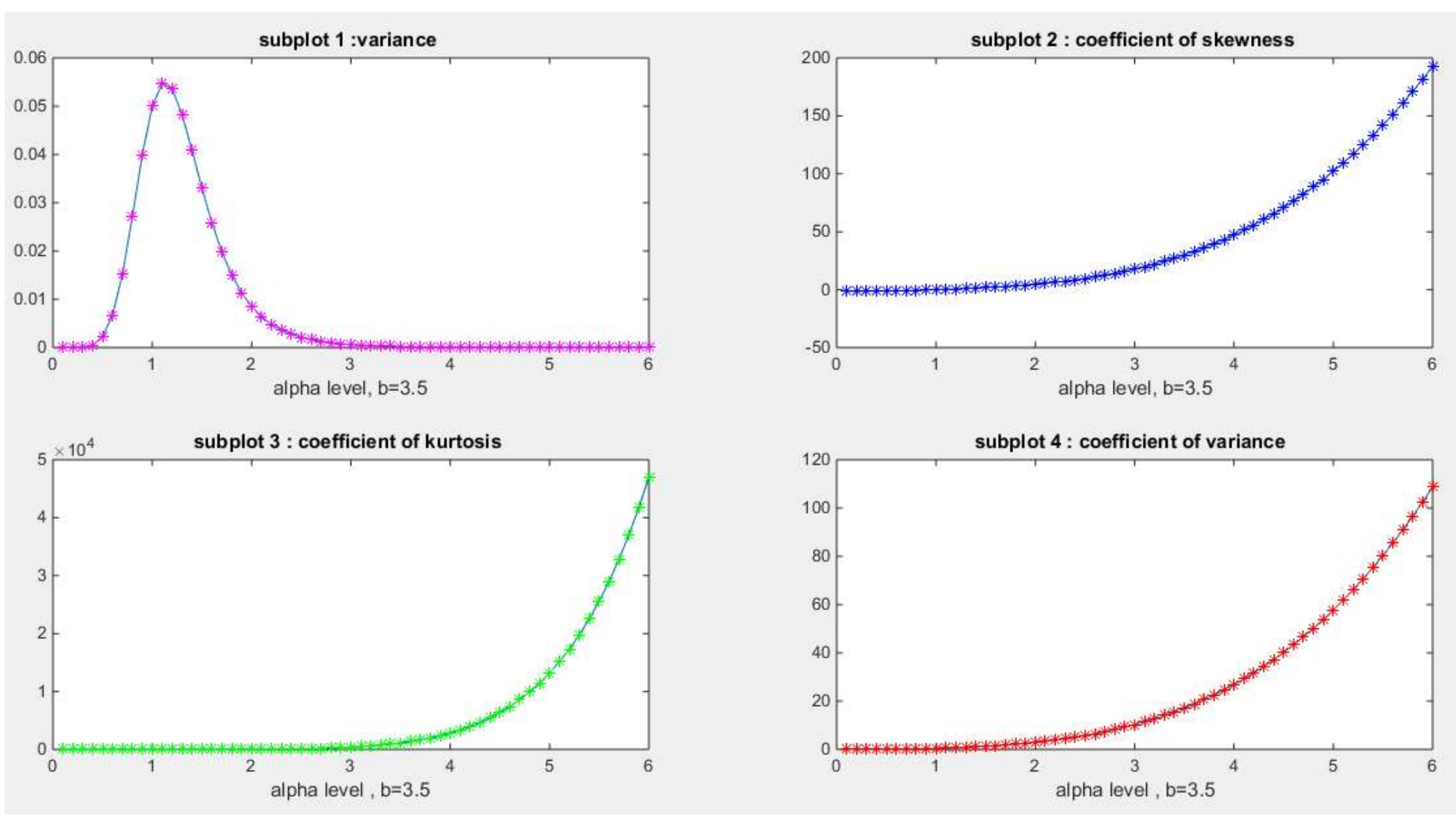

Fig. 9: the variance and different coefficients with alpha (from 0.1 to 6) & beta=3.5

## 1.3. Section

## Methods of Estimations

## Generalized Method of Moments:

### 1.3.1. The first and The second Raw Moments (method 1)

The generalized method of moments (GMM) has been developed by various authors, including Benson (1968) and the United States Water Resources Council (1967, 1981), as well as Bobé (1975), Hoshi and Burges (1981), Rao (1980), and Ashkar and Bobé (1986a, 1986b). These authors applied the well-known four methods of GMM to the Log Pearson Type 3 (LP) distribution. However, these methods can be utilized for any distribution and are not limited to the LP distribution.

As demonstrated by Shenton and Bowman (1977), maximum likelihood estimators are, in fact, moment estimators when dealing with the LP distribution. Additionally, Ashkar et al. (1988) introduced GMMs for estimating the parameters of the generalized gamma distribution.

GMMs utilize the raw moments of a distribution. The core principle of the method involves equating the sample moments with the population moments. Since the Median Base Unit Weibull (MBUW) has two parameters, the first and second moments will be employed to create a system of equations that can be solved numerically to determine the parameters.

Steps of the algorithm:

1- The non-central sample moments of order r for a given sample $y_1, y_2, \ldots, y_N$ are defined as in equation (19):

$$m'_r(y) = \frac{1}{N}\sum_{i=1}^{r} y_i^r \quad \ldots\ldots\ldots\ldots\ldots\ldots\ldots\ldots\ldots\ldots (19)$$

2- For this distribution, the non-central population moment has the following formulae equation (20):

$$\mu'_r = E(y^r) = \frac{6}{(2 + r\alpha^{\beta})(3 + r\alpha^{\beta})} \quad \ldots\ldots\ldots\ldots\ldots\ldots\ldots\ldots (20)$$

$$\mu'_1 = E(y) = \frac{6}{(2 + \alpha^{\beta})(3 + \alpha^{\beta})} \quad \ldots\ldots\ldots\ldots\ldots\ldots\ldots\ldots (21)$$

$$\mu'_2 = E(y^2) = \frac{6}{(2 + 2\alpha^{\beta})(3 + 2\alpha^{\beta})} \quad \ldots\ldots\ldots\ldots\ldots\ldots\ldots\ldots (22)$$

3- Equate the sample moment and the population moment for the first and second order as shown in equation (23).

$$m'_r(y) = \mu'_r \quad \ldots\ldots\ldots\ldots\ldots\ldots\ldots\ldots\ldots\ldots\ldots\ldots\ldots\ldots\ldots\ldots (23)$$

So $m'_1(y) = \mu'_1$ , this leads to the equation (24) . Let's call the right-hand side mom1:

$$\frac{1}{N}\sum_{i=1}^{r} y = \bar{y} = \frac{6}{(2+\alpha^{\beta})(3+\alpha^{\beta})} \quad \ldots\ldots\ldots\ldots\ldots\ldots\ldots\ldots\ldots\ldots\ldots (24)$$

So $m'_2(y) = \mu'_2$ , this leads to the equation (25) Let's call the right-hand side mom2:

$$\frac{1}{N}\sum_{i=1}^{r} y^2 = \frac{6}{(2+2\alpha^{\beta})(3+2\,\alpha^{\beta})} \ldots\ldots\ldots\ldots\ldots\ldots\ldots\ldots\ldots\ldots\ldots (25)$$

4- Differentiate equations 24-25 with respect to both parameters as shown in equations (26-29) and solve the system of equations using the Levenberg-Marquardt algorithm (LM) in equation (31).

$$\frac{\partial}{\partial\alpha}\left[\frac{6}{(2+\alpha^{\beta})(3+\alpha^{\beta})}\right] = -6\left(6+5\alpha^{\beta}+\alpha^{2\beta}\right)^{-2}\left(5\,\beta\alpha^{\beta-1}+2\,\beta\alpha^{2\beta-1}\right) \quad \ldots\ldots\ldots (26)$$

$$\frac{\partial}{\partial\beta}\left[\frac{6}{(2+\alpha^{\beta})(3+\alpha^{\beta})}\right] = -6\left(6+5\alpha^{\beta}+\alpha^{2\beta}\right)^{-2}\left(5\,\alpha^{\beta}\ln(\alpha)+2\,\alpha^{2\beta}\ln(\alpha)\right) \ldots\ldots\ldots (27)$$

$$\frac{\partial}{\partial\alpha}\left[\frac{6}{(2+2\alpha^{\beta})(3+2\,\alpha^{\beta})}\right] = -6\left(6+10\alpha^{\beta}+4\alpha^{2\beta}\right)^{-2}\left(10\,\beta\alpha^{\beta-1}+8\,\beta\alpha^{2\beta-1}\right) \ldots\ldots (28)$$

$$\frac{\partial}{\partial\beta}\left[\frac{6}{(2+2\alpha^{\beta})(3+2\,\alpha^{\beta})}\right] = -6\left(6+10\alpha^{\beta}+4\alpha^{2\beta}\right)^{-2}\left(10\,\alpha^{\beta}\ln(\alpha)+8\,\alpha^{2\beta}\ln(\alpha)\right) \ldots (29)$$

The objective function to be fitted is shown in equation 30

$$\mu'_r = E(y^r) = \frac{6}{(2+r\alpha^{\beta})(3+r\,\alpha^{\beta})} \ldots\ldots\ldots\ldots\ldots\ldots\ldots\ldots\ldots\ldots\ldots (30)$$

$$\theta^{(a+1)} = \theta^{(a)} + \left[J'(\theta^{(a)})J(\theta^{(a)}) + \lambda^{(a)}I^{(a)}\right]^{-1}\left[J'(\theta^{(a)})\right]\left(y-f(\theta^{(a)})\right) \ldots\ldots\ldots\ldots (31)$$

Where:

The Jacobian matrix is $\begin{bmatrix} \frac{\partial}{\partial\alpha}Mom_1 & \frac{\partial}{\partial\beta}Mom_1 \\ \frac{\partial}{\partial\alpha}Mom_2 & \frac{\partial}{\partial\beta}Mom_2 \end{bmatrix}$ and

$$\left(y - f\left(\theta^{(a)}\right)\right) = \begin{bmatrix} m_1' - \dfrac{6}{(2+\alpha^{\beta})(3+\alpha^{\beta})} \\ m_2' - \dfrac{6}{(2+2\alpha^{\beta})(3+2\alpha^{\beta})} \end{bmatrix}, where\ \boldsymbol{\theta}^{(\boldsymbol{a})} = \begin{bmatrix} \alpha^{(a)} \\ \beta^{(a)} \end{bmatrix}$$

$\theta^{(a)}$ is the parameters used in the first iteration which are the initial guess, then they are updated according to the sum of squares of errors.

LM algorithm is an iterative algorithm.

$J'\left(\theta^{(a)}\right)$ : this is the Jacobian function which is the first derivative of the objective function evaluated at the initial guess $\theta^{(a)}$ .

$\lambda^{(a)}$ is a damping factor that adjusts the step size in each iteration direction, the starting value usually is 0.001 and according to the sum square of errors (SSE) in each iteration this damping factor is adjusted:

$$SSE^{(a+1)} \geq SSE^{(a)} \quad so\ \ update: \quad \lambda_{updated} = 10 * \lambda_{old}$$

$$SSE^{(a+1)} < SSE^{(a)} \quad so\ \ update: \quad \lambda_{updated} = \frac{1}{10} * \lambda_{old}$$

$f\left(\theta^{(a)}\right)$ : is the objective function (non-central population moment, first and second order) evaluated at the initial guess.

$y$ : is the non-central sample moments, the first and second order.

Steps of the LM algorithm:

1- Start with the initial guess of parameters (alpha and beta).
2- Substitute these values in the objective function and the Jacobian.
3- Choose the damping factor, say lambda=0.001
4- Substitute in equation (LM equation (31)) to get the new parameters.
5- Calculate the SSE at these parameters and compare this SSE value with the previous one when using initial parameters to adjust for the damping factor.
6- Update the damping factor accordingly as previously explained.
7- Start new iteration with the new parameters and the new updated damping factor, i.e, apply the previous steps many times till convergence is achieved or a pre-specified number of iterations is accomplished.

The value of this quantity: $\left[J'\left(\theta^{(a)}\right)J\left(\theta^{(a)}\right) + \lambda^{(a)}I^{(a)}\right]^{-1}$ can be considered a good approximation to the variance – covariance matrix of the estimated parameters. Using delta method and squaring the Jacobian matrix is applied to derive the standard errors for the estimated parameters as illustrated below.

Other non-central raw moments can be used in the following variants of the generalized method of moments.

## 1.3.2. The Third and The Fourth Raw Moments (method 2)

Equating $m'_3(y) = \mu'_3$ and $m'_4(y) = \mu'_4$ gives this following system of equations (32-35). Let's call the functions to be differentiated: mom3 for (32-33) and mom4 for (34-35)

$$\frac{\partial}{\partial\alpha}\left[\frac{6}{(2+3\alpha^{\beta})(3+3\alpha^{\beta})}\right] = -6\left(6+15\alpha^{\beta}+9\alpha^{2\beta}\right)^{-2}\left(15\,\beta\alpha^{\beta-1}+18\,\beta\alpha^{2\beta-1}\right) \ldots\ldots(32)$$

$$\frac{\partial}{\partial\beta}\left[\frac{6}{(2+3\alpha^{\beta})(3+3\alpha^{\beta})}\right] = -6\left(6+15\alpha^{\beta}+9\alpha^{2\beta}\right)^{-2}\left(15\,\alpha^{\beta}\ln(\alpha)+18\,\alpha^{2\beta}\ln(\alpha)\right) \ldots(33)$$

$$\frac{\partial}{\partial\alpha}\left[\frac{6}{(2+4\alpha^{\beta})(3+4\,\alpha^{\beta})}\right] = -6\left(6+20\alpha^{\beta}+16\alpha^{2\beta}\right)^{-2}\left(20\,\beta\alpha^{\beta-1}+32\,\beta\alpha^{2\beta-1}\right) \ldots\ldots(34)$$

$$\frac{\partial}{\partial\beta}\left[\frac{6}{(2+4\alpha^{\beta})(3+4\,\alpha^{\beta})}\right] = -6\left(6+20\alpha^{\beta}+16\alpha^{2\beta}\right)^{-2}\left(20\,\alpha^{\beta}\ln(\alpha)+32\,\alpha^{2\beta}\ln(\alpha)\right) (35)$$

Apply LM algorithm where:

The Jacobian matrix is $\begin{bmatrix}\frac{\partial}{\partial\alpha}Mom_3 & \frac{\partial}{\partial\beta}Mom_3\\ \frac{\partial}{\partial\alpha}Mom_4 & \frac{\partial}{\partial\beta}Mom_4\end{bmatrix}$

$$\left(y-f\left(\boldsymbol{\theta}^{(a)}\right)\right) = \begin{bmatrix} m'_3 - \dfrac{6}{(2+3\alpha^{\beta})(3+3\alpha^{\beta})} \\ m'_4 - \dfrac{6}{(2+4\alpha^{\beta})(3+4\alpha^{\beta})}\end{bmatrix}, where\ \boldsymbol{\theta}^{(a)} = \begin{bmatrix}\alpha^{(a)}\\ \beta^{(a)}\end{bmatrix}$$

## 1.3.3.Skewness defined as quartile and log of the geometric mean (method 3):

Population skewness can be defined in equation (36) using the quantile function defined in equation (10) as

$$S_{pop} = \frac{Q(3/4) - 2Q\left(\frac{2}{4}\right) + Q(1/4)}{Q(3/4) - Q(1/4)} \ldots\ldots\ldots\ldots\ldots\ldots\ldots\ldots\ldots\ldots\ldots\ldots\ldots\ldots\ldots\ldots\ldots\ldots\ldots\ldots\ldots\ldots\ldots (36)$$

while the sample skewness is obtained from the sample by the empirical quantiles in equation (37).

$$S_{sample} = \frac{Q(3/4) - 2Q\left(\frac{2}{4}\right) + Q(1/4)}{Q(3/4) - Q(1/4)} \ldots \ldots \ldots \ldots \ldots \ldots \ldots \ldots \ldots \ldots \ldots \ldots \ldots \ldots \ldots \ldots \ldots \ldots \ldots \ldots . (37)$$

Equate equations (36) and (37), then differentiate equation 36 with respect to both parameters and apply LM algorithm in the same manner as in method 1 and 2.

Transformation of the observations into the log scale is used to obtain the log of the geometric mean which gives the following equation (38) (z is the log of the geometric mean or zero raw moment, i.e., $z_{sample}$ , which is the sample log geometric mean)

$$z_{sample} = \ln \prod_{i=1}^{n} y_i^{\frac{1}{n}} = \frac{1}{n} \sum_{i=1}^{n} \ln y_i = \sum_{i=1}^{n} \frac{1}{n} (\ln y_i) = \sum_{i=1}^{n} \ln y_i^{\frac{1}{n}} \ldots \ldots \ldots \ldots \ldots \ldots \ldots \ldots \ldots . \ldots (38)$$

Getting the expectation of the log of the observations gives the population log geometric mean, as in equation (39). Equate equation (37) and equation (38) to get equation (39):

$$Z_{pop} = E\{\ln y_i\} = \int_0^1 (\ln y_i)\, f_Y(y) dy = \frac{-5}{6}\, \alpha^{\beta} \ldots \ldots \ldots \ldots \ldots \ldots \ldots \ldots \ldots \ldots \ldots \ldots \ldots \ldots \ldots \ldots \ldots \ldots (39)$$

Equating equation (38) with equation (39), differentiate equation (39) with respect to both parameters and apply LM algorithm as in method 1 and 2. In other words, form the system of equations (40-43)

$$\frac{\partial}{\partial \alpha}\left[S_{pop}\right] =$$
$$= \frac{\left[Q^{'}(3/4) - 2Q^{'}\left(\frac{2}{4}\right) + Q^{'}(1/4)\right]\left[Q(3/4) - Q(1/4)\right] - \left[Q(3/4) - 2Q\left(\frac{2}{4}\right) + Q(1/4)\right]\left[Q^{'}(3/4) - Q^{'}(1/4)\right]}{\left[Q(3/4) - Q(1/4)\right]^2} \ldots (40)$$

$$\frac{\partial}{\partial \beta}\left[S_{pop}\right] =$$
$$= \frac{\left[Q^{'}(3/4) - 2Q^{'}\left(\frac{2}{4}\right) + Q^{'}(1/4)\right]\left[Q(3/4) - Q(1/4)\right] - \left[Q(3/4) - 2Q\left(\frac{2}{4}\right) + Q(1/4)\right]\left[Q^{'}(3/4) - Q^{'}(1/4)\right]}{\left[Q(3/4) - Q(1/4)\right]^2} \ldots (41)$$

$$\frac{\partial}{\partial \alpha}\left[Z_{pop}\right] = \frac{-5}{6} \beta\, \alpha^{\beta - 1} \ldots \ldots \ldots \ldots \ldots \ldots \ldots \ldots \ldots \ldots \ldots \ldots \ldots \ldots \ldots \ldots \ldots \ldots \ldots \ldots \ldots \ldots \ldots \ldots \ldots \ldots \ldots \ldots (42)$$

$$\frac{\partial}{\partial \beta}\left[Z_{pop}\right] = \frac{-5}{6}\, \alpha^{\beta} \ln \alpha \ldots \ldots \ldots \ldots \ldots \ldots \ldots \ldots \ldots \ldots \ldots \ldots \ldots \ldots \ldots \ldots \ldots \ldots \ldots \ldots \ldots \ldots \ldots \ldots \ldots \ldots . . (43)$$

The Jacobian matrix is $\begin{bmatrix} \frac{\partial}{\partial \alpha} S_{pop} & \frac{\partial}{\partial \beta} S_{pop} \\ \frac{\partial}{\partial \alpha} Z_{pop} & \frac{\partial}{\partial \beta} Z_{pop} \end{bmatrix}$

$$\left(y - f\left(\boldsymbol{\theta}^{(\boldsymbol{a})}\right)\right) = \begin{bmatrix} S_{sample} - \dfrac{Q(3/4) - 2Q\left(\frac{2}{4}\right) + Q(1/4)}{Q(3/4) - Q(1/4)} \\ z_{sample} - \dfrac{-5}{6}\,\alpha^{\beta} \end{bmatrix}, where\ \boldsymbol{\theta}^{(\boldsymbol{a})} = \begin{bmatrix} \alpha^{(a)} \\ \beta^{(a)} \end{bmatrix}$$

## 1.3.4.Skewness defined as decile and the log of the geometric mean ( method 4)

Method 4 is the same as method 3 with the difference being the definition of the sample and population skewness. In method 4 the skewness is defined as in equation (44) and follow the steps as in method 3, with the equation ( 45-48)

$$S = \frac{Q(9/10) - 2Q\left(\frac{1}{2}\right) + Q(1/10)}{Q(9/10) - Q(1/10)} \quad \dots\dots\dots\dots\dots\dots\dots\dots\dots\dots\dots\dots\dots\dots\dots\dots\dots\dots\dots\dots\dots\dots (44)$$

$$\frac{\partial}{\partial \alpha}\left[S_{pop}\right] =$$
$$= \frac{\left[Q^{'}(9/10) - 2Q^{'}\left(\frac{2}{4}\right) + Q^{'}(1/10)\right]\left[Q(9/10) - Q(1/10)\right] - \left[Q(9/10) - 2Q\left(\frac{2}{4}\right) + Q(1/10)\right]\left[Q^{'}(9/10) - Q^{'}(1/10)\right]}{\left[Q(9/10) - Q(1/10)\right]^{2}} \dots (45)$$

$$\frac{\partial}{\partial \beta}\left[S_{pop}\right] =$$
$$= \frac{\left[Q^{'}(9/10) - 2Q^{'}\left(\frac{2}{4}\right) + Q^{'}(1/10)\right]\left[Q(9/10) - Q(1/10)\right] - \left[Q(9/10) - 2Q\left(\frac{2}{4}\right) + Q(1/10)\right]\left[Q^{'}(9/10) - Q^{'}(1/10)\right]}{\left[Q(9/10) - Q(1/10)\right]^{2}} \dots (46)$$

$$\frac{\partial}{\partial \alpha}\left[Z_{pop}\right] = \frac{-5}{6}\beta\,\alpha^{\beta-1} \dots\dots\dots\dots\dots\dots\dots\dots\dots\dots\dots\dots\dots\dots\dots\dots\dots\dots\dots\dots\dots\dots\dots (47)$$

$$\frac{\partial}{\partial \beta}\left[Z_{pop}\right] = \frac{-5}{6}\,\alpha^{\beta}\ln\alpha \dots\dots\dots\dots\dots\dots\dots\dots\dots\dots\dots\dots\dots\dots\dots\dots\dots\dots\dots\dots\dots\dots (48)$$

The Jacobian matrix is $\begin{bmatrix} \frac{\partial}{\partial \alpha} S_{pop} & \frac{\partial}{\partial \beta} S_{pop} \\ \frac{\partial}{\partial \alpha} Z_{pop} & \frac{\partial}{\partial \beta} Z_{pop} \end{bmatrix}$

$$\left(y - f\left(\boldsymbol{\theta}^{(\boldsymbol{a})}\right)\right) = \begin{bmatrix} S_{sample} - \dfrac{Q(9/10) - 2Q\left(\frac{1}{2}\right) + Q(1/10)}{Q(9/10) - Q(1/10)} \\ z_{sample} - \dfrac{-5}{6}\,\alpha^{\beta} \end{bmatrix}, where\ \boldsymbol{\theta}^{(\boldsymbol{a})} = \begin{bmatrix} \alpha^{(a)} \\ \beta^{(a)} \end{bmatrix}$$

## 1.3.5. Method of Percentiles (method 5)

The use of percentiles for estimation has a longstanding and proven track record. Numerous authors have effectively employed this method, demonstrating that it is at least as efficient, if not more so, than Maximum Likelihood Estimation (MLE) and least squares (Bhatti et al., 2018; Dubey, 1967; Marks, 2005; Wang & Keats, 1995).

The percentile method used in this paper is applied to the 25th and 75th percentiles. The MBUW has the quantile or the percentile function defined in equation (10) substitute for u=0.25 and u=0.75 to obtain population 25th and 75th percentiles as shown in equations (49-50)

$$y_{25} = \left\{-.5\left(cos\left[\frac{cos^{-1}(1-2*0.25)}{3}\right] - \sqrt{3}\,sin\left[\frac{cos^{-1}(1-2*0.25)}{3}\right]\right) + .5\right\}^{\alpha^{\beta}} \quad \ldots\ldots\ldots\ldots\ldots\ldots\ldots\ldots.(49)$$

$$y_{75} = \left\{-.5\left(cos\left[\frac{cos^{-1}(1-2*0.75)}{3}\right] - \sqrt{3}\,sin\left[\frac{cos^{-1}(1-2*0.75)}{3}\right]\right) + .5\right\}^{\alpha^{\beta}} \quad \ldots\ldots(50)$$

$$y_{25} = c_{25}^{\alpha^{\beta}}\ , \qquad y_{75} = c_{75}^{\alpha^{\beta}} \ldots\ldots\ldots\ldots\ldots\ldots\ldots\ldots\ldots\ldots\ldots\ldots\ldots\ldots\ldots\ldots\ldots\ldots\ldots\ldots\ldots\ldots\ldots\ldots\ldots \quad (51)$$

Where $c_u = \left\{-.5\left(cos\left[\frac{cos^{-1}(1-2u)}{3}\right] - \sqrt{3}\,sin\left[\frac{cos^{-1}(1-2u)}{3}\right]\right) + .5\right\} \ldots\ldots\ldots\ldots\ldots\ldots\ldots\ldots.(52)$

The objective function to be minimized in the LM algorithm is the quantile function in equation (51). Differentiation of this objective function with respect to both parameters gives these equations:

$$\frac{\partial}{\partial\alpha}y_{25} = c_{25}^{\alpha^{\beta}}[\ \ln(c_{25})]\ \beta\ \alpha^{\beta-1} \ \ldots\ldots\ldots\ldots\ldots\ldots\ldots\ldots\ldots\ldots\ldots\ldots\ldots\ldots\ldots\ldots\ldots\ldots\ldots\ldots\ldots..(53)$$

$$\frac{\partial}{\partial\beta}y_{25} = c_{25}^{\alpha^{\beta}}[\ \ln(c_{25})]\ \left(\ \alpha^{\beta}\right)[\ln(\alpha)\ ] \ldots\ldots\ldots\ldots\ldots\ldots\ldots\ldots\ldots\ldots\ldots\ldots\ldots\ldots\ldots\ldots\ldots\ldots...(54)$$

$$\frac{\partial}{\partial\alpha}y_{75} = c_{75}^{\alpha^{\beta}}[\ \ln(c_{75})]\ \beta\ \alpha^{\beta-1} \ \ldots\ldots\ldots\ldots\ldots\ldots\ldots\ldots\ldots\ldots\ldots\ldots\ldots\ldots\ldots\ldots\ldots\ldots\ldots\ldots\ldots..(55)$$

$$\frac{\partial}{\partial\beta}y_{75} = c_{75}^{\alpha^{\beta}}[\ \ln(c_{75})]\ \left(\ \alpha^{\beta}\right)[\ln(\alpha)\ ] \ \ldots\ldots\ldots\ldots\ldots\ldots\ldots\ldots\ldots\ldots\ldots\ldots\ldots\ldots\ldots\ldots\ldots..(56)$$

The Jacobian matrix is $\begin{bmatrix} \frac{\partial}{\partial\alpha}y_{25} & \frac{\partial}{\partial\beta}y_{25} \\ \frac{\partial}{\partial\alpha}y_{75} & \frac{\partial}{\partial\beta}y_{75} \end{bmatrix}$ $\quad and \quad \left(y - f\left(\theta^{(a)}\right)\right) = \begin{bmatrix} y_{25} - c_{25}^{\alpha^{\beta}} \\ y_{75} - c_{75}^{\alpha^{\beta}} \end{bmatrix}$

Then use the LM algorithm as previously explained.

Any percentiles can be used. The most common to use are the 25th and 75th percentiles. (Ogana F. N., 2020) used many different combinations of percentiles and assessed the outcome for each combination.

## 1.3.6.Galton Skewness and Moors Kurtosis defined using octiles (method 6)

Galton defined skewness in equation 57 as

$$S_{Galton-pop} = \frac{Q(6/8) - 2Q(4/8) + Q(2/8)}{Q(6/8) - Q(2/8)} = \frac{Q(3/4) - 2Q\left(\frac{2}{4}\right) + Q(1/4)}{Q(3/4) - Q(1/4)} \ldots\ldots\ldots\ldots (57)$$

Moors defined kurtosis in equation 58 as

$$K_{Moors-pop} = \frac{Q(7/8) - Q\left(\frac{5}{8}\right) + Q\left(\frac{3}{8}\right) - Q(1/8)}{Q(6/8) - Q(2/8)} \ldots\ldots\ldots\ldots\ldots\ldots\ldots\ldots\ldots\ldots\ldots\ldots\ldots\ldots\ldots\ldots (58)$$

Population skewness and kurtosis are defined using the quantile function defined in equation 10 while the sample skewness and sample kurtosis are defined using the same equations but utilizing the empirical quantiles from the observations. The system of equations is defined as follows in equations (59-62)

$$\frac{\partial}{\partial\alpha}\left[S_{Galton-po}\ \right] = \\ = \frac{\left[Q'(6/8) - 2Q'\left(\frac{4}{8}\right) + Q'(2/8)\right]\left[Q(9/10) - Q(1/10)\right] - \left[Q(6/8) - 2Q\left(\frac{4}{8}\right) + Q(2/8)\right]\left[Q'(6/8) - Q'(2/8)\right]}{\left[Q(6/8) - Q(2/8)\right]^2} \ldots (59)$$

$$\frac{\partial}{\partial\beta}\left[S_{Galton-pop}\right] = \\ = \frac{\left[Q'(6/8) - 2Q'\left(\frac{4}{8}\right) + Q'(2/8)\right]\left[Q(9/10) - Q(1/10)\right] - \left[Q(6/8) - 2Q\left(\frac{4}{8}\right) + Q(2/8)\right]\left[Q'(6/8) - Q'(2/8)\right]}{\left[Q(6/8) - Q(2/8)\right]^2} \ldots (60)$$

$$\frac{\partial}{\partial\alpha}\left[K_{Moors-pop}\right] = \\ = \frac{\left[Q'(7/8) - Q'\left(\frac{5}{8}\right) + Q'\left(\frac{3}{8}\right) - Q'(1/8)\right]\left[Q(6/8) - Q(2/8)\right] - \left[Q(7/8) - Q\left(\frac{5}{8}\right) + Q\left(\frac{3}{8}\right) - Q(1/8)\right]\left[Q'(6/8) - Q'(2/8)\right]}{\left[Q(6/8) - Q(2/8)\right]^2} \ldots (61)$$

$$\frac{\partial}{\partial\beta}\left[K_{Moors-pop}\right] = \\ = \frac{\left[Q'(7/8) - Q'\left(\frac{5}{8}\right) + Q'\left(\frac{3}{8}\right) - Q'(1/8)\right]\left[Q(6/8) - Q(2/8)\right] - \left[Q(7/8) - Q\left(\frac{5}{8}\right) + Q\left(\frac{3}{8}\right) - Q(1/8)\right]\left[Q'(6/8) - Q'(2/8)\right]}{\left[Q(6/8) - Q(2/8)\right]^2} \ldots (62)$$

The Jacobian matrix is $\begin{bmatrix} \frac{\partial}{\partial\alpha} S_{Galton-pop} & \frac{\partial}{\partial\beta} S_{Galton-pop} \\ \frac{\partial}{\partial\alpha} K_{Moors-pop} & \frac{\partial}{\partial\beta} K_{Moors-po} \end{bmatrix}$

$$\left(y - f\left(\boldsymbol{\theta}^{(\boldsymbol{a})}\right)\right) = \begin{bmatrix} S_{Galton-sam} & -\dfrac{Q(6/8) - 2Q(4/8) + Q(2/8)}{Q(6/8) - Q(2/8)} \\ K_{Moors-sample} & -\dfrac{Q(7/8) - Q\left(\frac{5}{8}\right) + Q\left(\frac{3}{8}\right) - Q(1/8)}{Q(6/8) - Q(2/8)} \end{bmatrix},$$

$$where\ \boldsymbol{\theta}^{(\boldsymbol{a})} = \begin{bmatrix} \alpha^{(a)} \\ \beta^{(a)} \end{bmatrix}$$

## 1.3.7.Maximum Likelihood Estimation :

Let $y_1, y_2, \dots, y_n$ an observed random sample from MBUW distribution with parameters $\alpha, \beta$. The pdf is defined in equation (5). The likelihood function and the log-likelihood function are defined in equations (63) and (64) respectively:

$$L(\alpha, \beta; y) = \left(\frac{6}{\alpha^{\beta}}\right)^{n} \prod_{i=1}^{n} \left[1 - y_i^{\frac{1}{\alpha^{\beta}}}\right] \prod_{i=1}^{n} y_i^{\left(\frac{2}{\alpha^{\beta}} - 1\right)} \quad , \quad 0 < y < 1\, , \qquad \alpha\ \&\beta > 0 \dots \dots (63)$$

$$l(\alpha, \beta; y) = nln(6) - n\beta ln(\alpha) + \sum_{i=1}^{n} ln\left[1 - y_i^{\frac{1}{\alpha^{\beta}}}\right] + \left(\frac{2}{\alpha^{\beta}} - 1\right) \sum_{i=1}^{n} ln(y_i) \dots \dots \dots \dots (64)$$

The Maximum Likelihood (ML) equations are the first derivative of the log-likelihood function concerning each parameter as defined in equations (65-66):

$$\frac{dl(\alpha, \beta; y)}{d\alpha} = \frac{-\beta n}{\alpha} + \sum_{i=1}^{n} \frac{-y_i^{\frac{1}{\alpha^{\beta}}} (ln[y_i]) \left(-\beta \alpha^{-\beta-1}\right)}{1 - y_i^{\frac{1}{\alpha^{\beta}}}} - 2\beta \alpha^{-\beta-1} \sum_{i=1}^{n} ln(y_i) \dots \dots \dots \dots \dots \dots . . (65)$$

$$\frac{dl(\alpha, \beta; y)}{d\beta} = -nln\alpha + \alpha^{-\beta} \sum_{i=1}^{n} \frac{y_i^{\frac{1}{\alpha^{\beta}}} ln[\alpha] ln[y]}{1 - y_i^{\frac{1}{\alpha^{\beta}}}} - 2\alpha^{-\beta} [ln\alpha] \sum_{i=1}^{n} ln(y_i) \dots \dots \dots \dots \dots \dots \dots \dots (66)$$

The system of non-linear equation $\frac{dl(\alpha,\beta;y)}{d\alpha} = 0$ and $\frac{dl(\alpha,\beta;y)}{d\beta} = 0$ are solved numerically using optimization algorithm like Quasi-Newton method or Nelder Mead algorithm to obtain estimators for both parameters $\alpha$ and $\beta$. The expected information matrix $J_y$ is the negative expected value of the second derivative of the log-likelihood evaluated at the estimated parameter. Under some regularity conditions and with a large sample size, the inverse of this information matrix is the variance of the estimated parameter. This estimator is approximately normally distributed with

$$\sqrt{n} \left(\begin{bmatrix} \alpha \\ \beta \end{bmatrix} - \begin{bmatrix} \hat{\alpha} \\ \hat{\beta} \end{bmatrix}\right) \sim N\left(0, J_y^{-1}\right)$$

This is used to construct an approximate confidence interval for the parameter in the form of $\hat{\alpha} \pm z_{1-\frac{\gamma}{2}}\, se(\hat{\alpha})$. SE is the square root of the inverse of the expected information matrix. $z_{1-\frac{\gamma}{2}}$ is the quantile of standard normal distribution.

# Results

## Section 4: Some real data analysis:

The OECD is used. Some variables are analyzed to discover what distributions fit the data better. The data is available at:

https://stats.oecd.org/index.aspx?DataSetCode=BLI

***First data***: shown in table (1) (Dwelling Without Basic Facilities)

These observations measure the percentage of homes in the involved countries that lack essential utilities like indoor plumbing, central heating, and clean drinking water supplies.

Table (1): Dwelling without basic facilities

| **0.008** | **0.007** | **0.002** | **0.094** | **0.123** | **0.023** | **0.005** | **0.005** | **0.057** | **0.004** |
|---|---|---|---|---|---|---|---|---|---|
| **0.005** | **0.001** | **0.004** | **0.035** | **0.002** | **0.006** | **0.064** | **0.025** | **0.112** | **0.118** |
| **0.001** | **0.259** | **0.001** | **0.023** | **0.009** | **0.015** | **0.002** | **0.003** | **0.049** | **0.005** |
| **0.001** | **0.03** | **0.067** | **0.138** | **0.359** | | | | | |

***Second data*** : shown in table(2) (Quality of Support Network)

This data set explores the extent to which a person can rely on sources of support like family, friends, or community members in times of need and disparate. It is represented as a percentage of persons who found social support during times of crisis.

Table (2): Quality of Support Network

| **0.92** | **0.93** | **0.88** | **0.80** | **0.82** | **0.96** | **0.95** | **0.96** | **0.94** | **0.90** |
|---|---|---|---|---|---|---|---|---|---|
| **0.78** | **0.98** | **0.89** | **0.92** | **0.91** | **0.77** | **0.94** | **0.95** | **0.96** | **0.85** |

***Third data***: shown in table ( 3) ( Voter Turnout )

This dataset evaluates the percentage of eligible and qualified individuals who cast their votes in elections, reflecting the level of democracy in the country

Table (3): Voter turnout data set

| 0.92 | 0.76 | 0.88 | 0.68 | 0.47 | 0.53 | 0.66 | 0.62 | 0.85 | 0.64 |
|---|---|---|---|---|---|---|---|---|---|
| 0.69 | 0.75 | 0.79 | 0.58 | 0.70 | 0.81 | 0.63 | 0.67 | 0.73 | 0.53 |
| 0.77 | 0.55 | 0.57 | 0.90 | 0.63 | 0.79 | 0.82 | 0.78 | 0.68 | 0.49 |
| 0.66 | 0.53 | 0.72 | 0.87 | 0.45 | 0.86 | 0.68 | 0.65 | | |

***Fourth data***: shown in table (4) ( Flood Data)
This includes 20 observations regarding the maximum flood levels in the Susquehanna River at Harrisburg, Pennsylvania (Dumonceaux & Antle, 1973).

Table (4): Flood data set

| 0.26 | 0.27 | 0.3 | 0.32 | 0.32 | 0.34 | 0.38 | 0.38 | 0.39 | 0.4 |
|---|---|---|---|---|---|---|---|---|---|
| 0.41 | 0.42 | 0.42 | 0.42 | 0.45 | 0.48 | 0.49 | 0.61 | 0.65 | 0.74 |

***Fifth data***: shown in table (5) (Time between Failures of Secondary Reactor Pumps)(Maya et al., 2024, 1999)(Suprawhardana and Prayoto)

Table (5): time between failures data set

| 0.216 | 0.015 | 0.4082 | 0.0746 | 0.0358 | 0.0199 | 0.0402 | 0.0101 | 0.0605 |
|---|---|---|---|---|---|---|---|---|
| 0.0954 | 0.1359 | 0.0273 | 0.0491 | 0.3465 | 0.007 | 0.656 | 0.106 | 0.0062 |
| 0.4992 | 0.0614 | 0.532 | 0.0347 | 0.1921 | | | | |

***Sixth data***: shown in table (6) ( data to evaluate the factors concerning the unit capacity, data compare factors between algorithms like SC 16 and P3 )(Maya et al., 2024, 1999).

Table (6): unit capacity factors data set

| 0.853 | 0.759 | 0.866 | 0.809 | 0.717 | 0.544 | 0.492 | 0.403 | 0.344 |
|---|---|---|---|---|---|---|---|---|
| 0.213 | 0.116 | 0.116 | 0.092 | 0.07 | 0.059 | 0.048 | 0.036 | 0.029 |
| 0.021 | 0.014 | 0.011 | 0.008 | 0.006 | | | | |

## 4.1. Maximum Likelihood Estimation:

The analysis of the data involves fitting the beta distribution, Kumaraswamy distribution, Median Weibull distribution (MBUW), and median-based Rayleigh distribution (MBUR). These distributions are compared using several criteria: negative log-likelihood (2LL), Akaike Information Criterion (AIC), corrected Akaike Information

Criterion (CAIC), Bayesian Information Criterion (BIC), and Hannan-Quinn Information Criterion (HQIC). Additionally, the Kolmogorov-Smirnov (K-S) test is conducted, and its p-value is reported in relation to the null hypothesis ($H_0$), which states that the dataset follows the tested distribution. If the data does not conform to the distribution, the null hypothesis is rejected. Maximum Likelihood Estimation (MLE) is applied to fit the four distributions using Nelder Mead algorithm in MATLAB. Table 7 presents the descriptive statistics, while Figures 7-8 display the histogram and boxplot for each dataset, respectively.

These are the pdfs of the distributions used in the analysis of these 6 data sets,

1- Beta Distribution:

$$f(y;\alpha,\beta)=\frac{\Gamma(\alpha+\beta)}{\Gamma(\alpha)\Gamma(\beta)}y^{\alpha-1}(1-y)^{\beta-1}\,,0<y<1\,,\alpha>0,\ \ \beta>0$$

2- Kumaraswamy Distribution:

$$f(y;\alpha,\beta)=\alpha\beta y^{\alpha-1}(1-y^{\alpha})^{\beta-1}\,,0<y<1\,,\qquad \alpha>0,\ \ \beta>0$$

3- Median Based Unit Rayleigh:

$$f(y;\ \alpha)=\frac{6}{\alpha^2}\left[1-y^{\frac{1}{\alpha^2}}\right]y^{\left(\frac{2}{\alpha^2}-1\right)}\ ,\ \ 0<y<1\,,\qquad \alpha>0$$

Comparison tools are: (k) is the number of parameters while (n) is the number of observations.

$$AIC=-2MLL+2k\quad,\qquad CAIC=-2MLL+\frac{2k}{n-k-1}\quad,\qquad BIC=-2MLL+k\log(n)$$

$$HQIC=2\log\{\log(n)[k-2MLL]\}\ \text{ or }\ \ HQIC=-2\,MLL+2k\log\{\log(n)\}$$

$$KS-test=Sup_n\,|F_n-F|\ ,\ \ F_n=\frac{1}{n}\sum_{i=1}^{n}I_{x_i<x}$$

$$CVM-test=\frac{1}{12n}+\sum_{i=1}^{n}\left\{F(x_i)-\frac{2i-1}{2n}\right\}^2$$

$$AD-test=-n-\sum_{i=1}^{n}\left(\frac{2i-1}{n}\right)\{log[F(x_i)]+log[1-F(x_{n-i+1})]\}$$

Table (7) shows the descriptive statistics for each data set.

| | Dwellings | Quality support | Voter data | Flood data | Time Between failures | Unit capacity |
|---|---|---|---|---|---|---|
| Min | 0.001 | 0.77 | 0.45 | 0.26 | 0.0062 | 0.006 |
| Mean | 0.0475 | 0.9005 | 0.6911 | 0.4225 | 0.1578 | 0.2881 |
| St. dev. | 0.0777 | 0.064 | 0.1251 | 0.1244 | 0.1931 | 0.3181 |
| Skewness | 2.6628 | -0.9147 | -0.0456 | 1.1625 | 1.4614 | 0.8223 |
| Kurtosis | 10.9269 | 2.6716 | 2.2538 | 4.2363 | 3.9988 | 2.0246 |
| Q(1/4) | 0.004 | 0.865 | 0.62 | 0.33 | 0.0292 | 0.0308 |
| Q(2/4) | 0.009 | 0.92 | 0.68 | 0.405 | 0.0614 | 0.116 |
| Q(3/4) | 0.0623 | 0.95 | 0.78 | 0.465 | 0.21 | 0.531 |
| max | 0.359 | 0.98 | 0.92 | 0.74 | 0.656 | 0.866 |

In table 7 the descriptive analysis reveals that the second and third data sets exhibit significant negative skewness, characterized by a tail on the left side. In contrast, the other data sets show positive skewness, with a range of right tails; the first data set has a long tail, while the last data set has a shorter tail. The second, third, and last data sets are platykurtic, demonstrating negative excess kurtosis (kurtosis less than 3, indicating a thinner tail). Conversely, the other data sets are leptokurtic, exhibiting positive excess kurtosis (kurtosis greater than 3, indicating a fatter tail). Figures 10-11 show the histogram and boxplot for the datasets. The first, fourth and fifth datasets show some extreme values in the right upper tail. Table 8-13 show the metrics obtained from using MLE utilizing the Nelder Mead algorithm to fit the dataset respectively.

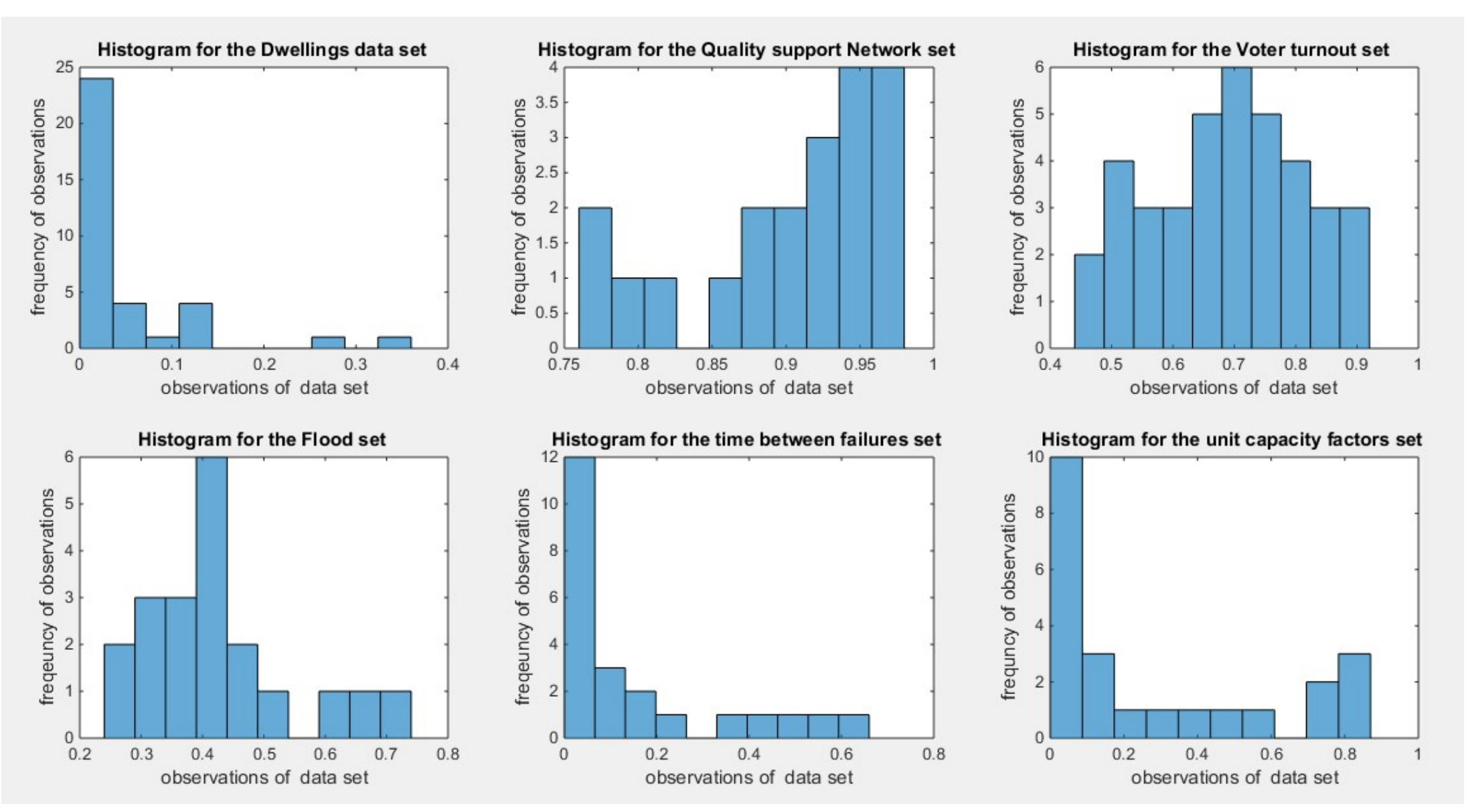

Fig.10 shows the histogram for each data set.

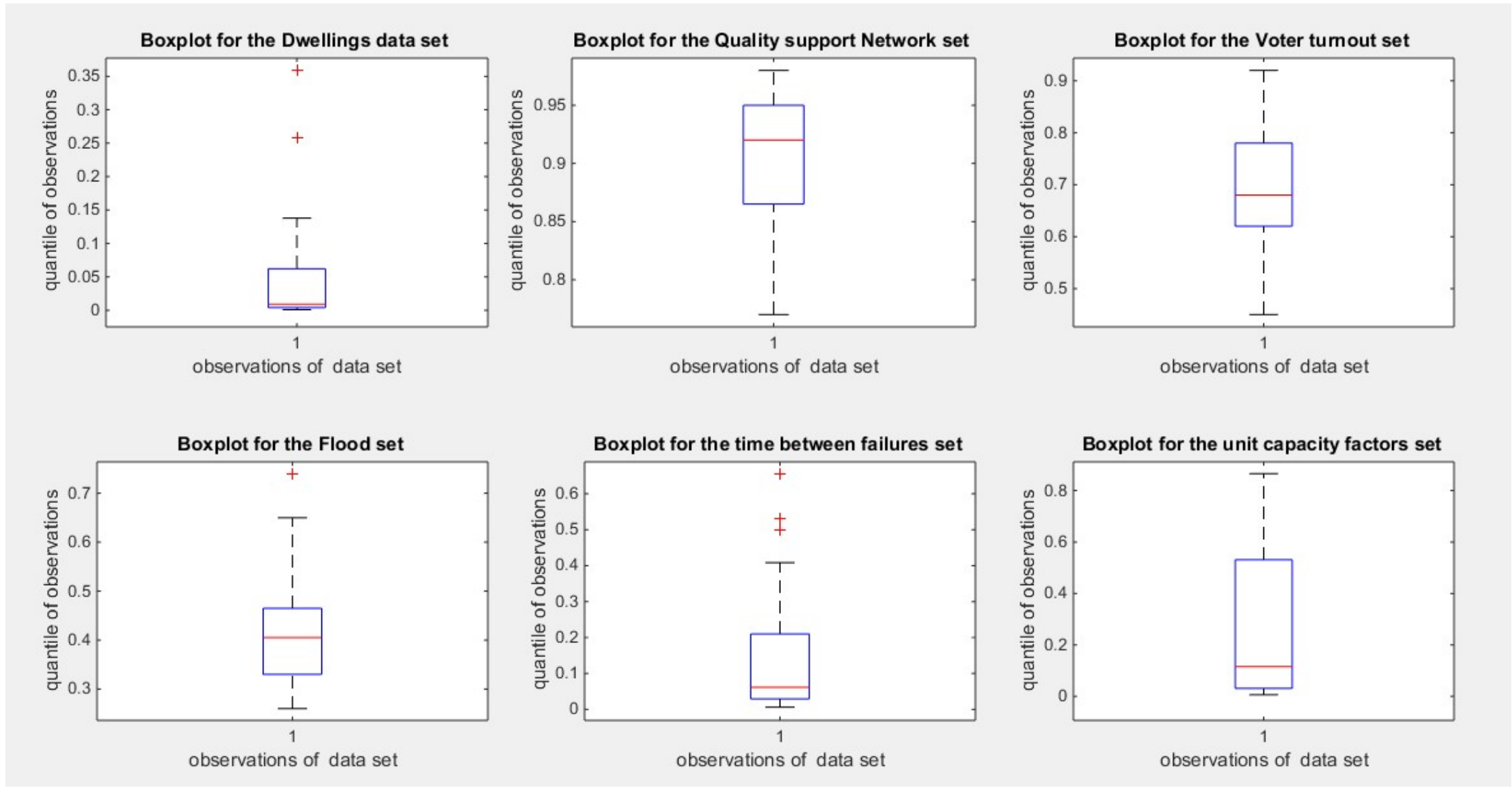

Fig.11 shows boxplot for each data set

## Table 8: metrics of the first data set (Dwelling Without Basic Facilities, n=35):

| | Beta | | Kumaraswamy | | MBUR | MBUW | |
|---|---|---|---|---|---|---|---|
| theta | $\alpha = 0.4798$ | | $\alpha = 0.5700$ | | 2.2834 | $\alpha = 6.636$ | |
| | $\beta = 9.3996$ | | $\beta = 6.2579$ | | | $\beta = 0.8726$ | |
| Var | 0.0222 | 0.3127 | 0.0074 | 0.1403 | 0.0011 | $3.5 * 10^6$ | $-2.5 * 10^5$ |
| | 0.3127 | 7.8562 | 0.1403 | 3.7854 | | $-2.5 * 10^5$ | $1.7 * 10^4$ |
| SE | 0.025 | | 0.015 | | 0.006 | 316.227 | |

|  | 0.474 | 0.329 |  | 22.039 |
|---|---|---|---|---|
| AIC | -153.3608 | -155.8059 | -146.585 | -144.585 |
| CAIC | -152.9858 | -155.4309 | -149.4638 | -144.21 |
| BIC | -150.2501 | -152.6952 | -145.0296 | -141.4743 |
| HQIC | -3.657 | -3.6878 | -6.0791 | -3.5422 |
| NLL | -78.6804 | -79.9029 | -74.2925 | -74.2925 |
| K-S | 0.1818 | 0.1570 | 0.1794 | 0.1794 |
| $H_0$ | Fail to reject | Fail to reject | Fail to reject | Fail to Reject |
| P-value | 0.1744 | 0.3202 | 0.1860 | 0.1860 |
| AD test | 1.1431 | 0.8327 | 2.3889 | 2.3889 |
| CVM | 0.2028 | 0.1372 | 0.3966 | 0.3966 |

Confidence interval for the estimator obtained from fitting MBUR is (2.27164 , 2.29516)

The 4 distributions fit the data because the null hypothesis for the K-S test is failed to be rejected for the 4 distributions. Kumaraswamy is the best followed by beta, MBUR and lastly MBUW. MBUR is special case of MBUW when b=2. Kumaraswamy has the largest negative AIC, CIA-corrected, BIC and HQIC compared to other distribution. The smallest values of KS-test, AD and CVM tests for Kumaraswamy distribution support this result. Figure (12) shows the fitted theoretical CDFs of the 4 distributions vs the empirical CDF. The horizontal axis shows the observations from the sample which are bounded on the unit interval. The variance of the estimators of MBUW is very large which leads to invalid estimation of the standard error. Using the GMM and the percentile estimator dramatically reduce the variance. Hence good estimation of the standard error and construction of the confidence interval.

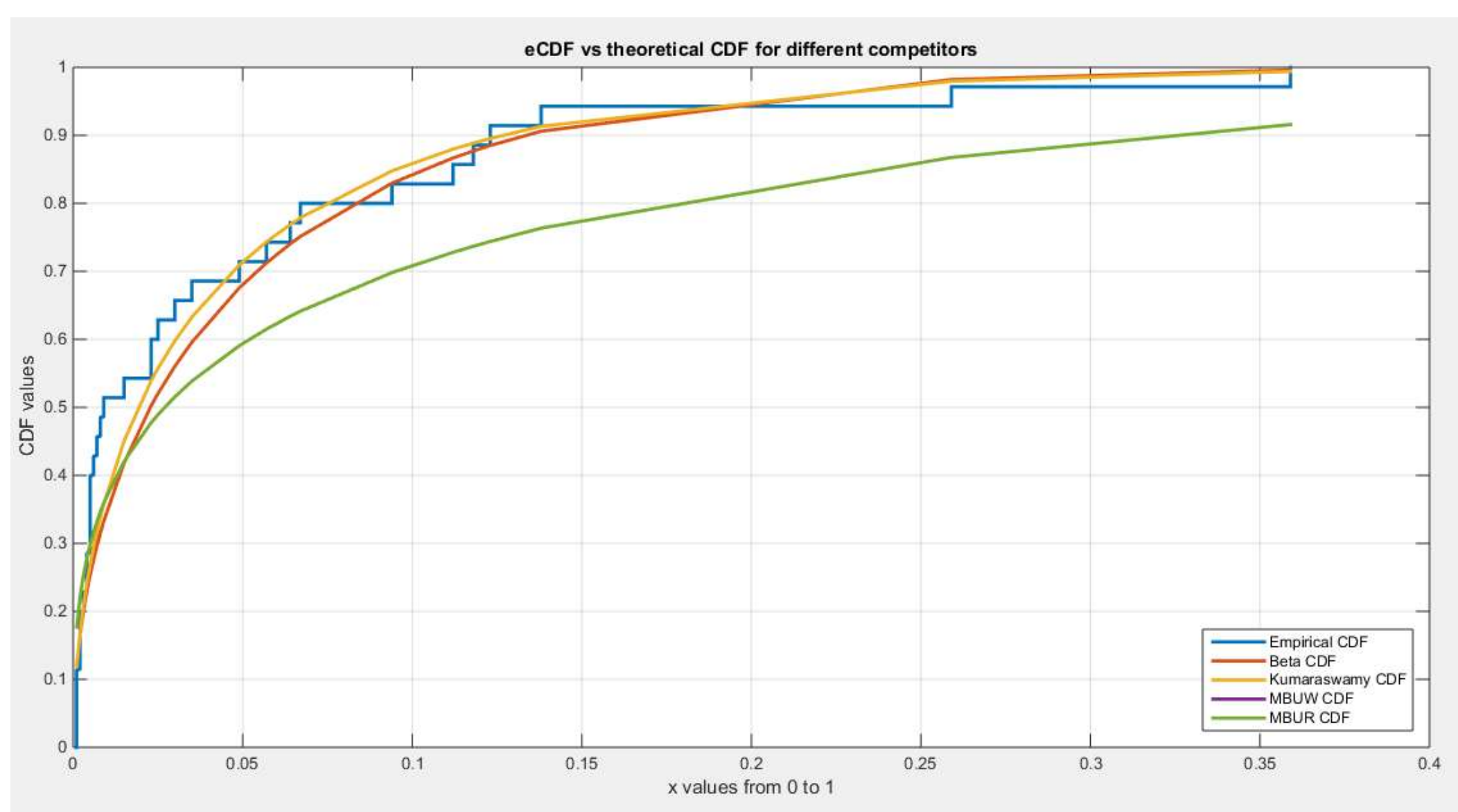

Fig. 12: shows the eCDF vs. theoretical CDF of the 4 distributions for the 1st data set (both curves of MBUR and MBUW are overlapped).

Table 9: Second data set (Quality of Support Network, n=20):

| | Beta | | Kumaraswamy | | MBUR | MBUW | |
|---|---|---|---|---|---|---|---|
| theta | $\alpha = 21.7353$ | | $\alpha = 16.5447$ | | 0.3591 | $\alpha = 0.257$ | |
| | $\beta = 2.4061$ | | $\beta = 2.772$ | | | $\beta = 1.5073$ | |
| Var | 86.4614 | 9.0379 | 15.7459 | 3.2005 | 0.0008 | | |
| | 9.0379 | 1.0646 | 3.2005 | 1.0347 | | | |
| SE | 2.079 | | 0.887 | | 0.006 | Cannot be estimated | |
| | 0.231 | | 0.227 | | | Cannot be estiamted | |
| AIC | -56.5056 | | -56.7274 | | -58.079 | -56.079 | |
| CAIC | -55.7997 | | -56.0215 | | -57.8567 | -55.3731 | |
| BIC | -54.5141 | | -54.7359 | | -57.0832 | -54.0875 | |
| HQIC | -2.4304 | | -2.4377 | | -4.6106 | -2.4163 | |
| NLL | -30.2528 | | -30.3637 | | -30.0395 | -30.0395 | |
| K-S | 0.0974 | | 0.0995 | | 0.1309 | 0.1309 | |
| $H_0$ | Fail to reject | | Fail to reject | | Fail to reject | Fail to Reject | |
| P-value | 0.9416 | | 0.9513 | | 0.8399 | 0.8399 | |
| AD | 0.3828 | | 0.3527 | | 0.3184 | 0.3184 | |
| CVM | 0.0566 | | 0.0498 | | 0.0407 | 0.0407 | |

Confidence interval for the estimator obtained from fitting MBUR is (0.346542 ,   0.371658)

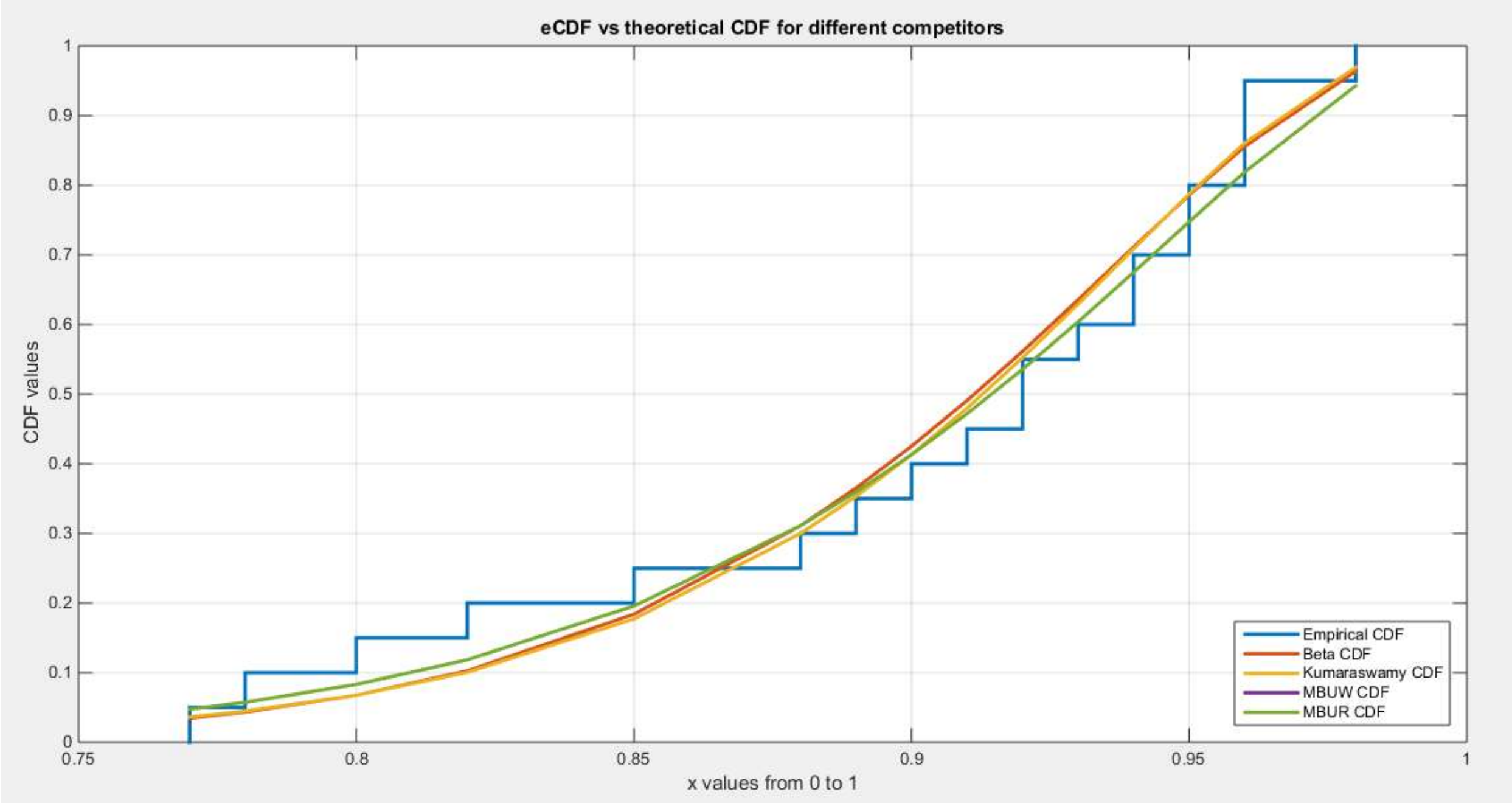

Fig. 13: shows the eCDF vs. theoretical CDF of the 4 distributions for the 2nd data set (both curves of MBUR and MBUW are overlapped).

The 4 distributions fit the data because the null hypothesis for the K-S test is failed to be rejected for the 4 distributions. MBUR is the best followed by Kumaraswamy, beta and lastly MBUW. MBUR is special case of MBUW when b=2. MBUR has the largest negative AIC, CIA-corrected, BIC and HQIC compared to other distribution. The smallest values of AD and CVM tests for MBUR distribution support this result. Figure (13) shows the fitted theoretical CDFs of the 4 distributions vs the empirical CDF. The horizontal axis shows the observations from the sample which are bounded on the unit interval. The variance of the estimators of MBUW is very large which leads to invalid estimation of the standard error. Using the GMM and the percentile estimator dramatically reduce the variance. Hence good estimation of the standard error and construction of the confidence interval.

Table 10: Third data set (Voter turnout , n=38):

<table>
<tr><td></td><td colspan="2">Beta</td><td colspan="2">Kumaraswamy</td><td>MBUR</td><td colspan="2">MBUW</td></tr>
<tr><td rowspan="2">theta</td><td colspan="2">$\alpha = 8.6959$</td><td colspan="2">$\alpha = 5.6224$</td><td rowspan="2">0.6832</td><td colspan="2">$\alpha = 0.5509$</td></tr>
<tr><td colspan="2">$\beta = 3.8673$</td><td colspan="2">$\beta = 4.5073$</td><td colspan="2">$\beta = 1.2779$</td></tr>
<tr><td rowspan="2">Var</td><td>6.4427</td><td>2.3588</td><td>0.7462</td><td>0.9005</td><td rowspan="2">0.0016</td><td></td><td></td></tr>
<tr><td>2.3588</td><td>0.9809</td><td>0.9005</td><td>1.6212</td><td></td><td></td></tr>
<tr><td rowspan="2">SE</td><td colspan="2">0.412</td><td colspan="2">0.140</td><td rowspan="2">0.006</td><td colspan="2">Cannot be estimated</td></tr>
<tr><td colspan="2">0.161</td><td colspan="2">0.207</td><td colspan="2">Cannot be estimated</td></tr>
<tr><td>AIC</td><td colspan="2">-47.9451</td><td colspan="2">-46.8575</td><td>-42.1377</td><td colspan="2">-40.1377</td></tr>
<tr><td>CAIC</td><td colspan="2">-47.6022</td><td colspan="2">-46.5146</td><td>-42.0266</td><td colspan="2">-39.7948</td></tr>
<tr><td>BIC</td><td colspan="2">-44.6699</td><td colspan="2">-43.5823</td><td>-40.5001</td><td colspan="2">-36.8625</td></tr>
<tr><td>HQIC</td><td colspan="2">-1.3488</td><td colspan="2">-1.3065</td><td>-3.6057</td><td colspan="2">-1.0231</td></tr>
<tr><td>LL</td><td colspan="2">25.9725</td><td colspan="2">25.4287</td><td>22.0688</td><td colspan="2">22.0688</td></tr>
<tr><td>K-S</td><td colspan="2">0.0938</td><td colspan="2">0.1048</td><td>0.1364</td><td colspan="2">0.1364</td></tr>
<tr><td>$H_0$</td><td colspan="2">Fail to reject</td><td colspan="2">Fail to reject</td><td>Fail to reject</td><td colspan="2">Fail to Reject</td></tr>
<tr><td>P-value</td><td colspan="2">0.8605</td><td colspan="2">0.7596</td><td>0.4401</td><td colspan="2">0.4401</td></tr>
<tr><td>AD</td><td colspan="2">0.2721</td><td colspan="2">0.3464</td><td>1.3262</td><td colspan="2">1.3262</td></tr>
<tr><td>CVM</td><td colspan="2">0.0416</td><td colspan="2">0.0559</td><td>0.2193</td><td colspan="2">0.2193</td></tr>
</table>

Confidence interval for the estimator obtained from fitting MBUR is (0.67144 , 0.69496)

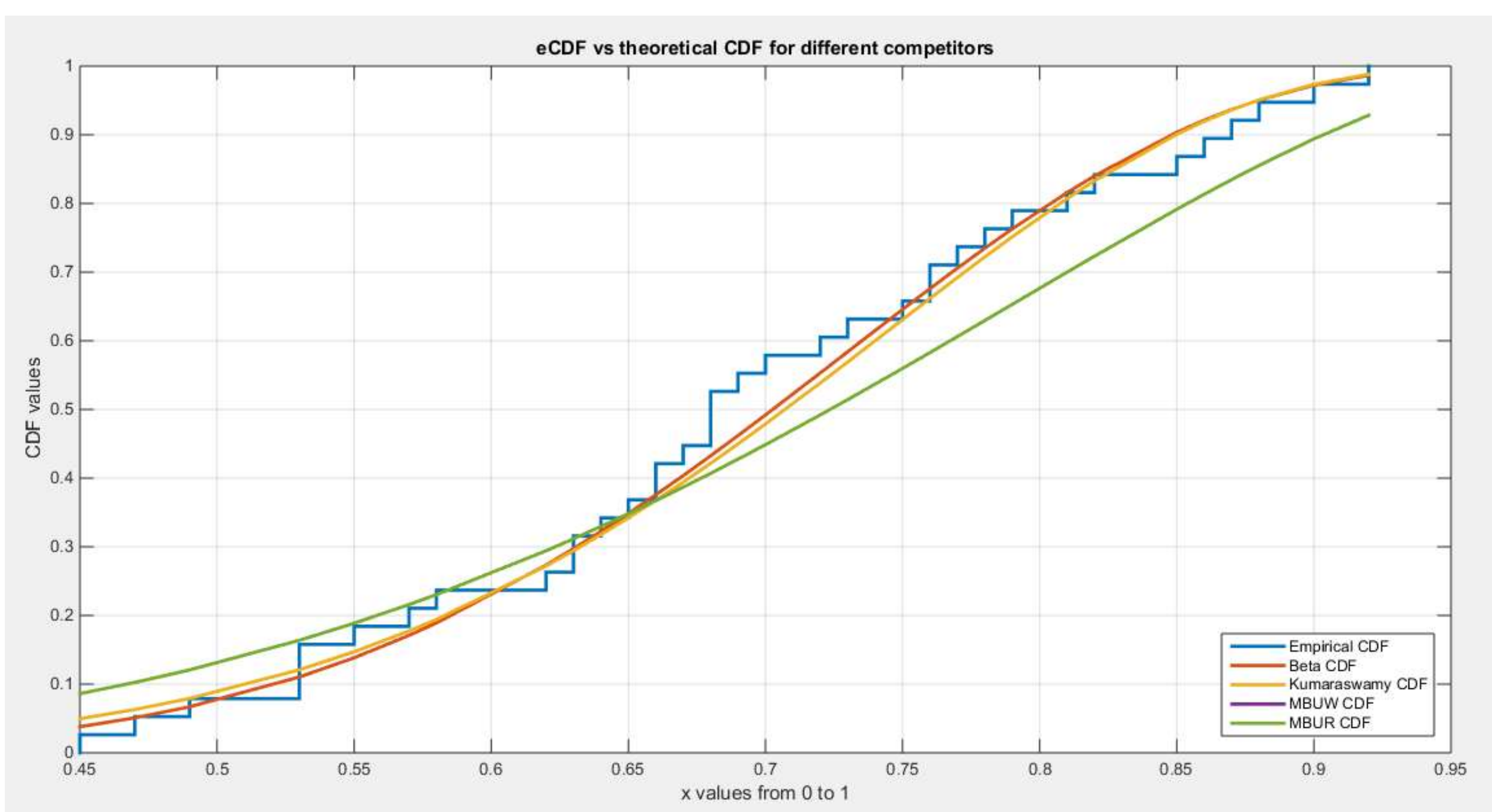

Fig. 14 shows the eCDF vs. theoretical CDF of the 4 distributions for the 3rd data set (both curves of MBUR and MBUW are overlapped).

The 4 distributions fit the data because the null hypothesis for the K-S test is failed to be rejected for the 4 distributions. Beta distribution is the best followed by Kumaraswamy, MBUR and lastly MBUW. MBUR is special case of MBUW when b=2. Beta distribution has the largest negative AIC, CIA-corrected, BIC and HQIC compared to other distribution. The smallest values of KS-test, AD and CVM tests for Beta distribution support this result. Figure (14) shows the fitted theoretical CDFs of the 4 distributions vs the empirical CDF. The horizontal axis shows the observations from the sample which are bounded on the unit interval. The variance of the estimators of MBUW is very large which leads to invalid estimation of the standard error. Using the GMM and the percentile estimator dramatically reduce the variance. Hence good estimation of the standard error and construction of the confidence interval.

Table 11: Fourth data set (Flood data , n=20):

| | Beta | | Kumaraswamy | | MBUR | MBUW | |
|---|---|---|---|---|---|---|---|
| theta | $\alpha = 6.8318$ | | $\alpha = 3.3777$ | | 1.0443 | $\alpha = 1.0932$ | |
| | $\beta = 9.2376$ | | $\beta = 12.0057$ | | | $\beta = 0.9719$ | |
| Var | 7.22 | 7.2316 | 0.3651 | 2.8825 | 0.007 | $1.66 * 10^4$ | $-1.7 * 10^5$ |
| | 7.2316 | 8.0159 | 2.8825 | 29.9632 | | $-1.7 * 10^5$ | $1.6 * 10^6$ |
| SE | 0.601 | | 0.135 | | 0.019 | 28.81 | |
| | 0.633 | | 1.224 | | | 28.81 | |
| AIC | -24.3671 | | -21.9465 | | -10.9233 | -8.9233 | |
| CAIC | -23.6613 | | -21.2407 | | -10.7011 | -8.2174 | |
| BIC | -22.3757 | | -19.9551 | | -9.9276 | -6.9319 | |
| HQIC | -0.9154 | | -0.737 | | -1.537 | 0.657 | |
| LL | 14.1836 | | 12.9733 | | 6.4617 | 6.4617 | |
| K-S | 0.2063 | | 0.2175 | | 0.3202 | 0.3202 | |
| $H_0$ | Fail to reject | | Fail to reject | | Fail to reject | Fail to Reject | |
| P-value | 0.3174 | | 0.2602 | | 0.0253 | 0.0253 | |
| AD | 0.7302 | | 0.9365 | | 2.7563 | 2.7563 | |
| CVM | 0.1242 | | 0.1653 | | 0.531 | 0.531 | |

Confidence interval for the estimator obtained from fitting MBUR is (1.004533 , 1.084067)

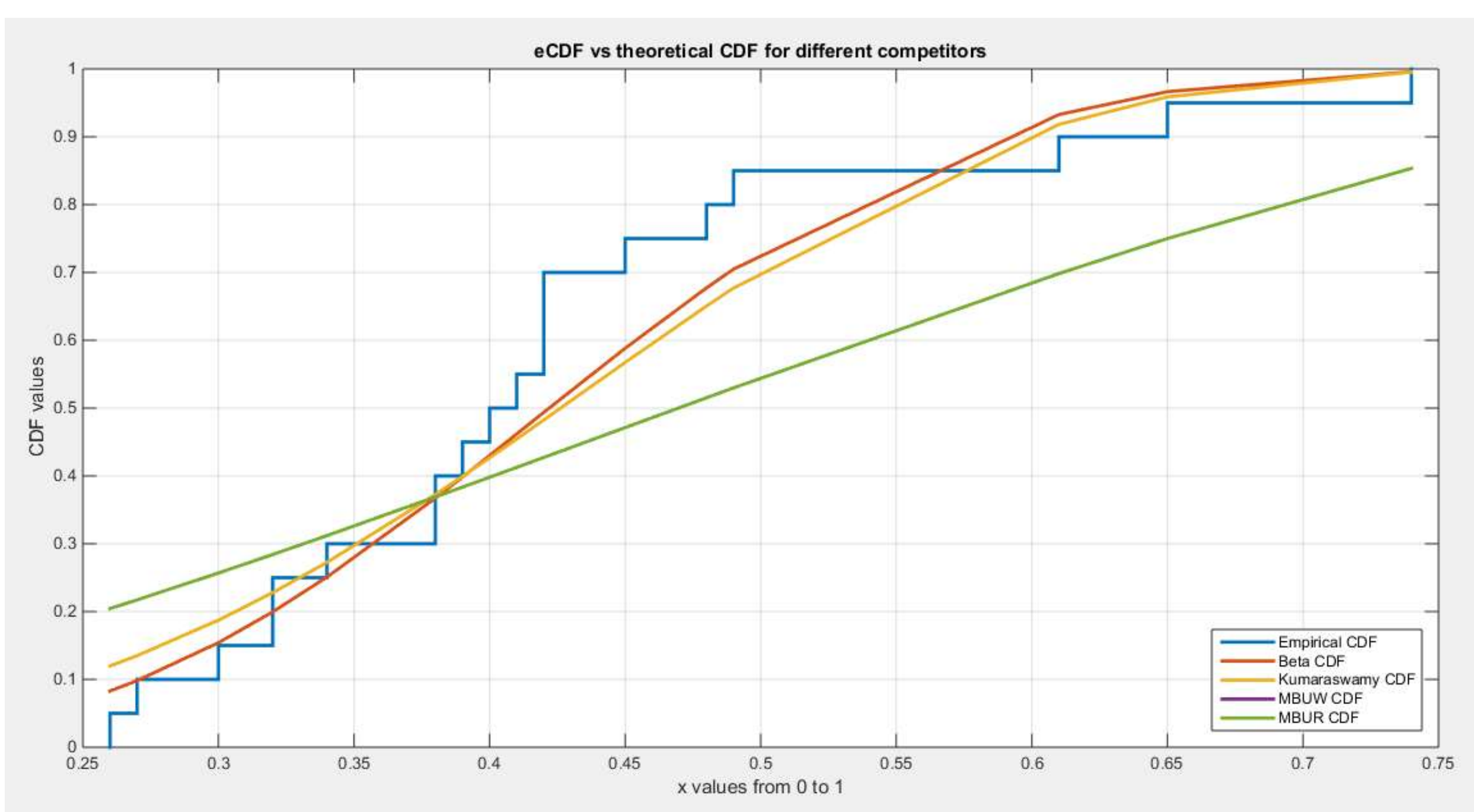

Fig. 15: shows the eCDF vs. theoretical CDF of the 4 distributions for the 4th data set (both curves of MBUR and MBUW are overlapped).

The 4 distributions fit the data because the null hypothesis for the K-S test is failed to be rejected for the 4 distributions. Beta distribution is the best followed by Kumaraswamy, MBUR and lastly MBUW. MBUR is special case of MBUW when b=2. Beta distribution has the largest negative AIC, CIA-corrected, BIC and HQIC compared to other distribution. The smallest values of KS-test, AD and CVM tests for Beta distribution support this result. Figure (15) shows the fitted theoretical CDFs of the 4 distributions vs the empirical CDF. The horizontal axis shows the observations from the sample which are bounded on the unit interval. The variance of the estimators of MBUW is very large which leads to invalid estimation of the standard error. Using the GMM and the percentile estimator dramatically reduce the variance. Hence good estimation of the standard error and construction of the confidence interval.

Table 12: Fifth data set (Time between Failures, n=23):

<table>
<tr><td></td><td colspan="2">Beta</td><td colspan="2">Kumaraswamy</td><td>MBUR</td><td colspan="2">MBUW</td></tr>
<tr><td rowspan="2">theta</td><td colspan="2">$\alpha = 0.6307$</td><td colspan="2">$\alpha = 0.6766$</td><td rowspan="2">1.7886</td><td colspan="2">$\alpha = 5.1285$</td></tr>
<tr><td colspan="2">$\beta = 3.2318$</td><td colspan="2">$\beta = 2.936$</td><td colspan="2">$\beta = 0.7113$</td></tr>
<tr><td rowspan="2">Var</td><td>0.071</td><td>0.2801</td><td>0.0198</td><td>0.1033</td><td rowspan="2">0.018</td><td>$1.02 * 10^7$</td><td>$-8.7 * 10^5$</td></tr>
<tr><td>0.2801</td><td>1.647</td><td>0.1033</td><td>0.9135</td><td>$-8.7 * 10^5$</td><td>$7.4 * 10^4$</td></tr>
<tr><td rowspan="2">SE</td><td colspan="2">0.056</td><td colspan="2">0.029</td><td rowspan="2">0.028</td><td colspan="2">665</td></tr>
<tr><td colspan="2">0.268</td><td colspan="2">0.199</td><td colspan="2">56.7</td></tr>
<tr><td>AIC</td><td colspan="2">-36.0571</td><td colspan="2">-36.6592</td><td>-37.862</td><td colspan="2">-35.862</td></tr>
<tr><td>CAIC</td><td colspan="2">-35.4571</td><td colspan="2">-36.0592</td><td>-37.6712</td><td colspan="2">-35.262</td></tr>
<tr><td>BIC</td><td colspan="2">-33.7861</td><td colspan="2">-34.3882</td><td>-36.7265</td><td colspan="2">-33.591</td></tr>
<tr><td>HQIC</td><td colspan="2">-1.4232</td><td colspan="2">-1.453</td><td>-3.699</td><td colspan="2">-1.4134</td></tr>
<tr><td>NLL</td><td colspan="2">-20.0285</td><td colspan="2">-20.3296</td><td>-19.9310</td><td colspan="2">-19.9310</td></tr>
<tr><td>K-S</td><td colspan="2">0.1541</td><td colspan="2">0.1393</td><td>0.1584</td><td colspan="2">0.1584</td></tr>
<tr><td>$H_0$</td><td colspan="2">Fail to reject</td><td colspan="2">Fail to reject</td><td>Fail to reject</td><td colspan="2">Fail to Reject</td></tr>
<tr><td>P-value</td><td colspan="2">0.5918</td><td colspan="2">0.7123</td><td>0.5575</td><td colspan="2">0.5575</td></tr>
<tr><td>AD</td><td colspan="2">0.6886</td><td colspan="2">0.5755</td><td>0.6703</td><td colspan="2">0.6703</td></tr>
<tr><td>CVM</td><td colspan="2">0.1264</td><td colspan="2">0.0989</td><td>0.1253</td><td colspan="2">0.1253</td></tr>
</table>

Confidence interval for the estimator obtained from fitting MBUR is (1.730528 , 1.846672)

The 4 distributions fit the data because the null hypothesis for the K-S test is failed to be rejected for the 4 distributions. MBUR distribution is the best followed by Kumaraswamy, Beta and lastly MBUW. MBUR is special case of MBUW when b=2. MBUR distribution has the largest negative AIC, CIA-corrected, BIC and HQIC compared to other distributions. The smallest values of KS-test, AD and CVM tests for MBUR distribution support this result. Figure (16) shows the fitted theoretical CDFs of the 4 distributions vs the empirical CDF. The horizontal axis shows the observations from the sample which are bounded on the unit interval. The variance of the estimators of MBUW is very large which leads to invalid estimation of the standard error. Using the GMM and the percentile estimator dramatically reduce the variance. Hence good estimation of the standard error and construction of the confidence interval.

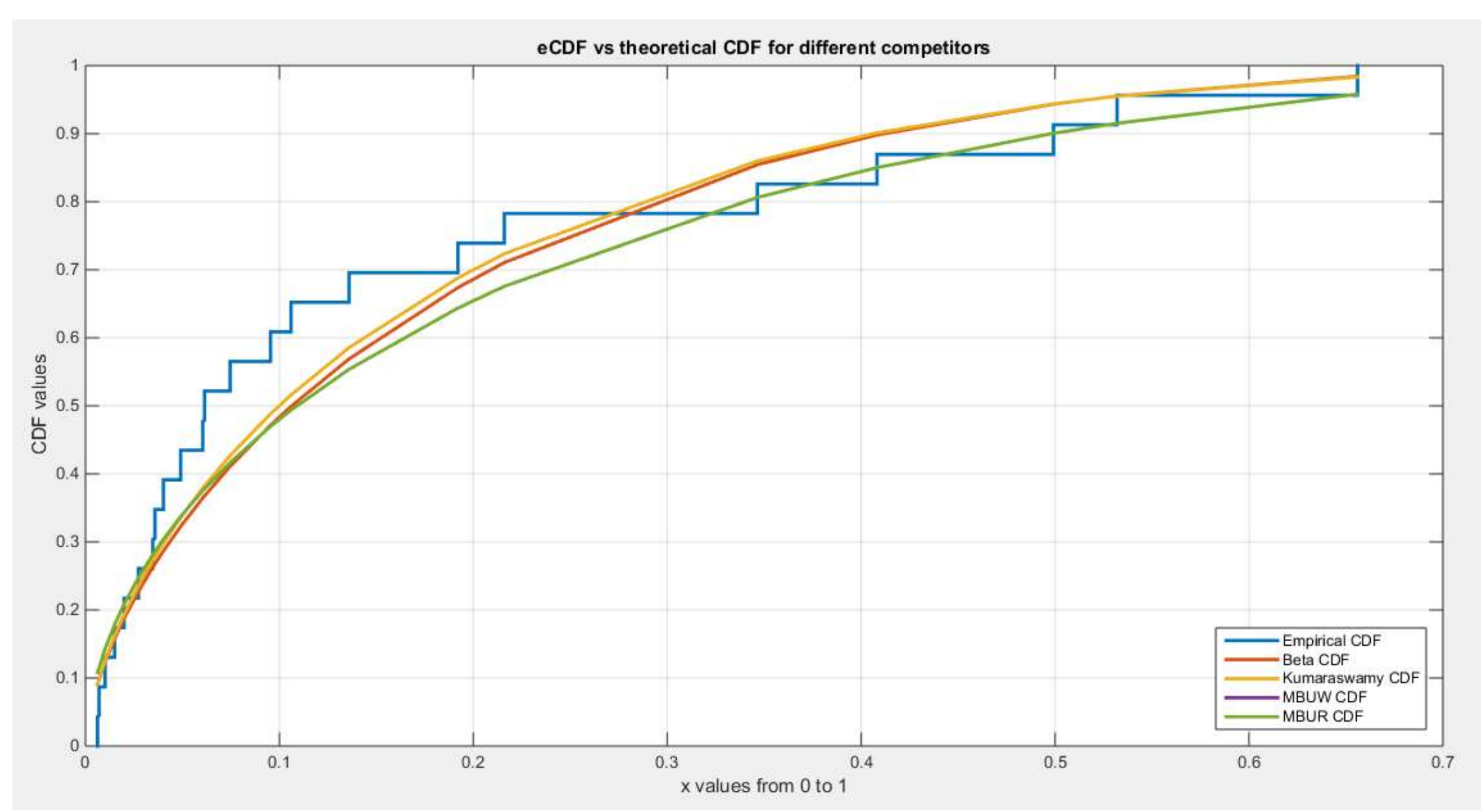

Fig. 16: shows the eCDF vs. theoretical CDF of the 4 distributions for the 5th data set (both curves of MBUR and MBUW are overlapped).

Table 13: Sixth data set (unit capacity data , n=23):

| | Beta | | Kumaraswamy | | MBUR | MBUW | |
|---|---|---|---|---|---|---|---|
| theta | $\alpha = 0.4869$ | | $\alpha = 0.5044$ | | 1.6243 | $\alpha = 3.7003$ | |
| | $\beta = 1.1679$ | | $\beta = 1.1862$ | | | $\beta = 0.7415$ | |
| Var | 0.0482 | 0.0960 | 0.0166 | 0.0274 | 0.0149 | $5.2 * 10^6$ | $-7.8 * 10^5$ |
| | 0.0960 | 0.2919 | 0.0274 | 0.1066 | | $-7.8 * 10^5$ | $1.2 * 10^5$ |
| SE | 0.046 | | 0.027 | | 0.025 | 475 | |
| | 0.113 | | 0.068 | | | 72.2 | |
| AIC | -15.2149 | | -15.3416 | | -13.2158 | -11.2158 | |
| CAIC | -14.6149 | | -14.7416 | | -13.0253 | -10.6158 | |
| BIC | -12.9439 | | -13.0706 | | -12.0803 | -8.9448 | |
| HQIC | 0.0461 | | 0.0329 | | -1.7728 | 0.5128 | |
| NLL | -9.6075 | | -9.6708 | | -7.6079 | -7.6079 | |
| K-S | 0.1836 | | 0.1790 | | 0.1518 | 0.1518 | |
| $H_0$ | Fail to reject | | Fail to reject | | Fail to reject | Fail to Reject | |
| P-value | 0.3742 | | 0.4051 | | 0.4074 | 0.4074 | |

| AD | 0.6886 | 0.5755 | 1.9075 | 1.9075 |
|---|---|---|---|---|
| CVM | 0.1264 | 0.0989 | 0.2033 | 0.2033 |

Confidence interval for the estimator obtained from fitting MBUR is (1.57245 , 1.67615)

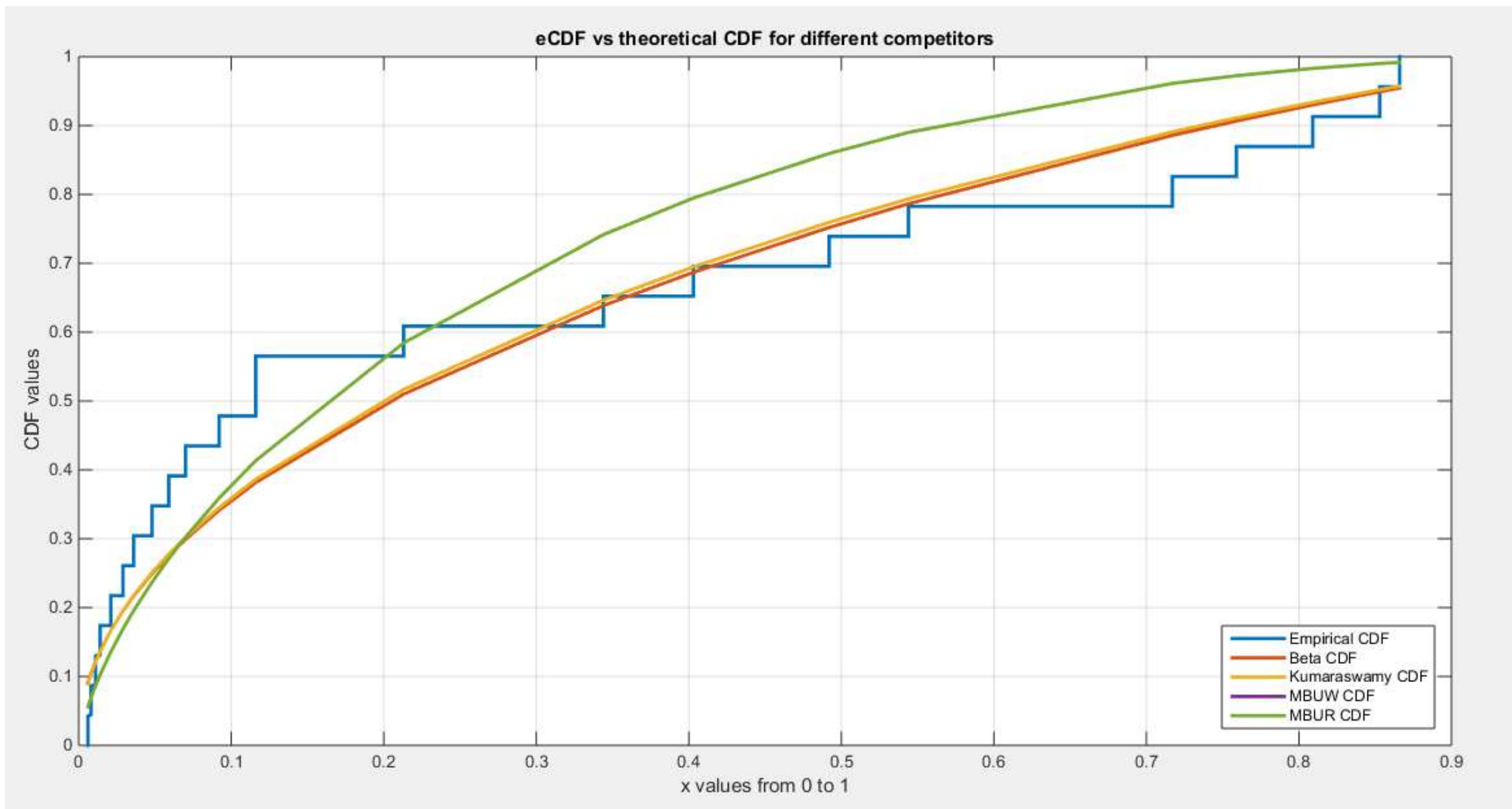

Fig. 17: shows the eCDF vs. theoretical CDF of the 4 distributions for the 6th data set (both curves of MBUR and MBUW are overlapped).

The 4 distributions fit the data because the null hypothesis for the K-S test is failed to be rejected for the 4 distributions. Kumaraswamy distribution is the best followed by beta, MBUR and lastly MBUW. MBUR is special case of MBUW when b=2. Kumaraswamy distribution has the largest negative AIC, CIA-corrected, BIC and HQIC compared to other distributions. The smallest values of KS-test, AD and CVM tests for kumaraswamy distribution support this result. Figure (17) shows the fitted theoretical CDFs of the 4 distributions vs the empirical CDF. The horizontal axis shows the observations from the sample which are bounded on the unit interval. The variance of the estimators of MBUW is very large which leads to invalid estimation of the standard error. Using the GMM and the percentile estimator dramatically reduce the variance. Hence good estimation of the standard error and construction of the confidence interval.

## 4.2. Generalized Method of Moments and Percentile Estimators

Tables 14-19 illustrate the comparisons between methods explained in section 3 for the respective datasets. Methods used for each dataset are illustrated with figures for the comparison between the methods. Figures 18-29 show the QQ plot, PP plot and the fitted PDF for each data set.

Table 14 : comparisons between methods for dwelling datasets

| | Method1 (1,2 mom) | | Method2 (3,4mom) | | Method3 (skew,geo.m) | | Method4 (skew,geo.m) | | Method5 (y25,y75) | | Method6 (G.skew,M.kurtosis) | |
|---|---|---|---|---|---|---|---|---|---|---|---|---|
| Alpha & (SE) | 5.8343(0.1677) | | 5.6178(0.1633) | | 2.3201(0.0205) | | 2.314(0.0161) | | 2.3507(0.0065) | | 2.4125(0.0553) | |
| Beta & (SE) | 1.4939(0.0882) | | 1.8436(0.0718) | | 1.9586(0.1678) | | 1.9585(0.1683) | | 2.2778(0.1684) | | 1.8291(0.1597) | |
| Var-cov matirx | 0.9846 | 0.0606 | 0.933 | 0.1171 | 0.0147 | -0.1205 | 0.0091 | -0.0948 | 0.0015 | 0.0105 | 0.1071 | 0.3092 |
| | 0.0606 | 0.2722 | 0.1171 | 0.1805 | -0.1205 | 0.9852 | -0.0948 | 0.9909 | 0.0105 | 0.9923 | 0.3092 | 0.8926 |
| SSE | 0.0000825 | | 0.0000345 | | 0.0472 | | 0.0015 | | 0.0000133 | | 0.0622 | |
| AD | 23.1033 | | 49.8776 | | 2.4095 | | 2.4479 | | 3.3013 | | 2.73 | |
| CVM | 4.837 | | 8.3129 | | 0.4018 | | 0.4115 | | 0.5909 | | 0.4811 | |
| KS | 0.6327 | | 0.816 | | 0.1804 | | 0.1821 | | 0.2852 | | 0.1933 | |
| H0 | Reject | | Reject | | Fail to reject | | Fail to reject | | reject | | Fail to reject | |
| P(KS) | <0.0001 | | <0.0001 | | 0.1813 | | 0.1731 | | 0.0015 | | 0.1277 | |
| Galton kewness | - | | - | | - | | - | | - | | 0.8283 | |
| Moore kurtosis | - | | - | | - | | - | | - | | 1.6288 | |
| Sig.a (p-value) | 34.7845(<0.0001) | | 34.4072(<0.0001) | | 113.0573(<0.0001) | | 143.7754(<0.0001) | | 363.6903(<0.0001) | | 43.6173(<0.0001) | |
| Sig.b ( p-value) | 16.9402(<0.0001) | | 25.6715(<0.0001) | | 11.6740(<0.0001) | | 11.6399(<0.0001) | | 13.5282(<0.0001) | | 11.4534(<0.0001) | |
| M1 , M2 | 0.0475 , 0.0081 | | - | | - | | - | | - | | - | |
| M3 , M4 | - | | 0.0021, 0.00063 | | - | | - | | - | | - | |
| Mean Log(y) | - | | - | | -4.3058 | | -4.3058 | | | | - | |
| Y25 , Y75 | - | | - | | - | | - | | 0.004 , 0.0623 | | - | |
| Sk.Quartiles | - | | - | | 0.8283 | | - | | - | | - | |
| Sk.decile | - | | - | | - | | 0.8689 | | - | | - | |
| Deter | 0.2643 | | 0.1547 | | 0 | | 0 | | 0.0013 | | 0 | |
| trace | 1.2568 | | 1.1136 | | 1 | | 1 | | 0.9937 | | 0.9997 | |
| Eigenvalues | 0.9897 , 0.2671 | | 0.9508 , 0.1627 | | 0 , 1 | | 0 , 1 | | 0.0014 , 0.9924 | | 0 , 0.9997 | |
| | **Eigenvalues and variance-covariance matrix before applying the delta method** | | | | | | | | | | | |
| **Eigenvalues** | **10.2654 , 732.92** | | **49.1593 , 837.283** | | **0.0369 , 1000** | | **0.0376 , 1000** | | **7.6253 , 998.6482** | | **0.2805 , 1000** | |
| **Var-cov matrix** | **715.4703** | **-110.931** | **805.459** | **-155.14** | **498.478** | **-499.979** | **495.6207** | **-499.96** | **175.3735** | **-371.62** | **574.4254** | **-494.309** |
| | **-110.931** | **27.7151** | **-155.14** | **80.983** | **-499.98** | **501.5586** | **-499.96** | **504.419** | **-371.62** | **830.9** | **-494.309** | **425.855** |

Table 15: comparisons between methods for Quality of support network datasets

| | Method1 (1,2 mom) | Method2 (3,4mom) | Method3 (skew,geo.m) | Method4 (skew,geo.m) | Method5 (y25,y75) | Method6 (G.skew,M.kurtosis) |
|---|---|---|---|---|---|---|
| Alpha & (SE) | 0.122(0.2223) | 0.1227(0.2236) | 0.1164(0.0462) | 0.0877(0.0793) | 0.1209(0.2077) | - |
| Beta & (SE) | 0.9709(0.2236) | 0.967(0.2236) | 1.0946(0.2186) | 1.4157(0.2085) | 0.9674(0.085) | - |
| Var-cov matrix | 0.9977 0.0005 | 0.9995 0.00014 | 0.0427 -0.2020 | 0.1257 -0.3307 | 0.8631 0.3410 | |
| | 0.0005 0.9996 | 0.00014 0.9998 | -0.2020 0.9557 | -0.3307 0.8696 | 0.3410 0.1446 | |
| SSE | 0.00000099 | 0.0000016 | 0.0186 | 0.0516 | 0.000000013 | - |
| AD | 0.3176 | 0.3202 | 2.1183 | 47.3955 | 0.3177 | - |
| CVM | 0.0415 | 0.0441 | 0.3029 | 4.5083 | 0.0413 | - |
| KS | 0.1293 | 0.1258 | 0.2311 | 0.7608 | 0.1297 | - |
| H0 | Fail to reject | Fail to reject | Fail to reject | reject | Fail to reject | - |
| P(KS) | 0.8498 | 0.8715 | 0.2017 | <0.00001 | 0.8477 | - |
| Galton Skewness | - | - | - | - | - | -0.2941 |
| Moore kurtosis | - | - | - | - | - | 1.4118 |
| Sig.a (p-value) | 0.5462( 0.2956) | 0.5487(0.2948) | 2.5181(0.0105) | 1.1066(0.1411) | 0.5821(0.2837) | - |
| Sig.b ( p-value) | 4.3427 (<0.0001) | 4.3251(<0.0001) | 5.0075(<0.0001) | 6.7892(<0.0001) | 11.3754(<0.0001) | - |
| M1 , M2 | 0.9005 , 0.8148 | - | - | - | - | - |
| M3 , M4 | - | 0.7405 , 0.6758 | - | - | - | - |
| Mean Log(y) | - | - | -0.1073 | -0.1073 | | - |
| Y25 , Y75 | - | - | - | - | 0.8650 , 0.95 | - |
| Sk.Quartiles | - | - | -0.2941 | - | - | - |
| Sk.decile | - | - | - | -0.5294 | - | - |
| Deter | 0.9973 | 0.9994 | - | 1.8*10^-17 | - | - |
| trace | 1.9973 | 1.9994 | 0.9984 | 0.9954 | 1.0078 | - |
| Eigenvalues | 0.9976 , 0.9997 | 0.9995 , 0.9999 | 0 , 0.9984 | 0 , 0.9954 | 0.9992 , 0.0086 | - |
| **Eigenvalues and variance-covariance matrix before applying the delta method** | | | | | | |
| **Eigenvalues** | **0.297 , 2.4361** | **0.1199 , 0.5221** | **1.6397 , 1000** | **4.6194 , 1000** | **0.7991 , 991.439** | - |
| **Var-cov matrix** | **1.1852 -1.0541** | **0.2422 -0.185** | **51.2444 216.9393** | **26.7567 146.7821** | **64.8661 243.6448** | - |
| | **-1.0541 1.5479** | **-0.185 0.3997** | **216.9393 950.3954** | **146.7821 977.8627** | **243.6448 927.372** | |

Table 16: comparisons between methods for voter datasets

| | Method1 (1,2 mom) | | Method2 (3,4mom) | | Method3 (skew,geo.m) | | Method4 (skew,geo.m) | | Method5 (y25,y75) | | Method6 (G.skew,M.kurtosis) | |
|---|---|---|---|---|---|---|---|---|---|---|---|---|
| Alpha & (SE) | 0.3602( 0.1619) | | 0.5572(0.1619) | | 0.1609(0.0339) | | 0.1793(0.0599) | | 0.5492(0.1453) | | 0.6811(0.1254) | |
| Beta & (SE) | 0.6347(0.1621) | | 1.3125(0.1620) | | 0.3292(0.1586) | | 0.3154(0.1507) | | 1.2783(0.0718) | | 1.385(0.1012) | |
| Var-cov matrix | 0.9961 | 0.00058 | 0.9958 | 0.0018 | 0.0436 | -0.2041 | 0.1362 | -0.3429 | 0.8026 | 0.3966 | 0.5973 | -0.4821 |
| | 0.00058 | 0.9985 | 0.0018 | 0.9971 | -0.2041 | 0.9558 | -0.3429 | 0.8632 | 0.3966 | 0.1960 | -0.4821 | 0.3891 |
| SSE | 0.00069 | | 0.000011 | | 0.1141 | | 0.0711 | | 0.00034 | | 0.2647 | |
| AD | 1.1728 | | 1.3581 | | 1.3736 | | 1.8384 | | 1.3492 | | 1.9416 | |
| CVM | 0.1622 | | 0.2275 | | 0.2000 | | 0.2982 | | 0.2252 | | 0.3207 | |
| KS | 0.1071 | | 0.1389 | | 0.1286 | | 0.1571 | | 0.1382 | | 0.1621 | |
| H0 | Fail to reject | | Fail to reject | | Fail to reject | | Fail to reject | | Fail to reject | | Fail to reject | |
| P(KS) | 0.4684 | | 0.4179 | | 0.2905 | | 0.1366 | | 0.4239 | | 0.1179 | |
| Galton Skewness | - | | - | | - | | - | | - | | 0.25 | |
| Moore kurtosis | - | | - | | - | | - | | - | | 1.5625 | |
| Sig.a (p-value) | 2.225 (0.0161) | | 3.4421(<0.0001) | | 4.7531(<0.0001) | | 2.9948(0.0024) | | 3.7793(<0.0001) | | 5.4328(<0.0001) | |
| Sig.b ( p-value) | 3.9154(<0.0001) | | 8.1025(<0.0001) | | 2.0757(0.0225) | | 2.0926(0.0217) | | 17.8014(<0.0001) | | 13.6869(<0.0001) | |
| M1 , M2 | 0.6911 , 0.4928 | | - | | - | | - | | - | | - | |
| M3 , M4 | - | | 0.3615 , 0.272 | | - | | - | | - | | - | |
| Mean Log(y) | - | | - | | -0.3862 | | -0.3862 | | - | | - | |
| Y25 , Y75 | - | | - | | - | | - | | 0.62 , 0.78 | | - | |
| Sk.Quartiles | - | | - | | 0.25 | | - | | - | | - | |
| Sk.decile | - | | - | | - | | 0.1098 | | - | | - | |
| Deter | 0.9945 | | 0.9929 | | 0 | | 0 | | 0 | | 0 | |
| trace | 1.9945 | | 1.9929 | | 0.9994 | | 0.9994 | | 0.9985 | | 0.9864 | |
| Eigenvalues | 0.9959, 0.9986 | | 0.9946 , 0.9984 | | 0 , 0.9994 | | 1*10^-16, 0.9994 | | 0.9985 , 0 | | 0.9864 , 0 | |
| **Eigenvalues and variance-covariance matrix before applying the delta method** | | | | | | | | | | | | |
| **Eigenvalues** | **1.3894 , 4.0747** | | **1.6473 , 5.4333** | | **0.6092 , 1000** | | **0.6077 , 1000** | | **1.4524 , 1000** | | **13.5604 , 1000** | |
| **Var-cov matrix** | **2.3828** | **-1.2964** | **2.2533** | **-1.388** | **444.0168** | **496.515** | **488.7263** | **499.562** | **63.526** | **241.102** | **56.2744** | **200.774** |
| | **-1.2964** | **3.0813** | **-1.388** | **4.8272** | **496.515** | **556.5925** | **499.562** | **511.881** | **241.102** | **937.926** | **200.774** | **957.286** |

Table 17: comparisons between methods for flood datasets

| | Method1 (1,2 mom) | | Method2 (3,4mom) | | Method3 (skew,geo.m) | | Method4 (skew,geo.m) | | Method5 (y25,y75) | | Method6 (G.skew,M.kurtosis) | |
|---|---|---|---|---|---|---|---|---|---|---|---|---|
| Alpha & (SE) | 1.0429(0.2229) | | 1.2265(0.2177) | | 1.0197(0.0465) | | 1.1478(0.0802) | | 1.1999(0.1622) | | 2.2838(0.0683) | |
| Beta & (SE) | 2.4960(0.2228) | | 2.4340(0.2174) | | 2.4066(0.2187) | | 1.2655(0.2086) | | 1.6970(0.1536) | | 1.9944(0.2129) | |
| Var-cov matrix | 0.9933 | 0.005 | 0.9475 | 0.0411 | 0.0432 | -0.2033 | 0.1287 | -0.3347 | 0.5260 | 0.4981 | 0.0933 | 0.2907 |
| | 0.005 | 0.9929 | 0.0411 | 0.9451 | -0.2033 | 0.9566 | -0.3347 | 0.8706 | 0.4981 | 0.4716 | 0.2907 | 0.9065 |
| SSE | 0.00046 | | 0.0001 | | 0.015 | | 0.0676 | | 0.0268 | | 0.5766 | |
| AD | 2.6701 | | 3.9985 | | 2.995 | | 2.472 | | 2.6484 | | 30.5302 | |
| CVM | 0.5078 | | 0.8405 | | 0.5936 | | 0.4523 | | 0.4921 | | 5.0019 | |
| KS | 0.3110 | | 0.3614 | | 0.3405 | | 0.2763 | | 0.2622 | | 0.8172 | |
| H0 | Fail to reject | | reject | | reject | | Fail to reject | | Fail to reject | | reject | |
| P(KS) | 0.0324 | | 0.0014 | | 0.0142 | | 0.0767 | | 0.0313 | | <0.0001 | |
| Galton Skewness | - | | - | | - | | - | | - | | -0.1111 | |
| Moore kurtosis | - | | - | | - | | - | | - | | 2 | |
| Sig.a (p-value) | 4.6799 (<0.0001) | | 5.6352(<0.0001) | | 21.9374(<0.0001) | | 14.3093(<0.0001) | | 7.3986(<0.0001) | | 33.4456(<0.0001) | |
| Sig.b ( p-value) | 11.2022(<0.0001) | | 11.1964(<0.0001) | | 11.0043(<0.0001) | | 6.0654(<0.0001) | | 11.0509(<0.0001) | | 9.3678(<0.0001) | |
| M1 , M2 | 0.4225 , 0.1932 | | - | | - | | - | | - | | - | |
| M3 , M4 | - | | - | | - | | - | | - | | - | |
| Mean Log(y) | - | | - | | -0.8989 | | -0.8989 | | - | | - | |
| Y25 , Y75 | - | | - | | - | | - | | 0.33 , 0.465 | | - | |
| Sk.Quartiles | - | | - | | -0.1111 | | - | | - | | - | |
| Sk.decile | - | | - | | - | | 0.3043 | | - | | - | |
| Deter | 0.9862 | | 0.8938 | | - | | 0 | | 0 | | 0 | |
| trace | 1.9862 | | 1.8926 | | 0.9998 | | 0.9993 | | 0.9977 | | 0.9998 | |
| Eigenvalues | 0.9981 , 0.9881 | | 0.9874 , 0.9052 | | 0 , 0.9998 | | 0 , 0.9993 | | 0.9977 , 0 | | 0 , 0.9998 | |
| **Eigenvalues and variance-covariance matrix before applying the delta method** | | | | | | | | | | | | |
| **Eigenvalues** | **1.9392 , 11.8641** | | **12.6151 , 94.7535** | | **0.2251 , 1000** | | **0.7164 , 1000** | | **2.3493 , 1000** | | **0.2298 , 1000** | |
| **Var-cov matrix** | **3.3599** | **-3.4759** | **27.538** | **-31.671** | **0.2935** | **-8.272** | **16.099** | **-123.02** | **18.636** | **-126.43** | **472.2203** | **-499.11** |
| | **-3.4759** | **10.4434** | **-31.671** | **79.831** | **-8.272** | **999.932** | **-123.02** | **984.617** | **-126.43** | **983.713** | **-499.11** | **528.01** |

Table 18: comparisons between methods for time between failure datasets

| | Method1 (1,2 mom) | | Method2 (3,4mom) | | Method3 (skew,geo.m) | | Method4 (skew,geo.m) | | Method5 (y25,y75) | | Method6 (G.skew,M.kurtosis) | |
|---|---|---|---|---|---|---|---|---|---|---|---|---|
| Alpha & (SE) | 2.8375 (0.2013) | | 4.383(0.1825) | | 1.9614(0.036) | | 1.9605(0.0447) | | 2.1876(0.034) | | 2.6225(0.0425) | |
| Beta & (SE) | 1.3949(0.1483) | | 0.8336(0.0309) | | 1.7439(0.2054) | | 1.7433(0.2036) | | 1.7442(0.2050) | | 1.9015(0.2041) | |
| Var-cov matrix | 0.9317 | 0.1645 | 0.766 | 0.0679 | 0.0298 | -0.1701 | 0.0460 | -0.2096 | 0.0266 | 0.1603 | 0.0416 | 0.1997 |
| | 0.1645 | 0.5061 | 0.0679 | 0.022 | -0.1701 | 0.9701 | -0.2096 | 0.9539 | 0.1603 | 0.9667 | 0.1997 | 0.9583 |
| SSE | 0.000018 | | 0.000004 | | 0.0775 | | 0.0438 | | 0.00029 | | 0.0044 | |
| AD | 1.4962 | | 0.5501 | | 0.6348 | | 0.6381 | | 0.8833 | | 7.0613 | |
| CVM | 0.2473 | | 0.0836 | | 0.1151 | | 0.1160 | | 0.1269 | | 1.4441 | |
| KS | 0.1792 | | 0.1229 | | 0.1523 | | 0.1529 | | 0.1396 | | 0.3724 | |
| H0 | Fail to reject | | Fail to reject | | Fail to reject | | Fail to reject | | Fail to reject | | Reject H0 | |
| P(KS) | 0.1754 | | 0.7767 | | 0.6071 | | 0.6021 | | 0.3776 | | 0.00041 | |
| Galton Skewness | - | | - | | - | | - | | - | | 0.6434 | |
| Moore kurtosis | - | | - | | - | | - | | - | | 2.1549 | |
| Sig.a (p-value) | 14.0978 (<0.0001) | | 24.0169 (<0.0001) | | 54.4636 (<0.0001) | | 43.8195 (<0.0001) | | 61.7045(<0.0001) | | 61.6402(<0.0001) | |
| Sig.b ( p-value) | 9.4036(<0.0001) | | 26.9516(<0.0001) | | 8.4914(<0.0001) | | 8.5602(<0.0001) | | 8.9911(<0.0001) | | 9.316(<0.0001) | |
| M1 , M2 | 0.1578 , 0.0606 | | - | | - | | - | | - | | - | |
| M3 , M4 | - | | 0.03 , 0.0162 | | - | | - | | - | | - | |
| Mean Log(y) | - | | - | | -2.6497 | | -2.6497 | | - | | - | |
| Y25 , Y75 | - | | - | | - | | - | | 0.0292 , 0.21 | | - | |
| Sk.Quartiles | - | | - | | 0.6434 | | - | | - | | - | |
| Sk.decile | - | | - | | - | | 0.7908 | | - | | - | |
| Deter | 0.4445 | | 0.0122 | | 4*10^-18 | | 0 | | 0.000036 | | 0 | |
| trace | 1.4379 | | 0.788 | | 0.9999 | | 0.9999 | | 0.9933 | | 0.9999 | |
| Eigenvalues | 0.9879,0.4499 | | 0.0159,0.7722 | | 0,0.9999 | | 0, 0.999 | | 0.0,0.9933 | | 0, 0.9999 | |
| **Eigenvalues and variance-covariance matrix before applying the delta method** | | | | | | | | | | | | |
| **Eigenvalues** | **12.0779 , 550.0674** | | **227.8278 , 984.15** | | **0.1071 , 1000** | | **0.1056 , 1000** | | **6.749 , 1000** | | **0.1144 , 1000** | |
| **Var-cov matrix** | **426.1776** | **-226.501** | **-70.496** | **-100.69** | **364.763** | **-481.29** | **364.43** | **-481.2** | **419.289** | **-489.45** | **638.788** | **-480.31** |
| | **-226.501** | **135.9676** | **-100.69** | **241.48** | **-481.29** | **635.34** | **-481.2** | **635.67** | **-489.45** | **587.46** | **-480.31** | **361.33** |

Table 19: comparisons between methods for capacity factors datasets.

| | Method1 (1,2 mom) | | Method2 (3,4mom) | | Method3 (skew,geo.m) | | Method4 (skew,geo.m) | | Method5 (y25,y75) | | Method6 (G.skew , .kurtosis) | |
|---|---|---|---|---|---|---|---|---|---|---|---|---|
| Alpha & (SE) | 1.6344(0.1956) | | 2.2742(0.2008) | | 1.8339(0.0385) | | 1.8335(0.0533) | | 2.5559(0.1202) | | 2.0054(0.1332) | |
| Beta & (SE) | 1.2007(0.1469) | | 0.3339(0.0839) | | 1.6527(0.2049) | | 1.6523(0.2016) | | 0.6914(0.1697) | | 1.6699(0.1602) | |
| Var-cov matrix | 0.88 | 0.2251 | 0.9274 | 0.1154 | 0.034 | -0.1813 | 0.0655 | -0.2473 | 0.3323 | 0.4691 | 0.4081 | 0.4908 |
| | 0.2251 | 0.4960 | 0.1154 | 0.1619 | -0.1813 | 0.9658 | -0.2473 | 0.9344 | 0.4691 | 0.6622 | 0.4908 | 0.5904 |
| SSE | 0.002 | | 0.00021 | | 0.1399 | | 0.0696 | | 0.0113 | | 0.1482 | |
| AD | 5.1399 | | 12.7946 | | 1.9005 | | 1.9002 | | 4.2367 | | 2.4234 | |
| CVM | 0.6761 | | 1.267 | | 0.1889 | | 0.1882 | | 0.5768 | | 0.2018 | |
| KS | 0.3455 | | 0.4664 | | 0.1373 | | 0.1373 | | 0.3179 | | 0.1578 | |
| H0 | reject | | reject | | Fail to reject | | Fail to reject | | Reject | | Fail to reject | |
| P(KS) | 0.0059 | | 0.000042 | | 0.3927 | | 0.393 | | 0.0145 | | 0.2707 | |
| Galton Skewness | - | | - | | - | | - | | - | | 0.6592 | |
| Moore kurtosis | - | | - | | - | | - | | - | | 1.0212 | |
| Sig.a (p-value) | 8.3556 (<0.0001) | | 11.3255(<0.0001) | | 47.6792(<0.0001) | | 34.3686(<0.0001) | | 21.2625(<0.0001) | | 15.0557(<0.0001) | |
| Sig.b ( p-value) | 8.1725(<0.0001) | | 3.9796(<0.0001) | | 8.0653(<0.0001) | | 8.1979(<0.0001) | | 4.0747(<0.0001) | | 10.4232(<0.0001) | |
| M1 , M2 | 0.2881 , 0.1798 | | - | | - | | - | | - | | - | |
| M3 , M4 | - | | 0.1307 , 0.1002 | | - | | - | | - | | - | |
| Mean Log(y) | - | | - | | -2.2015 | | -2.2015 | | - | | - | |
| Y25 , Y75 | - | | - | | - | | - | | 0.0308 , 0.531 | | - | |
| Sk.Quartiles | - | | - | | 0.6592 | | - | | - | | - | |
| Sk.decile | - | | - | | - | | 0.7384 | | - | | - | |
| Deter | 0.3862 | | 0.1369 | | 0 | | 6*10^-18 | | 5*10^-17 | | 0 | |
| trace | 1.3765 | | 1.0893 | | 0.9998 | | 0.9998 | | 0.9945 | | 0.9984 | |
| Eigenvalues | 0.9839 , 0.3925 | | 0.1449 , 0.9444 | | 0 , 0.9998 | | 0 , 0.9998 | | 0 , 0.9945 | | 0 , 0.9984 | |
| **Eigenvalues and variance-covariance matrix before applying the delta method** | | | | | | | | | | | | |
| **Eigenvalues** | **16.078 , 607.459** | | **55.5866 , 855.092** | | **0.1588 , 1000** | | **0.1538 , 1000** | | **5.5105 , 1000** | | **1.5686 , 1000** | |
| **Var-cov matrix** | **437.1831** | **-267.776** | **826.53** | **-148.39** | **311.824** | **-463.12** | **311.628** | **-463.04** | **923.699** | **-264.69** | **412.189** | **-491.29** |
| | **-267.776** | **186.354** | **-148.39** | **84.149** | **-463.12** | **688.33** | **-463.04** | **688.525** | **-264.69** | **81.811** | **-491.29** | **589.38** |

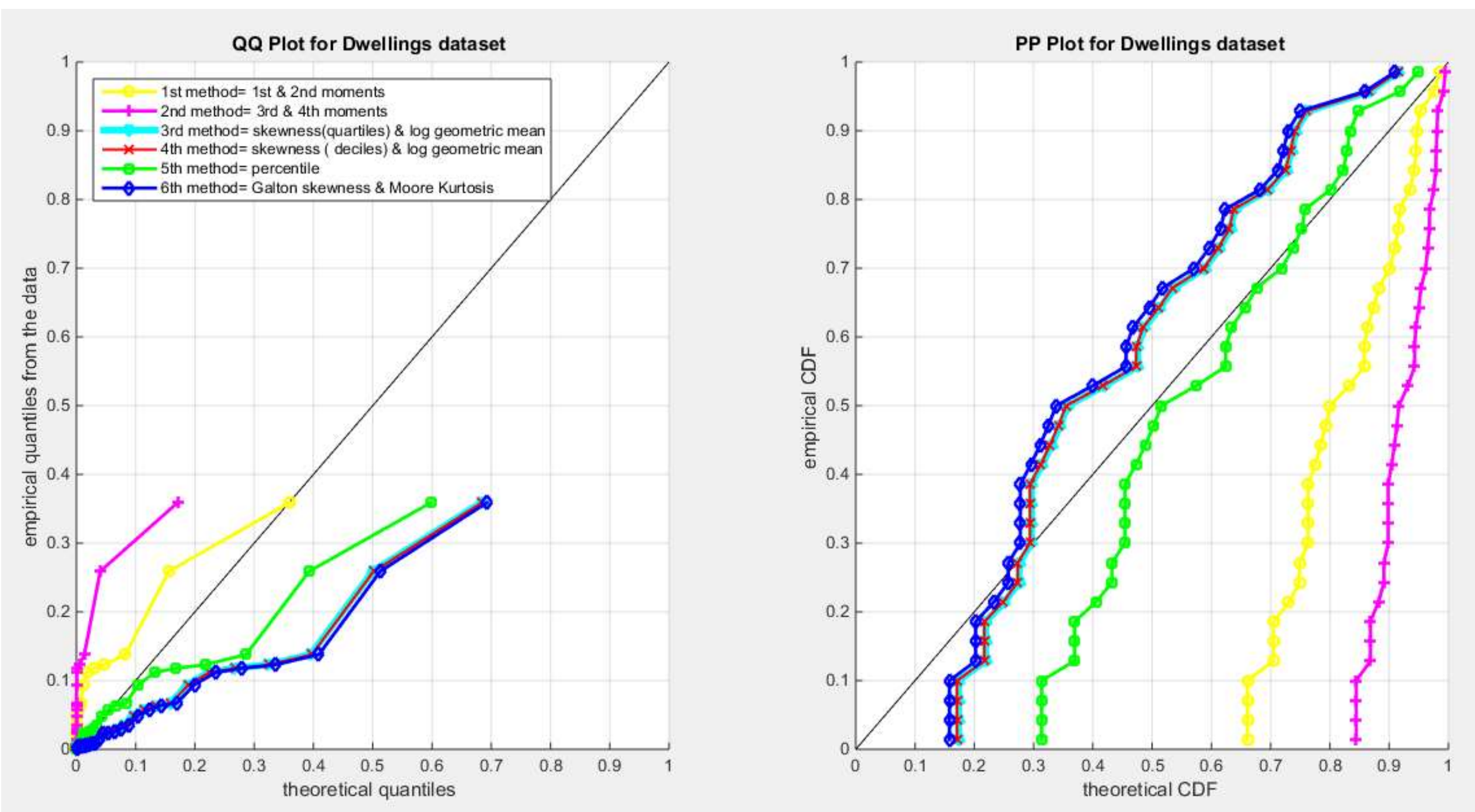

Fig.18 QQ plot and PP plot for dwellings dataset

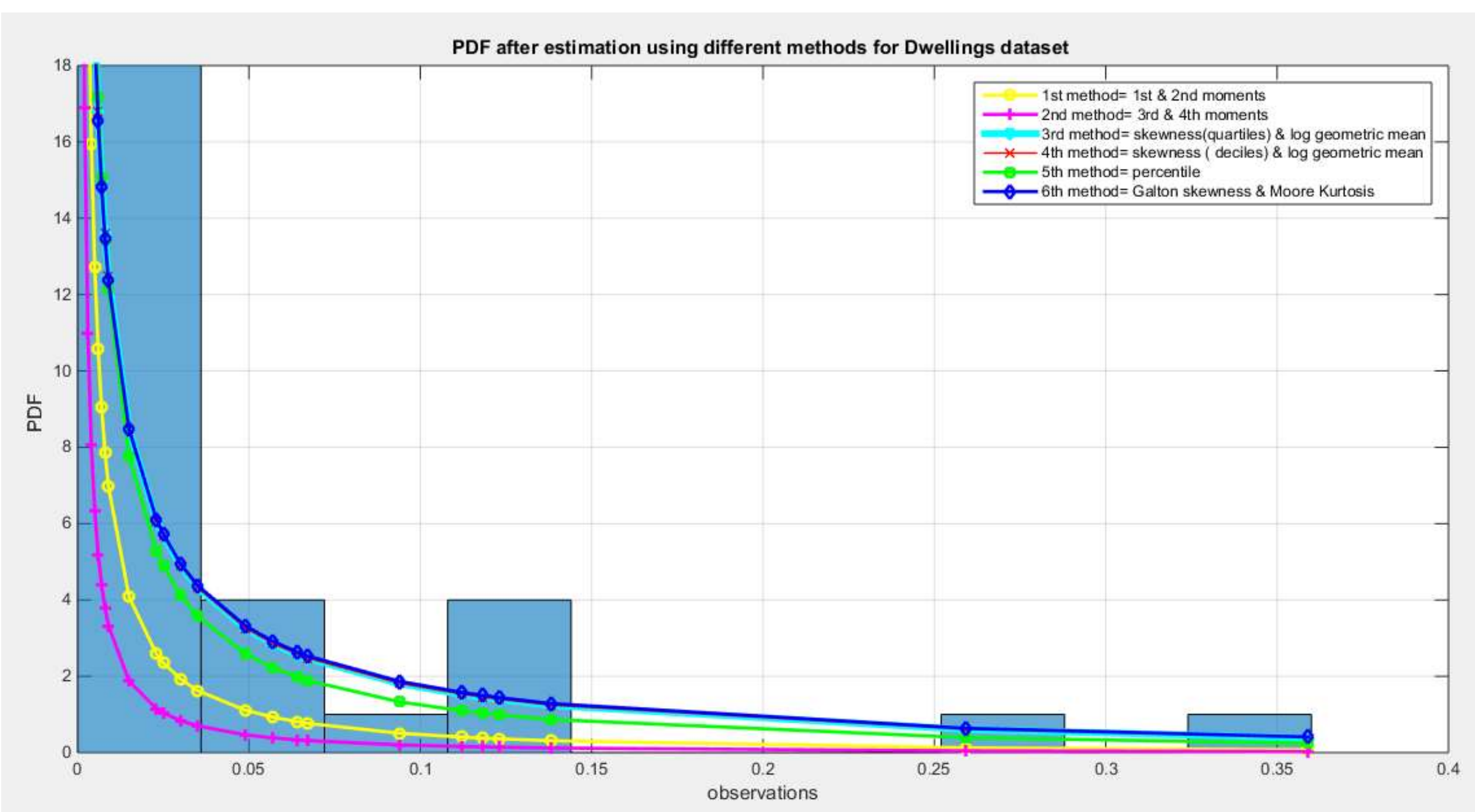

Fig.19 PDF after using different methods of estimation for dwelling dataset

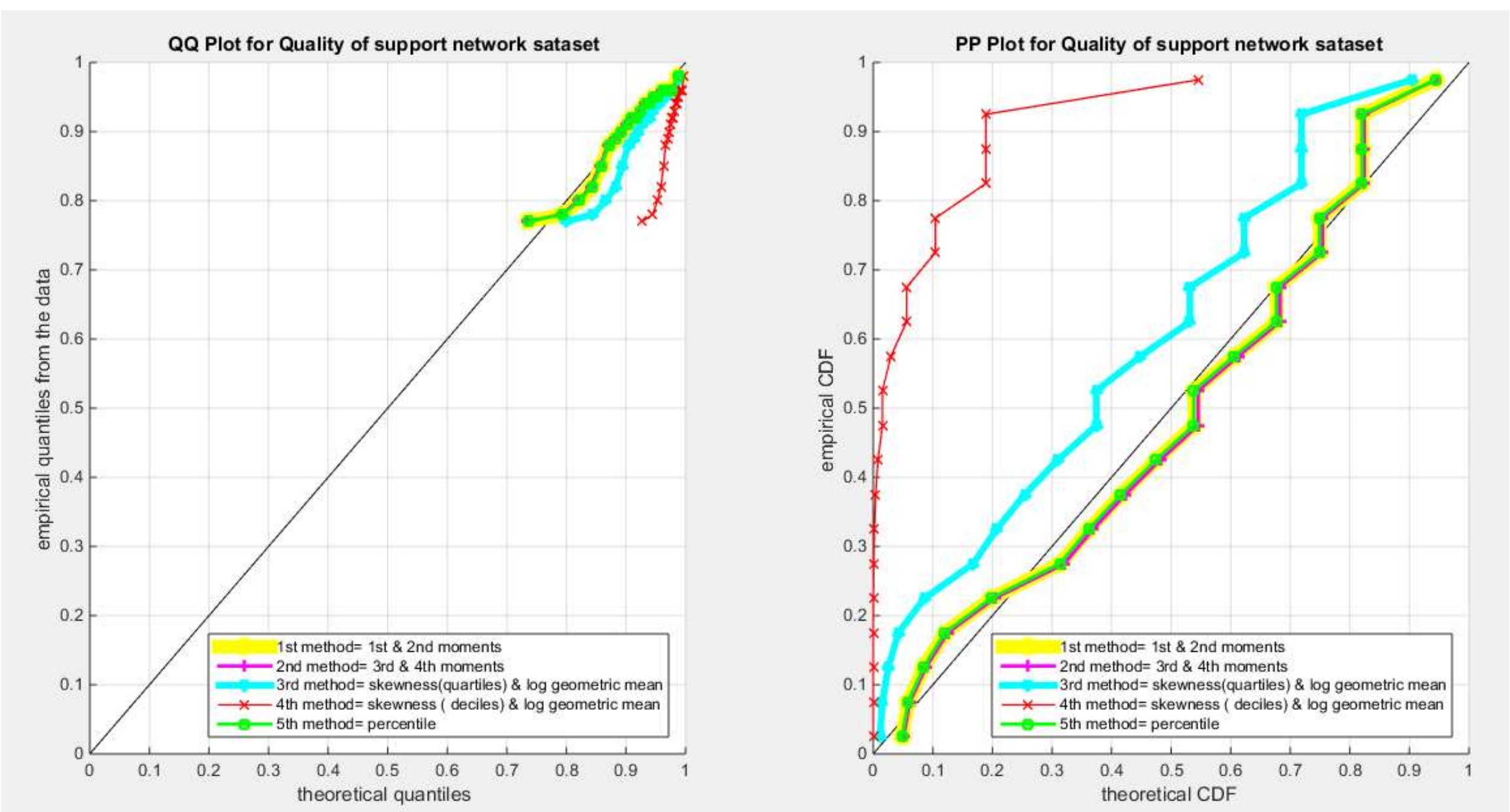

Fig.20 QQ plot and PP plot for Quality of support network dataset

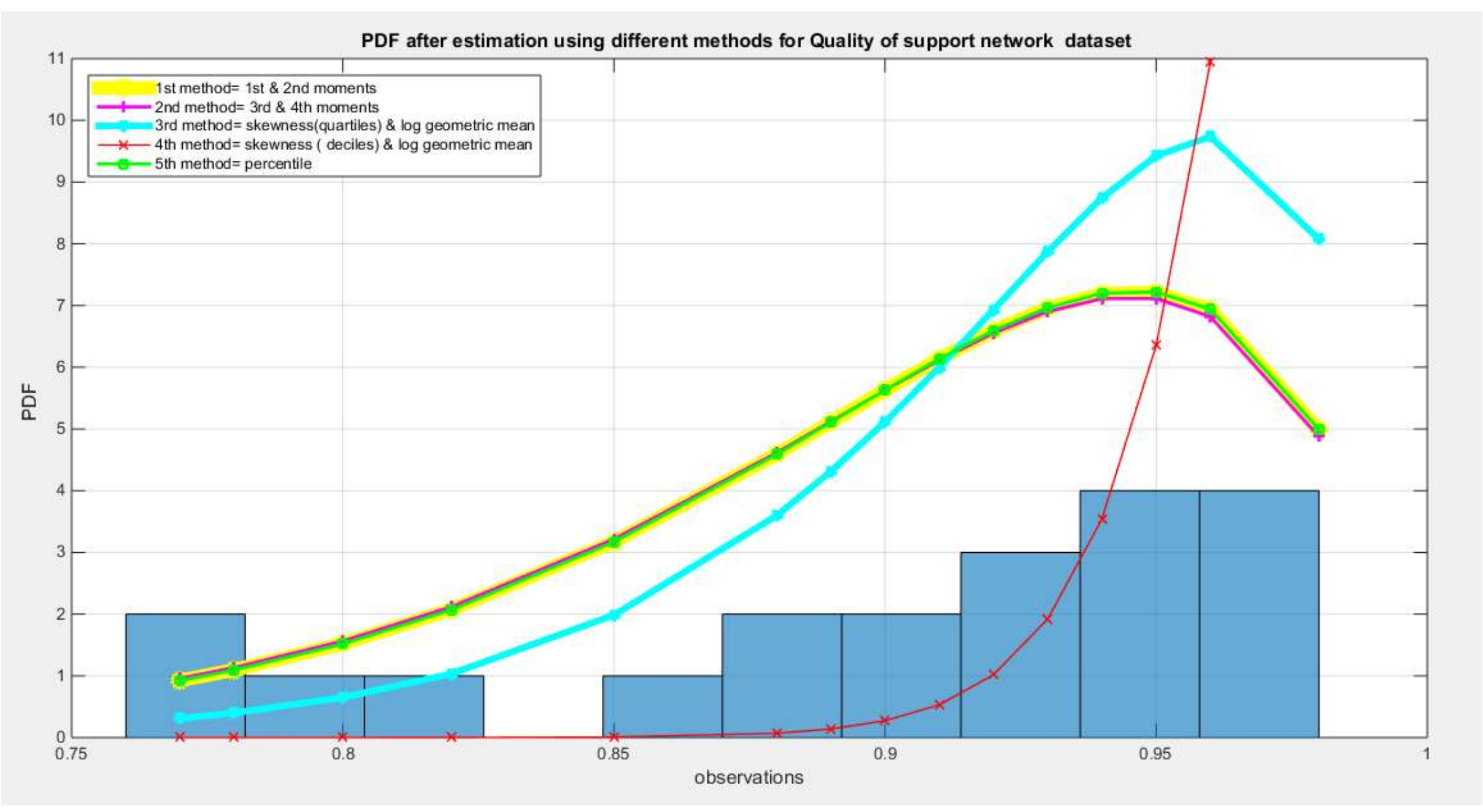

Fig.21 PDF after using different methods of estimation for Quality of support network dataset

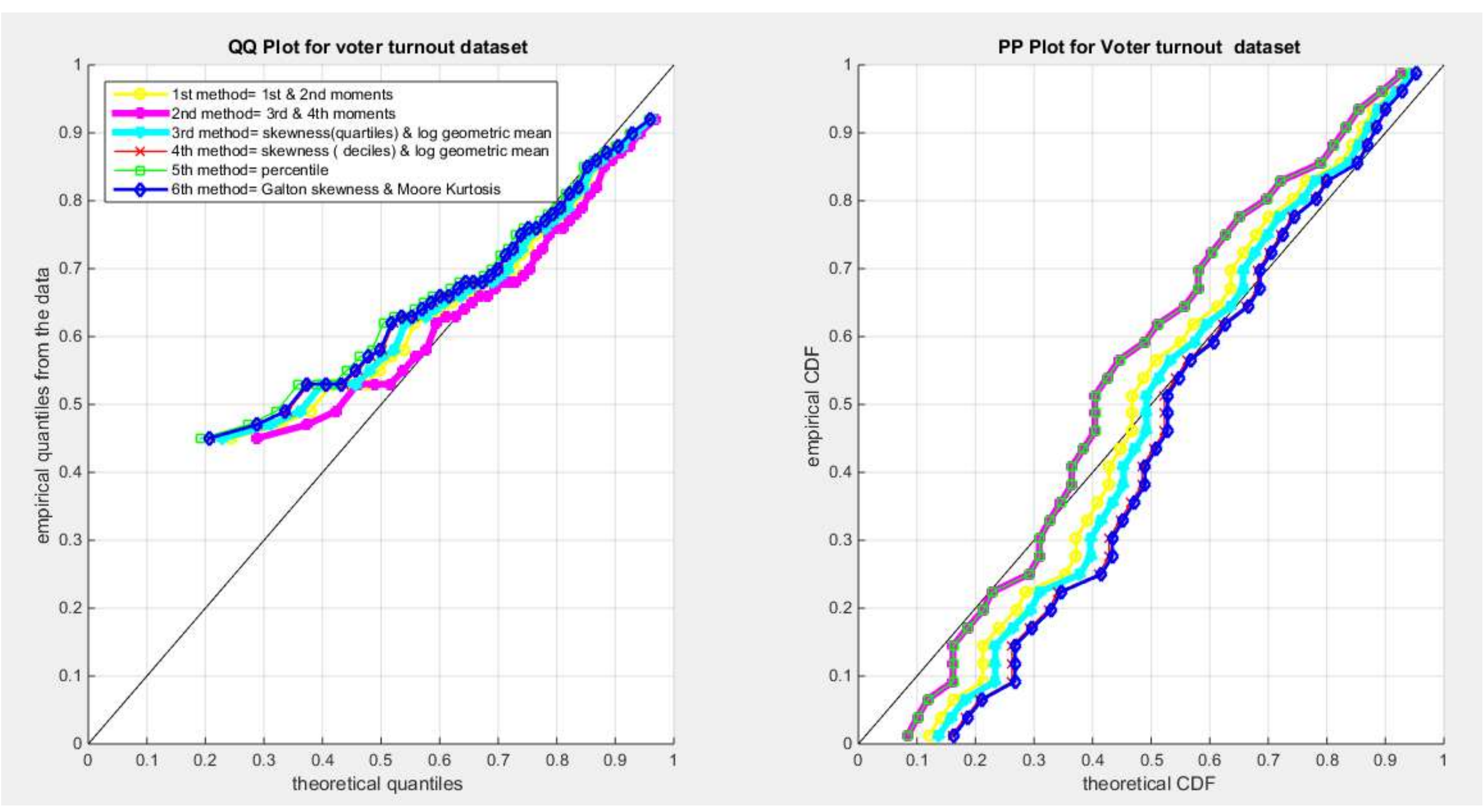

Fig.22 QQ plot and PP plot for voter turnout dataset

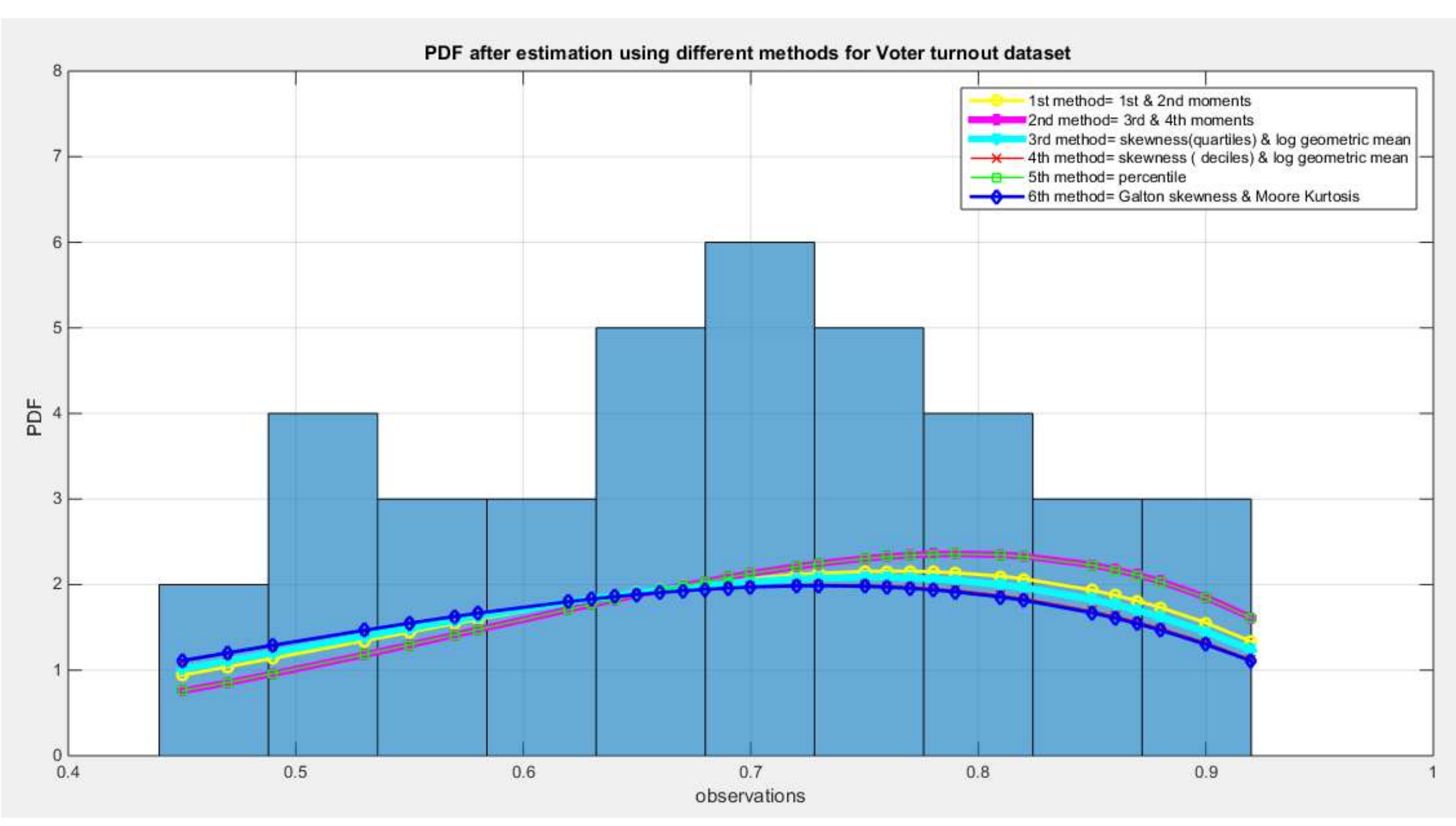

Fig. 23 PDF after using different methods of estimation for voter turnout dataset

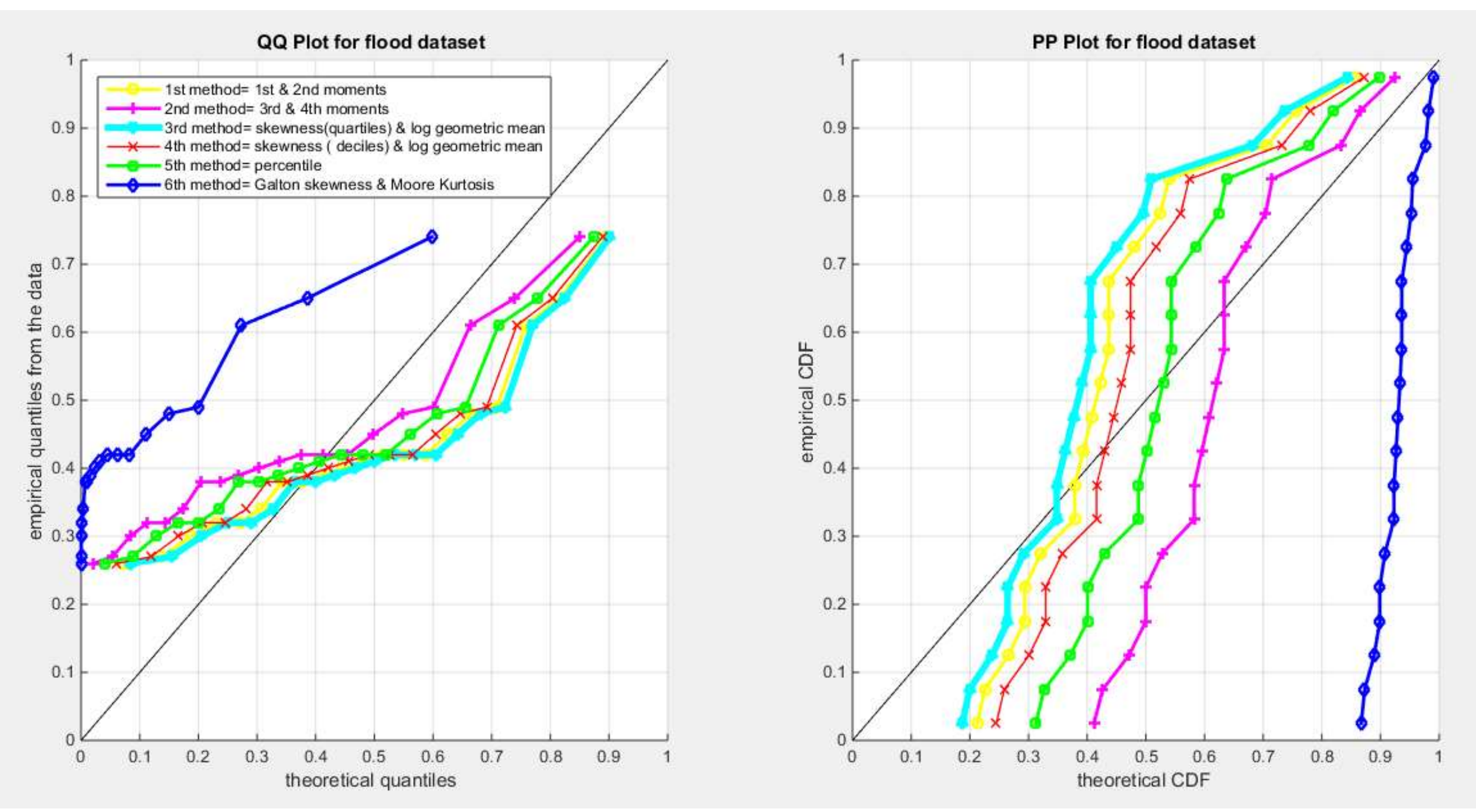

Fig.24 QQ plot and PP plot for flood dataset

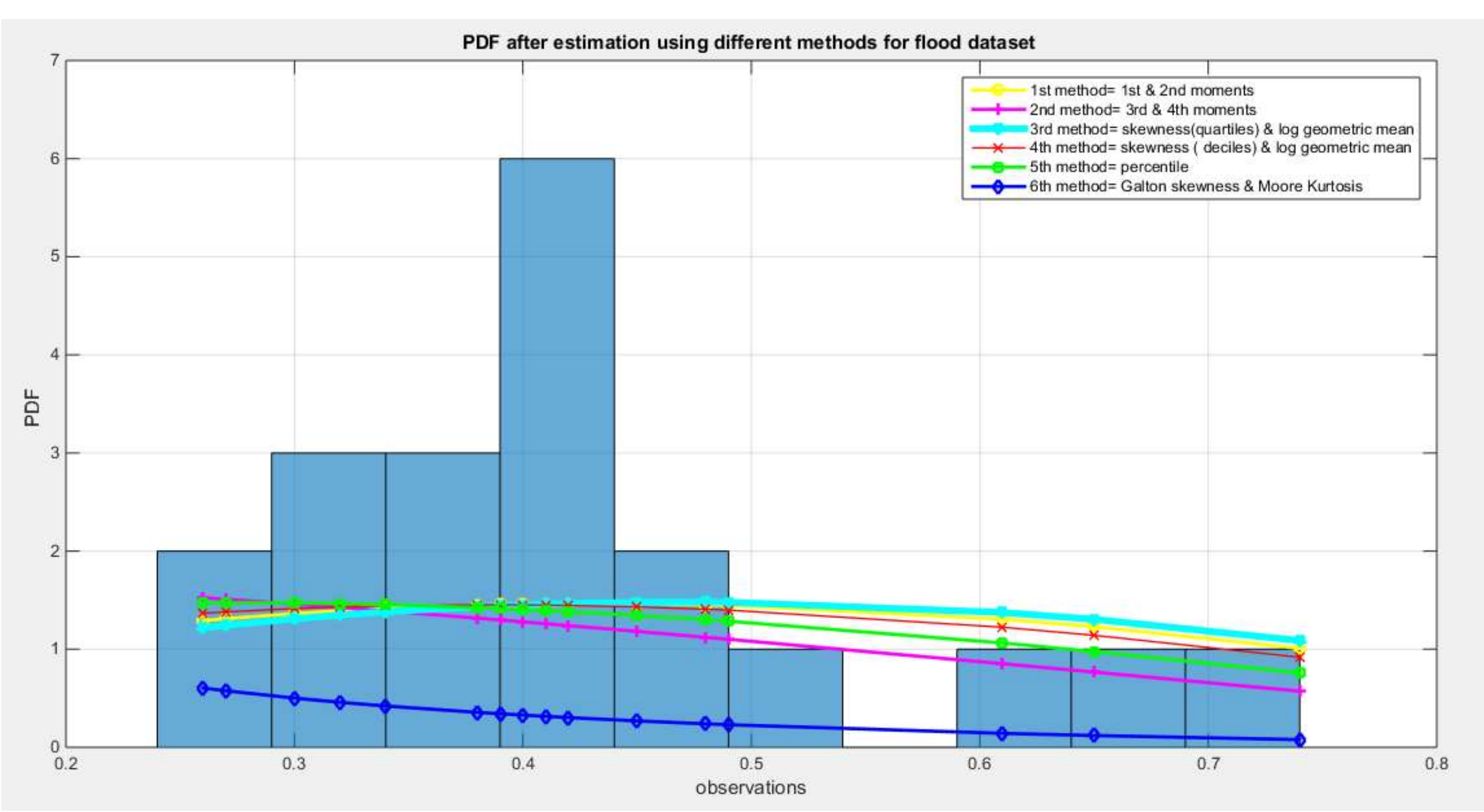

Fig. 25 PDF after using different methods of estimation for flood dataset

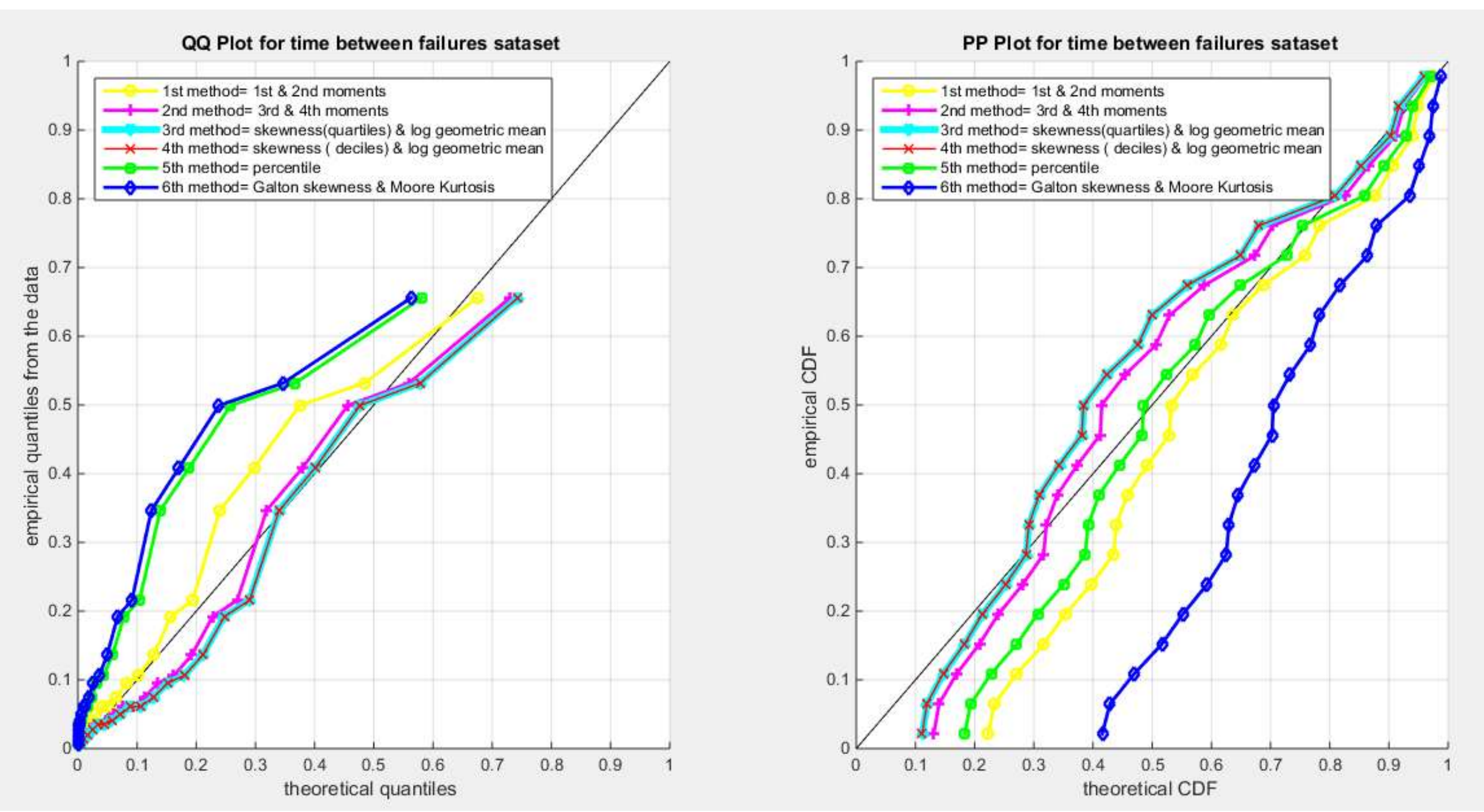

Fig.26 QQ plot and PP plot for time between failures dataset

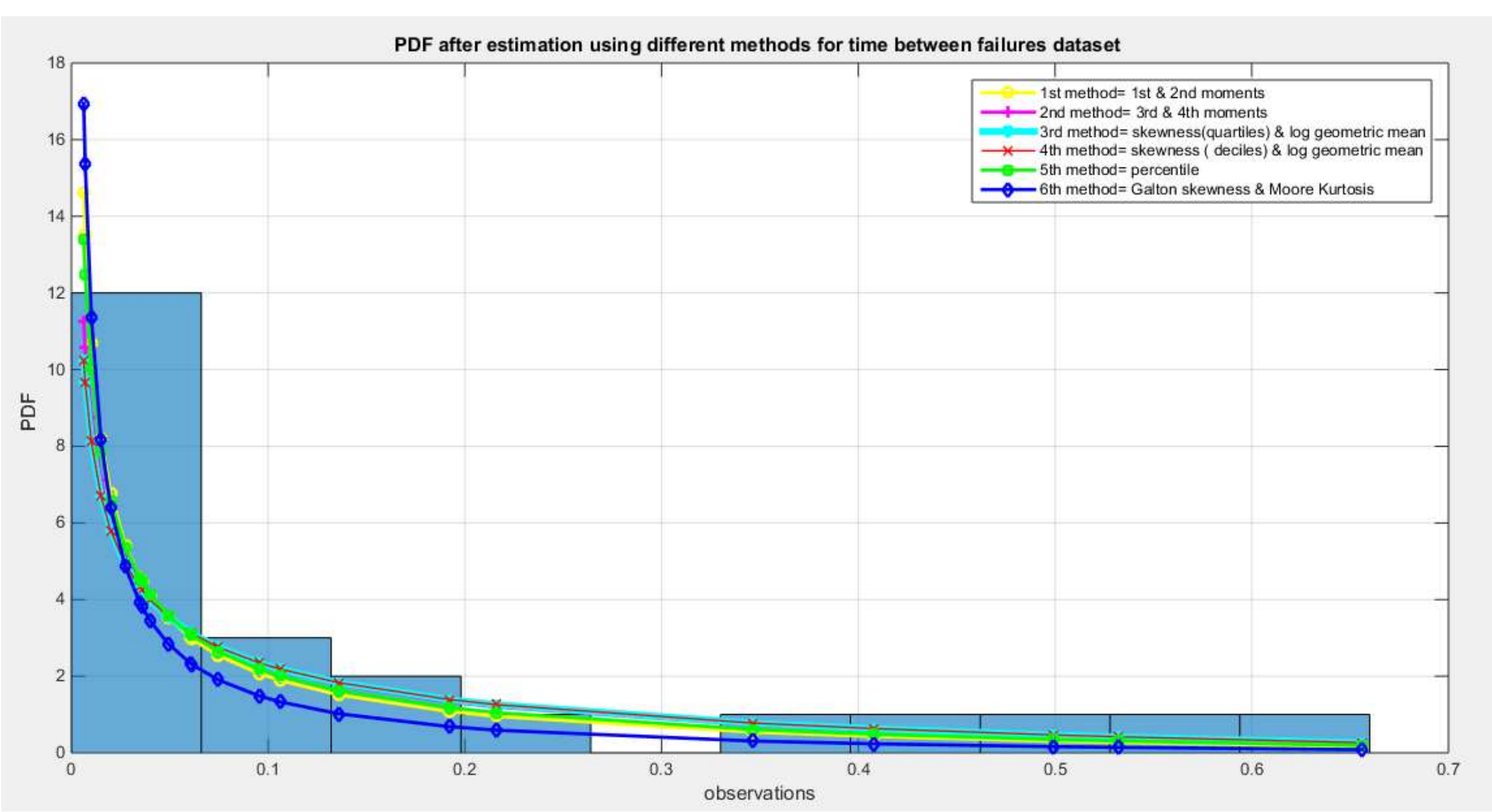

Fig 27 PDF after using different methods of estimation for time between failures dataset

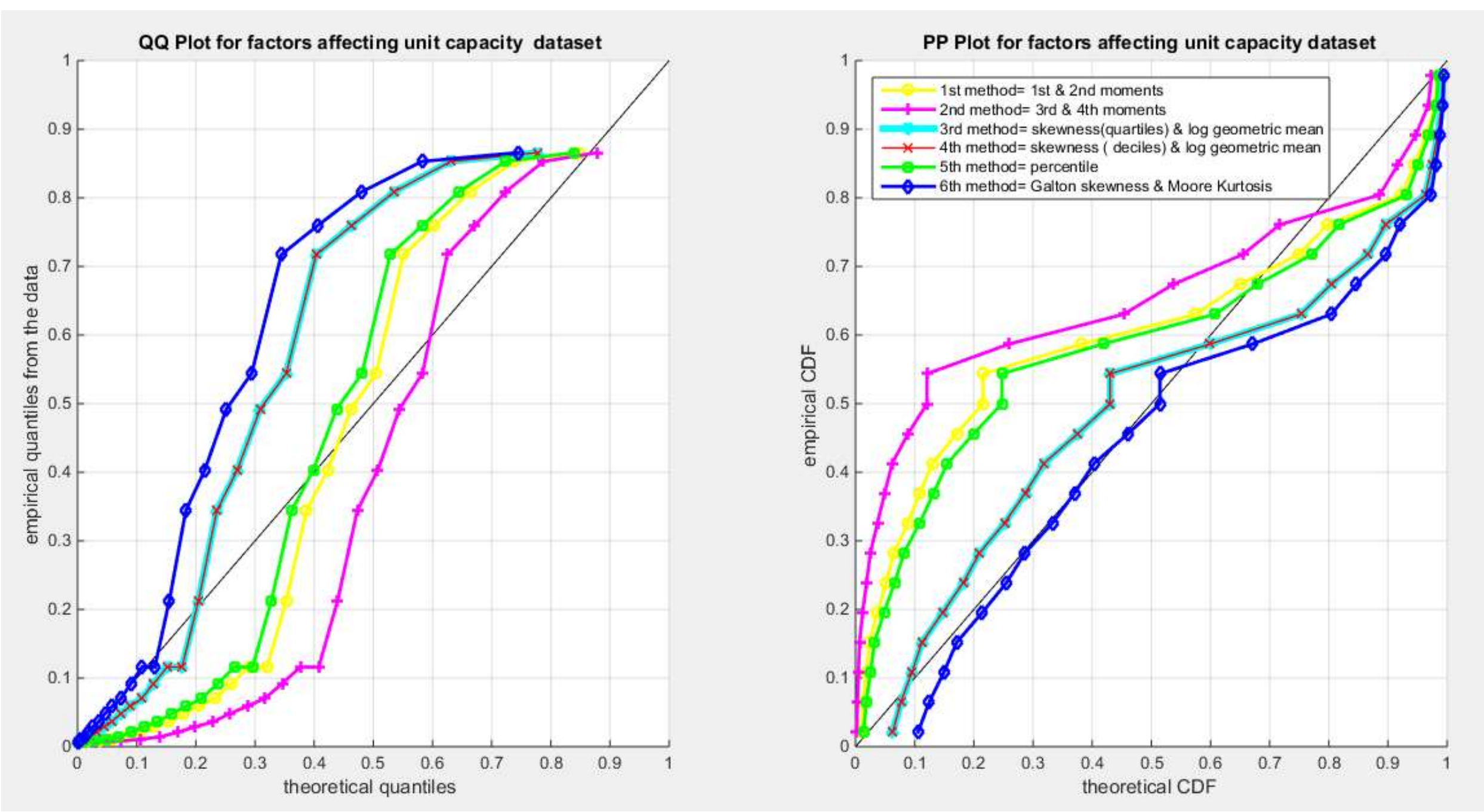

Fig.28 QQ plot and PP plot for unit capacity dataset

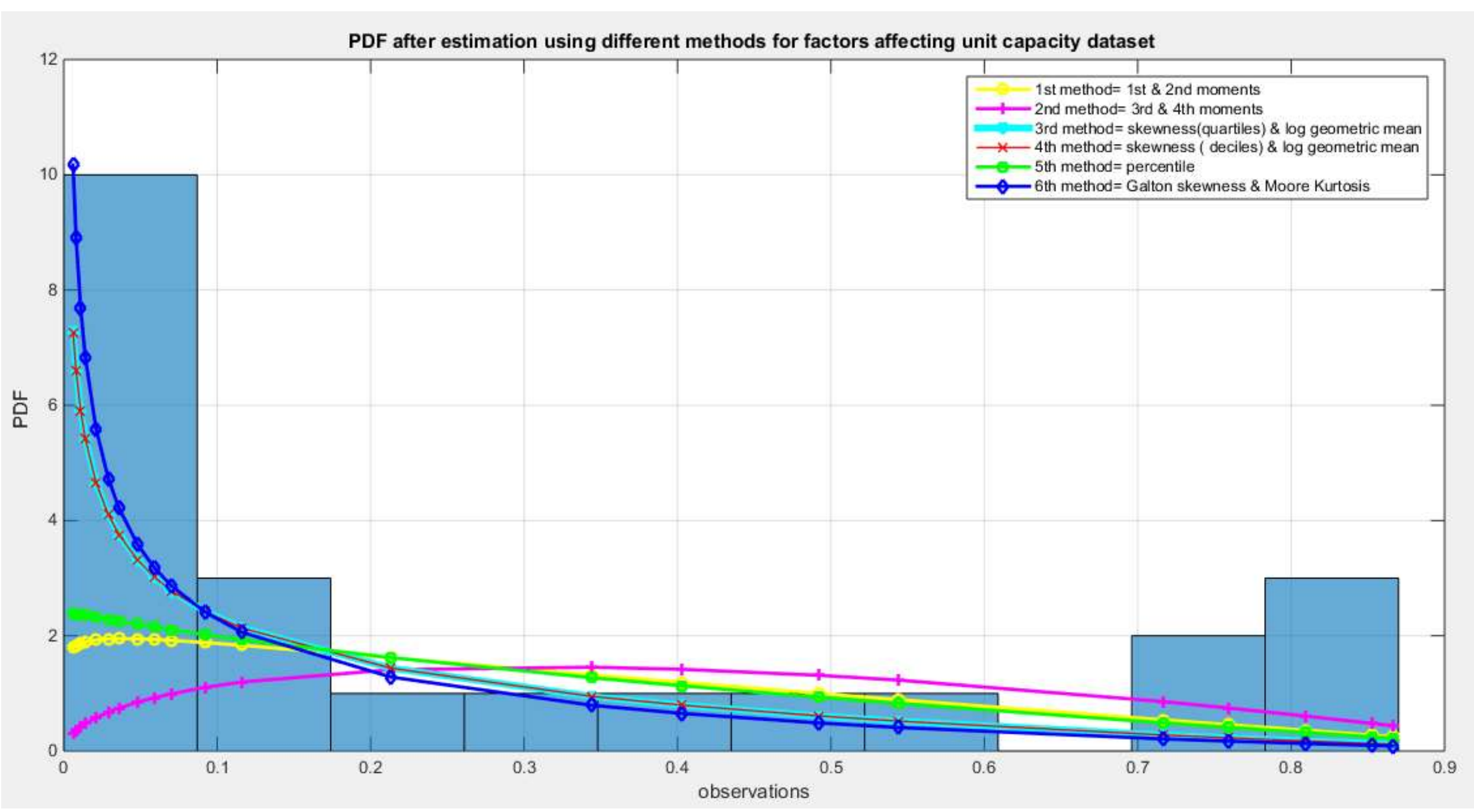

Fig 29 PDF after using different methods of estimation for factors affecting unit capacity dataset

# Discussion

## Section 5

### 5.1. Discussion of the results:

The analysis of the previous data sets indicates that the median-based unit Weibull distribution (MBUW) fits the data well. Although it ranks fourth compared to three other distributions based on comparison tools, the MBUW remains a novel distribution with notable advantages. It is derived from the two-parameter Weibull distribution and features a closed form for both the cumulative distribution function (CDF) and the quantile function. This characteristic makes it applicable in various fields, including reliability studies, time-to-event studies, and regression analysis involving a variable that follows the MBUW distribution in relation to covariates. It is important to emphasize that it has some advantages over the Beta distribution and Kumaraswamy distribution. It has a closed explicit quantile function in comparison with the beta distribution which requires a special function. It has an explicit function for the mean and variance in comparison to Kumaraswamy distribution which requires a special function. These advantages make the distribution a good candidate for a quantile regression model of the response variable with the covariates and for mean-based generalized regression models

In the analysis, maximum likelihood estimation (MLE) was employed. The optimization technique used was the derivative-free Nelder-Mead simplex algorithm. It is important to note that the variances were large across all data sets reaching infinity in some, complicating the estimation of standard errors and rendering them unreliable. Additionally, the variance-covariance matrices displayed near singularity, an issue that could be addressed using a more stable optimization method. This problem may be resolved through modified MLE or alternative estimation methods, such as the generalized method of moments (GMM) and its variants. Modifications to the percentile estimators could also prove beneficial.

The GMM and percentile methods were utilized to analyze the data. The author employed MLE estimators as initial guesses for some of these methods, which resulted in a significant reduction in the variance compared to the MLE method. GMM variants that use the first and second raw moments (method 1) or the third and fourth moments (method 2) demonstrated a good fit, as indicated by a small Kolmogorov-Smirnov (K-S) test value and a significant p-value supporting the model's fitness. Furthermore, the significance of each parameter was notably high in relation to the t-test because the variance obtained by the delta method was significantly reduced. Method 1 displays a good fit for all datasets except the dwellings and the capacity factors datasets. Method 2 reveals a good fit for all datasets except the dwellings, flood and capacity factors datasets. Method 3 illustrates a good fit for all datasets except the flood data set. Method 4 unveils a good fit for all datasets except the quality of support network dataset. Method 5 exhibit a good fit for all datasets except the dwellings and the capacity factors datasets. Method 6 shows a good fit for all datasets except the quality of network, flood and time between failures datasets. This also reflects the impact of the characteristics of the dataset on the method that demonstrates the good fit. The variance covariance matrix was calculated using the delta method and the function used in the delta method is the function to be minimized by the LM algorithm. The determinant of this variance covariance matrix for almost all method was zero. This may direct the use of relative rank method to be the element used for comparison between methods.

## 5.2. Ranking of Methods

### 5.2.1. Relative Rank

The traditional or standard ranks of methods are 1,2…,m. This ranking only shows the order of the methods. It fails to depict the exact positions of methods with one another. (Poudel,K.P & Cao,Q.V.,2013) proposed a new method of ranking, in which each method is given a relative rank, computed to display the relative position of the method. The relative rank is defined in equation 67 as

$$R_i = 1 + \frac{(m-1)(S_i - S_{min})}{(S_{max} - S_{min})} \quad \ldots\ldots\ldots (67)$$

$R_i$ is the relative rank of the method $i$ .

$S_i$ is the goodness of fit statistics obtained by method $i$.

$S_{min}$ is the minimum of the value of $S_i$

$S_{max}$ is the maximum of the value of $S_i$

In this ranking system, the best method has relative rank of 1 and the worst method has relative rank of m. Ranks of remaining methods are expressed as real numbers between 1 and m. Since the magnitude and not only the order of $S_i$ is taken into account, this ranking system should provide more information than the traditional ordinal rank. This system was used by (Sun et al., 2019). (Ogana, 2020) used the system and for each method the relative ranks were summed across the three goodness-of-fit (GOF) statistics to ascertain the best point for the estimator.

the system of relative rank; using the approach proposed by Ogana that is to sum across the 3 GOF statistics; was utilized to order the method for each data set. Tables 20-25 show comparison between methods showing a good fit as proposed by this relative rank system.

Table (20): comparison between methods after fitting the Dwelling dataset

| | | Dwelling dataset | | | Sum across the 3 goodness of fit | Ascending Sort of the sum from min to max | The Corresponding method ranked from the best to the worst |
|---|---|---|---|---|---|---|---|
| method | Number Of method | AD | CVM | KS | | | |
| M1&M2 | 1 | | | | | | |
| M3&,M4 | 2 | | | | | | |
| skew (quartile)&log geo. Mean | 3 | 2.4095 | 0.4018 | 0.1804 | | | |
| skew(decile) &log geo.Mean | 4 | 2.4479 | 0.4115 | 0.1821 | | | |
| P25 &P 75 | 5 | | | | | | |
| Galton & Moore | 6 | 2.73 | 0.4811 | 0.1933 | | | |
| MLE | 7 | 2.3889 | 0.3966 | 0.1794 | | | |
| | | | | | | | |
| Min of the goodness of fit | | 2.3889 | 0.3966 | 0.1794 | | | |
| Max of the goodness of fit | | 2.73 | 0.4811 | 0.1933 | | | |
| $R_i = 1 + \frac{(m-1)(S_i - S_{min})}{(S_{max} - S_{min})}$<br>Where :<br>$m = number\ of\ methods$<br>$S_i\ is\ the\ value\ of\ goodness\ of\ fit\ of\ the\ method$<br>$S_{min}\ is\ the\ minimum\ value\ of\ S_i$<br>$S_{max}\ is\ the\ maximum\ value\ of\ S_i$ | | | | | | | |
| | | | | | | | |
| | $R_{(3)}$ | 1.181179 | 1.184615 | 1.215827 | 3.581621 | 3 | Method 7 |
| | $R_{(4)}$ | 1.518909 | 1.528994 | 1.582734 | 4.630637 | 3.581621 | Method 3 |
| | | | | | | 4.630637 | Method 4 |
| | $R_{(6)}$ | 4 | 4 | 4 | 12 | 12 | Method 6 |
| | $R_{(7)}$ | 1 | 1 | 1 | 3 | | |

For each method showing a good fit calculate the AD, CVM and KS values. For each of these tests, evaluate the minimum and the maximum value. Then apply the formula in equation 67. Sum the results obtained from this equation for each method. Sort the results from the smallest to the largest which corresponds to the sorting from the best to the worst as illustrated for each dataset in each table.

Table (21): comparison between methods after fitting quality support dataset:

| | | Quality support | | | Sum across the 3 goodness of fit | Ascending Sort of the sum from min to max | The Corresponding method ranked from the best to the worst |
|---|---|---|---|---|---|---|---|
| method | Number Of method | AD | CVM | KS | | | |
| M1&M2 | 1 | 0.3176 | 0.0415 | 0.1293 | | | |
| M3&,M4 | 2 | 0.3202 | 0.0441 | 0.1258 | | | |
| skew(quartile)&log geo mean | 3 | 2.1183 | 0.3029 | 0.2311 | | | |
| skew(decile) &log geo.Mean | 4 | | | | | | |
| P25 &P 75 | 5 | 0.3177 | 0.0413 | 0.1297 | | | |
| Galton & Moore | 6 | | | | | | |
| MLE | 7 | 0.3184 | 0.0407 | 0.1309 | | | |
| | | | | | | | |
| Min of the goodness of fit | | 0.3176 | 0.0407 | 0.1258 | | | |
| Max of the goodness of fit | | 2.1183 | 0.3029 | 0.2311 | | | |
| $R_i = 1 + \frac{(m-1)(S_i - S_{min})}{(S_{max} - S_{min})}$<br>Where :<br>$m = number\ of\ methods$<br>$S_i\ is\ the\ value\ of\ goodness\ of\ fit\ of\ the\ method$<br>$S_{min}\ is\ the\ minimum\ value\ of\ S_i$<br>$S_{max}\ is\ the\ maximum\ value\ of\ S_i$ | R(1) | 1 | 1.012204 | 1.132953 | 3.145158 | | |
| | R (2) | 1.0057755 | 1.051869 | 1 | 3.057644 | | |
| | R (3) | 5 | 5 | 5 | 15 | 3.057644 | Method2 |
| | | | | | | 3.145158 | Method 1 |
| | R (5) | 1.0002221 | 1.009153 | 1.148148 | 3.157524 | 3.157524 | Method 5 |
| | R (6) | | | | | 3.195509 | Method 7 |
| | R (7) | 1.0017771 | 1 | 1.193732 | 3.195509 | 15 | Method 3 |

Table (23): comparison between methods after fitting the voter dataset:

| | | Voter | | | Sum across the 3 goodness of fit | Ascending Sort of the sum from min to max | The Corresponding method ranked from the best to the worst |
|---|---|---|---|---|---|---|---|
| method | Number Of method | AD | CVM | KS | | | |
| M1&M2 | 1 | 1.1728 | 0.1622 | 0.1071 | | | |
| M3&,M4 | 2 | 1.3581 | 0.2275 | 0.1389 | | | |
| skew (quartile)&log geo. Mean | 3 | 1.3736 | 0.2000 | 0.1286 | | | |
| skew(decile) &log geo.Mean | 4 | 1.8384 | 0.2982 | 0.1571 | | | |
| P25 &P 75 | 5 | 1.3492 | 0.2252 | 0.1382 | | | |
| Galton & Moore | 6 | 1.9416 | 0.3207 | 0.1621 | | | |
| MLE | 7 | 1.3262 | 0.2193 | 0.1364 | | | |
| | | | | | | | |
| Min of the goodness of fit | | 1.3262 | 0.2193 | 0.1364 | | | |
| Max of the goodness of fit | | 49.8776 | 8.3129 | 0.816 | | | |
| $R_i = 1 + \frac{(m-1)(S_i - S_{min})}{(S_{max} - S_{min})}$<br>Where :<br>$m = number\ of\ methods$<br>$S_i\ is\ the\ value\ of\ goodness\ of\ fit\ of\ the\ method$<br>$S_{min}\ is\ the\ minimum\ value\ of\ S_i$<br>$S_{max}\ is\ the\ maximum\ value\ of\ S_i$ | $R_{(1)}$ | 1 | 1 | 1 | 3 | 3 | Method 1 |
| | $R_{(2)}$ | 2.44615 | 3.471924 | 4.469091 | 10.38717 | 8.343487 | Method 3 |
| | $R_{(3)}$ | 2.567118 | 2.430915 | 3.345455 | 8.343487 | 9.555068 | Method 7 |
| | $R_{(4)}$ | 6.194589 | 6.148265 | 6.454545 | 18.7974 | 10.15428 | Method 5 |
| | $R_{(5)}$ | 2.376691 | 3.384858 | 4.392727 | 10.15428 | 10.38717 | Method 2 |
| | $R_{(6)}$ | 7 | 7 | 7 | 21 | 18.7974 | method 4 |
| | $R_{(7)}$ | 2.19719 | 3.161514 | 4.196364 | 9.555068 | 21 | method 6 |

Table (24): comparison between methods after fitting flood dataset:

| | | flood | | | Sum across the 3 goodness of fit | Ascending Sort of the sum from min to max | The Corresponding method ranked from the best to the worst |
|---|---|---|---|---|---|---|---|
| method | Number of method | AD | CVM | KS | | | |
| M1&M2 | 1 | 2.6701 | 0.5078 | 0.3110 | | | |
| M3&,M4 | 2 | | | | | | |
| skew (quartile)&log geo. Mean | 3 | | | | | | |
| skew(decile) &log geo.Mean | 4 | 2.472 | 0.4523 | 0.2763 | | | |
| P25 &P 75 | 5 | 2.6484 | 0.4921 | 0.2622 | | | |
| Galton & Moore | 6 | | | | | | |
| MLE | 7 | 2.7563 | 0.531 | 0.3202 | | | |
| | | | | | | | |
| Min of the goodness of fit | | 2.4566 | 0.4474 | 0.2694 | | | |
| Max of the goodness of fit | | 2.7563 | 0.531 | 0.3202 | | | |
| $R_i = 1 + \frac{(m-1)(S_i - S_{min})}{(S_{max} - S_{min})}$<br>Where :<br>$m = number\ of\ methods$<br>$S_i$ *is the value of goodness of fit of the method*<br>$S_{min}$ *is the minimum value of* $S_i$<br>$S_{max}$ *is the maximum value of* $S_i$ | R(1) | 3.090397 | 3.115629 | 3.524138 | 9.730164 | | |
| | | | | | | | |
| | | | | | | | |
| | R (4) | 1 | 1 | 1.72931 | 3.72931 | 3.72931 | Method 4 |
| | R (5) | 2.861414 | 2.517154 | 1 | 6.378568 | 6.378568 | Method 5 |
| | | | | | | 9.730164 | Method 1 |
| | R (7) | 4 | 4 | 4 | 12 | 12 | Method 7 |

Table (25): comparison between methods after fitting the time between failure data:

| | | time between failure | | | Sum across the 3 goodness of fit | Ascending Sort of the sum from min to max | The Corresponding method ranked from the best to the worst |
|---|---|---|---|---|---|---|---|
| method | Number Of method | AD | CVM | KS | | | |
| M1&M2 | 1 | 1.4962 | 0.2473 | 0.1792 | | | |
| M3&,M4 | 2 | 0.5501 | 0.0836 | 0.1229 | | | |
| skew (quartile)&log geo. Mean | 3 | 0.6348 | 0.1151 | 0.1523 | | | |
| skew(decile) &log geo.Mean | 4 | 0.6381 | 0.1160 | 0.1529 | | | |
| P25 &P 75 | 5 | 0.8833 | 0.1269 | 0.1396 | | | |
| Galton & Moore | 6 | | | | | | |
| MLE | 7 | 0.6703 | 0.1253 | 0.1584 | | | |
| | | | | | | | |
| Min of the goodness of fit | | 0.6566 | 0.1214 | 0.1562 | | | |
| Max of the goodness of fit | | 1.8605 | 0.3943 | 0.2471 | | | |
| $R_i = 1 + \frac{(m-1)(S_i - S_{min})}{(S_{max} - S_{min})}$ | R(1) | 6 | 6 | 6 | 18 | | |
| Where : | R (2) | 1 | 1 | 1 | 3 | 3 | Method 2 |
| $m = number\ of\ methods$ | R (3) | 1.447627 | 1.962126 | 3.611012 | 7.020765 | 7.020765 | Method 3 |
| $S_i\ is\ the\ value\ of\ goodness\ of\ fit\ of\ the\ method$ | R (4) | 1.465067 | 1.989615 | 3.664298 | 7.118981 | 7.118981 | Method 4 |
| $S_{min}\ is\ the\ minimum\ value\ of\ S_i$ | R (5) | 2.760913 | 2.322541 | 2.483126 | 7.566581 | 7.566581 | Method 5 |
| | | | | | | 8.061664 | Method 7 |
| $S_{max}\ is\ the\ maximum\ value\ of\ S_i$ | R (7) | 1.635239 | 2.273671 | 4.152753 | 8.061664 | 18 | Method 1 |

Table (26): Comparison between methods after fitting the unit capacity factors:

| | | unit capacity factor | | | Sum across the 3 goodness of fit | Ascending Sort of the sum from min to max | The Corresponding method ranked from the best to the worst |
|---|---|---|---|---|---|---|---|
| method | Number of method | AD | CVM | KS | | | |
| M1&M2 | 1 | | | | | | |
| M3&,M4 | 2 | | | | | | |
| skew (quartile)&log geo. Mean | 3 | 1.9005 | 0.1889 | 0.1373 | | | |
| skew(decile) &log geo.Mean | 4 | 1.9002 | 0.1882 | 0.1373 | | | |
| P25 &P 75 | 5 | | | | | | |
| Galton & Moore | 6 | 2.4234 | 0.2018 | 0.1578 | | | |
| MLE | 7 | 0.6703 | 0.1253 | 0.1518 | | | |
| | | | | | | | |
| Min of the goodness of fit | | 0.6703 | 0.1253 | 0.1401 | | | |
| Max of the goodness of fit | | 2.4258 | 0.2786 | 0.2046 | | | |
| $R_i = 1 + \frac{(m-1)(S_i - S_{min})}{(S_{max} - S_{min})}$<br>Where :<br>$m = number\ of\ methods$<br>$S_i\ is\ the\ value\ of\ goodness\ of\ fit\ of\ the\ method$<br>$S_{min}\ is\ the\ minimum\ value\ of\ S_i$<br>$S_{max}\ is\ the\ maximum\ value\ of\ S_i$ | | | | | | | |
| | | | | | | | |
| | $R_{(3)}$ | 3.105185 | 3.494118 | 1 | 7.599303 | 5.121951 | Method 7 |
| | $R_{(4)}$ | 3.104672 | 3.466667 | 1 | 7.571338 | 7.571338 | Method 4 |
| | | | | | | 7.599303 | Method 3 |
| | $R_{(6)}$ | 4 | 4 | 4 | 12 | 12 | Method 6 |
| | $R_{(7)}$ | 1 | 1 | 3.121951 | 5.121951 | | |

### 5.2.2. Delta Method

Using the delta method (Rao.1973) to estimate the variance covariance of each method encountered a lot of problems. The variance of each method is defined according to the delta method as $H_g(\theta)VH_g'(\theta)$. V matrix is the inverse of the Hessian matrix obtained from the LM algorithm used to estimate the parameters, and the H matrix is the partial derivatives of the parameters within the equations used in each method or the Jacobian used in the LM algorithm. Some methods, using the delta approach yielded estimators with variance –covariance matrix whose determinant is zero. This needs to search for functions other than the equations used in the system or to account for the dependency between these functions within the method. This could be a topic for future research.

The variance-covarince matrix for each method with the determinant, trace and eigenvalues for each matrix are illustrated in Tables 14-19.

# Section 6:

## Conclusion:

This new distribution effectively addresses the limitations of the Weibull distribution, particularly its absence of bathtub and unimodal shapes. Furthermore, it is defined over the unit interval, making it ideal for fitting proportions and ratios. With a well-defined quantile function, this distribution is fully compatible with parametric quantile regression, enabling conditioning on the median or any other quantile. This is a significant advantage, as the mean is often an inadequate measure of central tendency for highly skewed distributions. The closed form expression of its mean and variance makes it a good candidate for mean based regression model if needed.

The innovative approach utilizing the GMM method and percentile estimator significantly enhances the parameter estimation process for the MBUW distribution. It computes both parameters simultaneously by solving differential equations using iterative techniques such as the Levenberg-Marquardt algorithm. It provides more stable and reliable estimates with reduced variance. These estimators serve as robust initial guesses for other methods requiring numerical computations.

## Future work:

Estimating the parameters of the Weibull distribution can be challenging, especially when both parameters are unknown. One effective approach is to use Bayesian statistics. Alternative methods include modified maximum likelihood estimation and a variant of the Generalized method of moments (GMMs) that incorporates kurtosis and the log of the geometric mean, or the Fisher coefficient of skewness and kurtosis, which the author did not utilize. Other techniques that can be explored include least squares, weighted least squares, and the maximum product of spacing method.

**Declarations:**

**Ethics approval and consent to participate**

Not applicable.

**Consent for publication**

Not applicable

**Availability of data and material**

Not applicable. Data sharing does not apply to this article as no datasets were generated or analyzed during the current study.

**Competing interests**

The author declares no competing interests of any type.

**Funding**

No funding resources. No funding roles in the design of the study and collection, analysis, and interpretation of data and in writing the manuscript are declared.

**Authors' contribution**

AI carried the conceptualization by formulating the goals, and aims of the research article, formal analysis by applying the statistical, mathematical, and computational techniques to synthesize and analyze the hypothetical data, carried the methodology by creating the model, software programming and implementation, supervision, writing, drafting, editing, preparation, and creation of the presenting work.

**Acknowledgment**

Not applicable